\documentclass[12pt,tightenlines,eqsecnum,floats,showpacs,nofootinbib,amsmath,amssymb,aps,prd]{revtex4}

  \usepackage{natbib}
  \usepackage{rotating}
  \usepackage{amsmath,amssymb}
 \usepackage{amsthm}
  \usepackage{graphicx}
  \usepackage{verbatim}
  \usepackage[utf8]{inputenc}
  \usepackage{lmodern}
  \usepackage[english]{babel}
  \usepackage{makeidx}
   
\usepackage{multirow}

\def\be{\nopagebreak[3]\begin{equation}}
\def\ee{\end{equation}}
\def\ba{\nopagebreak[3]\begin{eqnarray}}
\def\ea{\end{eqnarray}}

\def\SU(2){\rm{SU(2)}}
\def\su(2){\rm {su(2)}}
\def\SL(2,C){\rm {SL(2,C)}}
\def\U(1){\rm U(1)}
\def\Ar{\rm Ar}

\def\Ab{\bar{\cal A}}

\def\ub{\underbar}

\def\LA{\ub{A}}

\def\lp{\ell_{\rm Pl}}

\def\Diff{\rm Diff}

\def\rcr{\rho_{\rm max}}

\def\R{\cal R}

\def\Al{\mathfrak{A}}

\def\S{{\cal S}}

\def\gps{\mathbf{\Gamma}}
\def\Hk{\mathcal{H}_{\rm kin}}
\def\Hkg{{\mathcal{H}}^{\rm kin}_{\rm grav}}
\def\Hdiff{{\mathcal{H}}^{\rm diff}}

\def\H{\mathcal{H}}
\def\phy{\mathrm{phy}}
\def\Hp{\mathcal{H}_{\rm phy}}

\def\R{\mathcal{R}}

 \makeindex

\begin{document}
  
  \title{From General Relativity to Quantum Gravity}
\author{Abhay Ashtekar}
\email{ashtekar@gravity.psu.edu} \affiliation{Institute for
Gravitation and the Cosmos \& Physics
  Department, Penn State, University Park, PA 16802, U.S.A.}
\author{Martin Reuter}
\email{reuter@thep.physik.uni-mainz.de} \affiliation{Institute of Physics, University of Mainz, Staudingerweg 7, D-55099 Mainz, Germany}
\author{Carlo Rovelli}
\email{rovelli@cpt.univ-mrs.fr} \affiliation{Centre de Physique Th\'eorique,Universit\'e de Marseille, F-13288 Marseille, France}

\begin{abstract}
In general relativity (GR), spacetime geometry is
no longer just a background arena but a physical and dynamical
entity with its own degrees of freedom. We present an overview of
approaches to quantum gravity in which this central feature of GR is
at the forefront. However, 
the short distance dynamics in the quantum theory are quite
different from those of GR and classical spacetimes and gravitons
emerge only in a suitable limit.
Our emphasis is on communicating the key strategies, the main
results and open issues. 
In the spirit of this volume, we focus on a few avenues that have
led to the most significant advances over the past 2-3 decades.%
\footnote{To appear in \emph{General Relativity and Gravitation: A Centennial Survey}, commissioned by the International Society for General Relativity and Gravitation and to be published by Cambridge University Press. Abhay Ashtekar served as the `coordinating author' and combined the three contributions.}
\end{abstract}

\pacs{04.60Pp, 04.60.Ds, 04.60.Nc, 03.65.Sq}
\maketitle

\section{Introduction}
\label{s1}

The necessity of reconciling general relativity (GR) with quantum
physics was recognized by Einstein \cite{ae} already in 1916 when he
wrote:
\begin{quote}
\sl{``Nevertheless, due to the inner-atomic movement of electrons, atoms
would have to radiate not only electro-magnetic but also
gravitational energy, if only in tiny amounts. As this is hardly
true in Nature, it appears that quantum theory would have to modify
not only Maxwellian electrodynamics, but also the new theory of
gravitation.''} 
\end{quote}
Yet, almost a century later, we still do not have a satisfactory
reconciliation. Why is the problem so difficult? An obvious response
is that this is because there are no observations to guide us.
However, this cannot be the entire story because, if there are no
observational constraints, one would expect an overabundance of
theories, not scarcity!

The viewpoint in approaches discussed in this Chapter is that the
primary obstacle is rather that, among fundamental forces of Nature,
gravity is special: it is encoded in the very geometry of spacetime.
This is a central feature of GR, a crystallization of the
equivalence principle that lies at the heart of the theory.
Therefore, one argues, it should be incorporated at a fundamental
level in a viable quantum theory. The perturbative treatments which
dominated the field since the 1960s ignored this aspect of gravity.
They assumed that the underlying spacetime can be taken to be a
continuum, endowed with a smooth background geometry, and the
quantum gravitational field can be treated as any other quantum
field on this background. But the resulting quantum GR turned out to
be non-renormalizable; the strategy failed by its own criteria. The
new strategy is to free oneself of the background spacetime that
seemed indispensable for formulating and addressing physical
questions; the goal is to lift this anchor and learn to sail the
open
seas. 
This task requires novel mathematical techniques and conceptual
frameworks. From the perspective of this Chapter, we do not yet have
a satisfactory quantum gravity theory primarily because serious
attempts to meet these challenges squarely are relatively recent.
However, as our overview will illustrate, the community \emph{has}
made notable advances towards this goal in recent years.

In this Chapter, we will focus on two main programs, each of which
in turn has two related but distinct parts: i) \emph{Loop Quantum
Gravity} (LQG) whose Hamiltonian or canonical framework is well
suited for cosmological issues, and whose Spinfoam or covariant
framework is geared to address scattering theory
\cite{alrev,crbook,ttbook,PoS,perez}; and, ii) the \emph{Asymptotic
Safety} paradigm which includes the Effective Average Action
Framework with its functional renormalization group equation in the
continuum, and the Causal Dynamical Triangulation Approach in which
one uses numerical simulations a la lattice gauge theory
\cite{reviews1,reviews2,reviews3,agjl-rev}. (String Theory is
discussed in Chapter 12 and other approaches in the Introduction to
Part IV.)

A common theme in these programs is that their starting point is the
physical, dynamical spacetime geometry of GR. However, as will be
clear from the detailed discussion, this does not imply a
conventional quantization of GR. In LQG, for example, the
fundamental quanta of geometry \index{quanta of geometry}
are one dimensional, polymer-like
excitations over nothing, rather than gravitons, the wavy
undulations over a continuum background. In particular, classical
general relativity is recovered only in an appropriate
coarse-grained limit. Another common theme is that these programs
first focus on geometry and rely on non-perturbative effects
---rather than specific matter couplings--- to cure the ultraviolet
difficulties of perturbative quantum GR. The viewpoint is that the
short distance behavior of quantum geometry is qualitatively
different from that suggested by the continuum picture and it would
be more efficient to first develop a detailed understanding of
quantum geometry in the Planck regime and then couple matter in a
second stage. In the Asymptotic Safety scenario, for example, this
strategy was successfully implemented first for pure gravity, and
incorporating certain matter fields afterwards did not change the
basic picture \cite{percacci}. \index{QCD}
Finally, as in QCD, the first
priority in these programs is to uncover and explore qualitatively
new, non-perturbative features of quantum gravity by focusing on
just one interaction, rather than on achieving unification. Such
features have already emerged. Examples are: a quantum resolution of
singularities of GR \cite{asrev,PoS},
finiteness of microstates of black hole and cosmological horizons
\cite{alrev,blvrev}, and effective dimension reduction in the Planck
regime \cite{reviews1,reviews2,reviews3,agjl-rev}.

Although these programs share several common elements, there are
also some key differences in the underlying viewpoints. Let us begin
with Asymptotic Safety. Recall first that, although GR is
perturbatively non-renormalizable, there does exist a well-developed
and powerful effective field theory \cite{donoghue} 
\index{effective field theory} which, for
example, has been applied with remarkable success to the long
standing problem of equations of motion of compact binaries in GR
\cite{rothstein}. However, this theory abandons the idea of handling
the Planck regime and focuses on low energy processes. Asymptotic
Safety can be thought of as a specific \emph{ultraviolet (UV)
completion} of this effective field theory using Wilson's
generalized notion of renormalization \cite{wilson}. The idea is to
avoid the notorious proliferation of undetermined couplings in the
UV faced by perturbative GR by using a reliable strategy that has
already been successfully tested in well understood,
\emph{perturbatively non-renormalizable} field theories where one
can, so to say, `renormalize the non-renormalizable' \cite{gaw-kup}.
This success suggests that a state of `peaceful coexistence' of
perturbative divergences with Asymptotic Safety may be possible also
for gravity \cite{reviews1,reviews2,reviews3,max-pert}.

In LQG the guiding principle is rather different. The viewpoint is
that, just as Riemannian geometry is essential to the formulation of
general relativity, an appropriate \emph{quantum} Riemannian
geometry should underlie a viable theoretical account of space,
time, and gravitation that does not disregard quantum theory. 
\index{quantum Riemannian geometry} To
meet this goal, a specific quantum theory of geometry was
constructed in detail drawing motivation from geometric structures
that underlie the phase space of GR \cite{alrev,crbook,ttbook,PoS}.
In this and subsequent constructions one makes a heavy use of
non-perturbative techniques that have already been successful in
gauge theories but with a crucial twist: now there is no reference
to a background metric. This requirement of background independence
\index{background independence|(}
is surprisingly powerful and leads to a unique kinematical framework
\cite{lost,cf} on which dynamics of the quantum theory is being built.
While the Hamiltonian LQG has broad similarities with the older
Wheeler-DeWitt (WDW) theory \cite{jw}, the quantum nature of underlying
geometry makes a key difference leading, for example, to a natural
resolution of classical singularities in cosmological models
\cite{asrev}. Similarly, Spinfoams provide transition amplitudes
that are UV finite to any order in a natural expansion. Furthermore,
a \emph{positive} cosmological constant provides a natural mechanism
for regulating their infrared (IR) behavior \cite{PoS,perez}. 
\index{cosmological constant} 

Thus, although both LQG and Asymptotic 
Safety programs have similar goals, the physical concepts and 
mathematical techniques used in subsequent analysis are quite 
different. In particular, because the quantum geometry underlying LQG is fundamentally discrete, the physical degrees of freedom terminate at the Planck scale, much like in string theory. In the Asymptotic Safety program, on the other hand, there is no kinematic reason that would prevent degrees of freedom at arbitrarily small scale. A first reading of the flow equations suggests that there are physical degrees of freedom at any scale, all the way to the infinitely small. However, 
it is the fixed point action that determines the physical degrees of freedom in this approach. Non-perturbative renormalizability indicates that these are fewer than what one would expect classically and the mean field considerations 
indicate that there are at most as many as in a theory in 2 space-time dimensions. A more thorough understanding of the fixed point is necessary to settle this important question in the Asymptotic Safety program.

This Chapter is organized as follows. Section \ref{s2} provides a
broad brush overview of the two programs. Since this volume is
likely to draw readership from diverse quarters, we have made a
special attempt to make the sub-sections self-contained. Thus a
reader interested only in Asymptotic Safety can skip sections
\ref{s2.2} and \ref{s2.3} and a reader interested only in LQG can
skip section \ref{s2.1} without loss of continuity. Section \ref{s3}
discusses illustrative applications to cosmology of the very early
universe, black holes physics and scattering theory. While advances
over the past decade are encouraging, a large number of issues
remain. These are discussed in section \ref{s4}.

\section{Frameworks}
\label{s2}

This section is divided into three parts. The first summarizes the
main ideas and results in the Asymptotic Safety program, the second,
in Hamiltonian LQG and the third in Spinfoams.

\subsection{Asymptotic Safety}
\label{s2.1}
\index{asymptotic safety|(}
Since GR is not renormalizable in the standard perturbation theory,
it is commonly argued that a satisfactory microscopic quantum theory
of the gravitational interaction cannot be set up within the realm
of quantum field theory without adding further symmetries, extra
dimensions or new principles such as holography.
In contrast, the Asymptotic Safety program \cite{wein} retains
quantum field theory without such additions as the theoretical arena
and instead abandons the traditional techniques of perturbative
renormalization. Moreover, as we will see, in a certain sense it
even abandons the standard notion of `quantization' because its
starting point is not a given classical model to be promoted to a
quantum theory.

Rather, in its modern incarnation, this program may be thought of as
a systematic \emph{search strategy among theories that are already
`quantum';} it identifies the `islands' of physically acceptable
theories in the `sea' of unacceptable ones plagued by short distance
pathologies. Since the approach is based on Wilson's generalized
notion of renormalization \cite{wilson} and the use of functional
renormalization group (RG) equations, \index{RG flow}
concepts from statistical
field theory play an important role. They provide a unified
framework for approaching the problem with both continuum and
discrete methods. In this section we discuss two such complementary
approaches within the Asymptotic Safety paradigm: the Effective
Average Action (EAA) with its Functional renormalization group
Equation (FRGE) \cite{mr}, and Causal Dynamical Triangulations (CDT)
\cite{al}.

\subsubsection{The Functional Renormalization Group}
\index{functional renormalization group FRG|(} 

The goal of the Asymptotic Safety program consists in giving a
mathematically precise meaning to, and actually computing functional
integrals over `all' spacetime metrics of the form $\int
\mathcal{D}\tilde{g}_{\mu\nu} \,  \exp \big({\rm i}
S[\tilde{g}_{\mu\nu}]\big)$, or
\begin{align}
 Z&= \int \mathcal{D}\tilde{g}_{\mu\nu}\, e^{-S[\tilde{g}_{\mu\nu}]}\,,
 \label{as.2}
\end{align}
from which all quantities of physical interest can be deduced then.
Here $S[\tilde{g}_{\mu\nu}]$ denotes the classical or, more
appropriately, the bare action. It is required to be diffeomorphism
invariant, but is kept completely arbitrary otherwise. In general it
differs from the usual Einstein-Hilbert action. This generality is
essential in the Asymptotic Safety scenario: the viewpoint is that
the functional integral would exist only for a certain class of
actions $S$ and the task is to identify this class.

\index{effective average action|(}\index{EAA|see{effective average action}}
Following the approach proposed in \cite{mr} one attacks this
problem in an indirect way: rather than dealing with the integral
per se, one interprets it as the solution of a certain differential
equation, a functional renormalization group equation, or `FRGE'.
The advantage is that, contrary to the functional integral, the FRGE
is manifestly well defined. It can be seen as an `evolution
equation' in a mathematical sense, defining an infinite dimensional
dynamical system in which the RG scale plays the role of time.
\index{RG flow} Loosely speaking, this reformulation replaces the problem of
defining functional integrals by the task of finding evolution
histories of the dynamical system that extend to {\it infinitely
late times}. According to the Asymptotic Safety conjecture the
dynamical system possesses a fixed point which is approached at late
times, yielding well defined, fully extended evolutions, which in
turn tell us how to construct (or `renormalize') the functional
integral.

Let us start by explaining the passage from the functional integrals
to the FRGE. Recall that in trying to put the integrals on a solid
basis one is confronted with a number of obstacles:\\
{\bf (i)} As in every field theory, difficulties arise since one
tries to quantize {\it infinitely many degrees of freedom}.
Therefore, at the intermediate steps of the construction one keeps
only finitely many of them by introducing cutoffs at very small and
very large distances, $\Lambda^{ -1}$ and $k^{-1}$, respectively. We
shall specify their concrete implementation in a moment. The
ultraviolet (UV) and infrared (IR) cutoff scales $\Lambda$ and $k$,
respectively, have the dimension of a mass, and the original system
is recovered for $\Lambda\rightarrow \infty$, $k\rightarrow 0$.\\
{\bf (ii)} Conceptually, the most severe problem one encounters when
quantizing the gravitational field, one which is not shared by any
conventional matter field theory, is the requirement of {\it
background independence}: 
no particular spacetime (such as Minkowski space, say) should be given a privileged status. \index{background independence|)}
Rather, the
geometry of spacetime should be determined dynamically. In the
approach to Asymptotic Safety along the lines of \cite{mr} this
problem is dealt with by following the spirit of DeWitt's background
field method \cite{dewittbook} and introducing a (classical,
non-dynamical) background metric $\bar{g}_{\mu\nu}$ which, however,
is kept absolutely arbitrary.
One then decomposes the integration variable as
$\tilde{g}_{\mu\nu}\equiv \bar{g}_{\mu\nu} + \tilde{h}_{\mu\nu}$,
and interprets $\mathcal{D}\tilde{g}_{\mu\nu}$ as an integration
over the nonlinear fluctuation, $\mathcal{D}\tilde{h}_{\mu\nu}$. In
this way one arrives at a conceptually easier task, the quantization
of the matter-like field $\tilde{h}_{\mu\nu}$ in a generic, but
classical background $\bar{g}_{\mu\nu}$. The availability of the
background metric is crucial at various stages of the construction
of an FRGE. However the final physical results do not depend on the
choice of a specific background.\\
{\bf (iii)} As in every gauge field theory, the {\it redundancy of
gauge-equivalent field configurations} (diffeomorphic metrics) has
to be carefully accounted for. Here we employ the Faddeev-Popov
method and add a gauge fixing term $S_{\text{gf}}\propto \int
\sqrt{\bar{g}}\bar{g}^{\mu\nu}F_ {\mu}F_{\nu}$ to $S$ where
$F_{\mu}\equiv F_{\mu}(\tilde{h};\bar{g})$ is chosen such that the
condition $F_{\mu}=0$ picks a single representative from each gauge
orbit. The resulting volume element on orbit space, the
Faddeev-Popov determinant, we express as a functional integral over
Grassmannian ghost fields $\tilde{C}^{\mu}$ and
$\tilde{\bar{C}}_{\mu}$, governed by an action $S_{\text{gh}}$. In
this way the original integral \eqref{as.2} gets replaced by
$\widetilde{Z}[\bar{\Phi}]= \int \mathcal{D}\tilde{\Phi}\, \exp\big(
{-S_{\text{tot}}[\tilde{\Phi},\bar{\Phi}]}\big) $. Here the total
bare action $S_{\text{tot}}\equiv S+S_{\text{gf}}+S_{\text {gh}}$
depends on the dynamical fields
$\tilde{\Phi}\equiv(\tilde{h}_{\mu\nu},\tilde{C}^{\mu},\tilde{\bar{C}}_{\mu})$,
the background fields $\bar{\Phi}\equiv (\bar{g}_{\mu\nu})$, and
possibly also on (both dynamical and background) matter fields,
which for simplicity are not included here.

Using the gauge fixed and regularized integral we can compute
arbitrary ( $\bar{\Phi}$-dependent!) expectation values $\langle
\mathcal{O}(\tilde{\Phi})\rangle\equiv \widetilde{Z}^{-1}\int
\mathcal{D}\tilde{\Phi}\,
\mathcal{O}(\tilde{\Phi})\,e^{-S_{\text{tot}}[\tilde{\Phi},\bar{\Phi}]}$, for instance $n$-point functions where $\mathcal{O}$ consists of
strings $\tilde{\Phi}(x_1)\tilde{\Phi}(x_2)\cdots
\tilde{\Phi}(x_n)$. For $n=1$ we use the notation $\Phi\equiv
\langle\tilde{\Phi}\rangle\equiv (h_{\mu\nu},
C^{\mu},\bar{C}_{\mu})$, i.e. the elementary field expectation
values are $h_{\mu\nu}\equiv \langle \tilde{h}_{\mu\nu}\rangle$,
$C^{\mu}\equiv\langle \tilde{C}^{\mu}\rangle$ and
$\bar{C}_{\mu}\equiv\langle \tilde{\bar{C }}_{\mu}\rangle$. Thus the
full dynamical metric has the expectation value $g_{\mu\nu}\equiv
\langle \tilde{g}_{\mu\nu}\rangle=\bar{g}_{\mu\nu}+ h_{\mu\nu}$.

The dynamical laws which govern the expectation value $\Phi(x)$ have
an elegant description in terms of the {\it effective action}
$\Gamma$. It is a functional depending on $\Phi$ similar to the
classical $S[\Phi]$ to which it reduces in the classical limit.
Requiring stationarity, $S$ yields the classical field equation
$(\delta S\slash \delta \Phi)[\Phi_{\text{class}}]=0$, while
$\Gamma$ gives rise to a quantum mechanical analog satisfied by the
expectation values, the {\it effective field equation }
$(\delta\Gamma\slash\delta \Phi)[\langle\tilde{\Phi}\rangle] =0$.
If, as in the case at hand, $\Gamma \equiv
\Gamma[\Phi,\bar{\Phi}]\equiv\Gamma[h_{\mu\nu},C^{\mu},
\bar{C}_{\mu};\bar{g}_{\mu\nu}]$ depends also on background fields,
the solutions to this equation inherit this dependence and so
$h_{\mu\nu}\equiv \langle \tilde{h}_{\mu\nu}\rangle$ functionally
depends on $\bar{g}_{\mu\nu}$. Technically, $\Gamma$ is obtained
from a functional integral with $S_{\text{tot}}$ replaced by
$S^{J}_{\text{tot} }\equiv S_{\text{tot}}-\int{\rm d} x\,
J(x)\tilde{\Phi}(x)$. The new term couples the dynamical fields to
an external, classical source, $J(x)$, and repeated functional
differentiation $(\delta\slash\delta J)^n$ of $\ln
\widetilde{Z}[J,\bar{\Phi}]$ yields the $n$-point functions. In
particular, $\Phi=\delta \ln \widetilde{Z}\slash\delta J$. It is a
standard result that $\Gamma[\Phi,\bar{\Phi}]$ equals exactly the
Legendre transform of $\ln \widetilde{Z}[J,\bar{\Phi}]$, at fixed
background fields $\bar{\Phi}$. The importance of $\Gamma$ also
resides in the fact that \emph{it is the generating functional of
special $n$-point functions from which all others can be easily
reconstructed.} Therefore, finding $\Gamma$ in some quantum field
theory is often considered equivalent to completely `solving' this
theory.

To calculate $\Gamma[\Phi,\bar{\Phi}]$ it is advantageous to employ
a gauge breaking condition $F_{\mu}$ which fixes a gauge belonging
to the distinguished class of the so called {\it background gauges}.
To see the benefit, recall that the original gauge transformations
read $\delta \tilde{g }_{\mu\nu}=\mathcal{L}_{v} \tilde{g}_{\mu\nu}$
where $\mathcal{L}_v$ denotes the Lie derivative w.r.t. the vector
field $v$. When we decompose $\tilde{g}_{\mu\nu}= \bar{g}_{\mu\nu}
+\tilde{h}_{\mu\nu}$ we can distribute the gauge variation of
$\tilde{g}_{\mu\nu}$ in different ways over $\bar{g}_{\mu\nu}$ and
$\tilde{h}_{\mu\nu}$. In particular this gives rise to what is known
as {\it quantum gauge transformations} ($\boldsymbol{\delta^{{\rm
Q}}}\tilde{h}_{\mu\nu}= \mathcal{L}_{v}(\bar{g}_{\mu\nu}+
\tilde{h}_{\mu\nu})$, $\boldsymbol{\delta^{{\rm Q}}}
\bar{g}_{\mu\nu}=0$) and {\it background gauge transformations}
($\boldsymbol{\delta^{{\rm B}}} \tilde{h}_{\mu\nu}
=\mathcal{L}_{v}\tilde{h}_{\mu\nu}$, $\boldsymbol{\delta ^{{\rm
B}}}\bar{g}_{\mu\nu}=\mathcal{L}_v \bar{g}_{\mu\nu}$). Since the
functional integral is defined by fixing an externally prescribed
background metric, $\bar{g}_{\mu\nu}$, we must ensure invariance
under the `ordinary' or `true' gauge transformations the
Faddeev-Popov method deals with. Hence it is the
$\boldsymbol{\delta^{{\rm Q}}}$-invariance which needs to be
gauge-fixed by the condition $F_{\mu}=0$. Interestingly enough,
there exist $F_{\mu}$'s, a variant of the harmonic coordinate
condition, for example, which indeed  fix the
$\boldsymbol{\delta^{{\rm Q}}}$-transformations, but at the same
time are {\it invariant under $\boldsymbol{\delta^{{\rm
B}}}$-transformations}: $\boldsymbol{\delta^{{\rm
 B}}}F_{\mu}=0$. They implement the background gauges, and from now on
we assume  that we employ one of those. Then, as a consequence, the
effective action {\it $\Gamma[\Phi,\bar{\Phi}]$ is invariant under
background gauge transformations} which include the ghosts:
$\boldsymbol{\delta^{{\rm B}}} \Gamma[\Phi,\bar{\Phi}]=0$ for all
$\boldsymbol{\delta^{{\rm B} }}\Phi=\mathcal{L}_{v}\Phi$,
$\boldsymbol{\delta^{{\rm B}}}\bar{\Phi}=
\mathcal{L}_{v}\bar{\Phi}$. We emphasize that this property should
not be confused with another notion of `gauge independence' which
the above $\Gamma[ \Phi,\bar{\Phi}]$ actually does {\it not} have:
It is not independent of which particular $F_{\mu}$ is picked from
the class with $\boldsymbol{\delta^{{\rm B}}}F_{\mu}=0$. This
$F_{\mu}$-dependence will disappear only at the level of
observables.

Turning now to the concept of a {\it functional renormalization
group equation} recall that  the above definition of $\Gamma$ is
based on the functional integral regularized in the IR and UV, hence
it depends on the corresponding cutoff  scales: $\Gamma\equiv
\Gamma_{k,\Lambda}[\Phi,\bar{\Phi}]$. It is this object for which we
derive a FRGE, more precisely a closed evolution equation governing
its dependence on the IR cutoff scale $k $. This is possible only if
the IR regularization is implemented appropriately, as
in the so called {\it effective average action} (EAA) \cite{wet-eq}.

The EAA is related to the modified integral, 
$\int \mathcal{D}\tilde{\Phi} \, e^{-S_{\text{tot}}^J}\, \,
e^{-\Delta S_{k}[\tilde{\Phi },\bar{\Phi}]} \,\equiv\,
Z_{k,\Lambda}[J,\bar{\Phi}]$ whose second exponential factor in the
integrand, containing the {\it cutoff action} $\Delta S_k$, is
designed to achieve the IR regularization. To see how this works,
assume the integration variable $\tilde{\Phi}=(\tilde{h},\tilde{C},
\tilde{\bar{C}})$ is expanded in terms of  eigenfunctions
$\varphi_{p}$ of the covariant tensor Laplacian related to the
background metric, $\bar{D }^2\equiv \bar{g}^{\mu\nu} \bar{D}_{\mu}
\bar{D}_{\nu}$. Writing $-\bar{D}^ 2\varphi_p=p^2 \varphi_p$ we
have, symbolically, $\tilde{\Phi}(x)=\sum_p\,\alpha_p \varphi_p(x)$.
The $\alpha_p$'s are generalized Fourier coefficients, and so the
functional integration over $\tilde{\Phi}$ amounts to integrating
over all $\alpha_p$:
\begin{align}
Z_{k,\Lambda}[J,\bar{\Phi}]= \prod_{p^2\in[0,\Lambda^2]}
\int_{-\infty} ^{\infty}{\rm d}\alpha_p \,
\exp\big(-S^{J\,\prime}_{\text{tot}}[\{\alpha \},\bar{\Phi}]\big)
\label{as.2c}
\end{align}
Here $S^{J\,\prime}_{\text{tot}}$ equals
$S^J_{\text{tot}}[\tilde{\Phi},\bar{\Phi}]+\Delta
S_k[\tilde{\Phi},\bar{\Phi}]$ with the expansion for $\tilde{\Phi}$
inserted. In \eqref{as.2c} we implemented the UV regularization by
retaining only eigenfunctions (or `modes') corresponding to $-\bar
{D}^2$-eigenvalues (or squared `momenta') smaller than $\Lambda^2$.
The IR contributions, i.e. those corresponding to eigenvalues
between $p^ 2=0$ and about $p^2=k^2$ are cut off smoothly instead,
namely by a $p ^2$-dependent suppression factor arising from $\Delta
S_k$. To obtain a structurally simple FRGE, $\Delta S_k$ should be
chosen quadratic in the dynamical fields. Usually one sets $\Delta
S_k=\frac{1}{2}\int {\rm d}x \,
\tilde{\Phi}\mathcal{R}_k\tilde{\Phi}$ with an operator
$\mathcal{R}_k \propto k^2 R^{(0)}(-\bar{D}^2\slash k^2)$ containing
a dimensionless function $R^{(0)}$. In the $-\bar{D}^2$-basis we
have then $\Delta S_k \propto k^2 \sum_{p} R^{(0)}(p^2\slash
k^2)\alpha_p^2$ which shows that $\Delta S_k$ represents a kind of
$p^2$-dependent mass term: A mode with eigenvalue $p^2$ acquires a
$(\text{mass})^2$ of the order $k^2 R^{(0)}(p^2\slash k^2)$. We
require $R^{(0)}(p^2\slash k^2)$ to have the qualitative properties
of a smeared step function which, around $p^2\slash k^2\approx 1$ ,
drops smoothly from $R^{(0)}=1$ for $p^2\slash k^2 \lesssim 1$ to
$R^ {(0)}=0$ for $p^2\slash k^2 \gtrsim 1$. This achieves precisely
the desired IR regularization: In the product over $p^2$ in
\eqref{as.2c}, $\Delta S_k$ equips all $\int {\rm d}\alpha_p
$-integrals pertaining to the {\it low momentum modes}, i.e. those
with $p^2\in[0,k^2]$, with a Gaussian suppression factor $e^{- k^2
\alpha_p^2}$ since for such eigenvalues $R^{(0)}(p^2\slash
k^2)\approx 1$. The {\it high momentum modes}, having
$p^2\in[k^2,\Lambda^2]$, yield $R^{(0)}(p^2\slash k^2)\approx 0$ and
so they remain unaffected by $\Delta S_k$. At least on a flat
background, low (high) momentum modes $\varphi_p(x)$ have long
(short) wavelengths. Therefore, when one lowers $k$ from $k=\Lambda$
down to $k=0$ one `un-suppresses' modes of increasingly long
wavelengths, thus proceeding from the UV to the IR. (In FRGE jargon,
this is called the `integrating out' of the high momentum modes
since in older approaches the low momentum modes were completely
discarded, rather than just suppressed.) This process of encoding
the contribution of an increasing number of modes in a scale
dependent, or `running' functional is precisely a {\it
renormalization in the modern sense} due to Wilson \cite{wilson}.

The effective average action, $\Gamma_{k,\Lambda}[\Phi,\bar{\Phi}]$,
is defined to be the Legendre transform of $\ln
Z_{k,\Lambda}[J,\bar{\Phi}]$ given by \eqref{as.2c}, with respect to
$J$, for $k$, $\Lambda$, and $\bar{\Phi}$ fixed (and with $\Delta
S_k[\Phi,\bar{\Phi}]$ subtracted from the result of the
transformation, which is not essential here). As for the $\Phi$,
$\bar{\Phi}$-arguments, we stress that the modes classified low or
high momentum are only those of the {\it fluctuation} field,
$\tilde{\Phi}$. The externally prescribed background and source
fields $ \bar{\Phi}(x)$ and $J(x)$, which are also present under the
integral defining $Z_{k,\Lambda}[J,\bar{\Phi}]$, have nonzero
Fourier coefficients for all $p^2\in[0,\Lambda^2]$ in general, they
may contain both high and low momentum components. As a consequence,
the same is true for the $\Phi$-argument of the EAA, since $J$ and
$\Phi=\delta \ln Z_{k,\Lambda}\slash \delta J$ are
Legendre-conjugates of one another.

The EAA, $\Gamma_{k,\Lambda}[\Phi,\bar{\Phi}]$, has a number of
important features not realized in other functional RG approaches:\\
{\bf (i)} Since no fluctuation modes are taken into account in the
$k=\Lambda\rightarrow \infty$ limit, the EAA approaches the bare
(i.e., un-renormalized) action, $\Gamma_{\Lambda,\Lambda}=
S_{\text{tot}}$. In the limit $k\rightarrow 0$, it yields the
standard  \index{RG flow}
effective action (with an UV cutoff).\\
{\bf (ii) } It satisfies a closed FRGE, and can be computed by
integrating this FRGE towards low $k$, with the initial condition
$\Gamma_{\Lambda,\Lambda}=S_{\text{tot}}$ at $k=\Lambda$.\\
{\bf (iii)} The functional $\Gamma_{k,\Lambda}[\Phi,\bar{\Phi}]$ is
invariant under background gauge transformations
$\boldsymbol{\delta^{{\rm B}} }$ for all values of the cutoffs. 
This property is preserved by the FRGE: the RG evolution does not
generate $\boldsymbol{\delta^{{\rm B}}}$-noninvariant terms.\\
{\bf (iv)} The FRGE continues to be well behaved when the UV cutoff
is removed ($\Lambda\rightarrow\infty)$. Denoting solutions to the
UV cutoff-free FRGE by $\Gamma_k[\Phi,\bar{\Phi}]$, it reads:
\begin{align}
k \partial_k \Gamma_k [\Phi,\bar{\Phi}] &= \frac{1}{2}\, \text{STr}
\Big[ \big(\Gamma_k^{(2)} [\Phi,\bar{\Phi}]
+\mathcal{R}_k[\bar{\Phi}] \big )^{-1} k \partial_k
\mathcal{R}_k[\bar{\Phi}]\Big] \label{as.3}
\end{align}
Here $\text{STr}$ denotes the functional supertrace, and
$\Gamma^{(2)}_k$ stands for the matrix of second functional
derivatives of $\Gamma_k$ with respect to $\Phi$ at fixed
$\bar{\Phi}$. Since $R^{(0)}$ is essentially a step function, the
derivative $\partial_k \mathcal{R}_k$ is nonzero only in a thin
shell of momenta near $p^2=k^2$, and so the supertrace on the RHS of
\eqref{as.3} receives contributions only from such momenta. As a
result, it is perfectly finite both in the IR and the UV, and this
is why sending $\Lambda \rightarrow \infty$ was unproblematic.\\
{\bf (v)} $\Gamma_k$ is closely related to a generating functional
for field averages over finite domains of size $k^{-1}$; hence the
name EAA \cite{wet-eq}. Thanks to this property, when treated as a
{\it classical} action $\Gamma_{k}$ can provide an effective field
theory description of {\it quantum} physics involving typical
momenta near $k$. This property has been exploited in numerous
applications of the EAA to particle and condensed matter physics,
but it plays no role in the present context. Rather, it is its
interpolating property between $S$ and $\Gamma_{k=0}$ which is
instrumental in the Asymptotic Safety program.

The arena in which the RG dynamics takes place is the infinite
dimensional {\it theory space}, $\mathcal{T}$. 
It consists of all
well behaved action functionals $(\Phi,\bar{\Phi}) \mapsto
A[\Phi,\bar{\Phi}]$ which depend on a given set of fields and are
invariant under some symmetry group possibly. In metric gravity
$\mathcal{T}$ comprises arbitrary $\boldsymbol{ \delta^{{\rm B}}}$
invariant functionals $A[g_{\mu\nu},\bar{g}_{\mu\nu},
C^{\mu},\bar{C}_{\mu}]$. The RHS of the FRGE \eqref{as.3} defines a
vector field $\boldsymbol{\beta}$ on $\mathcal{T}$. 
Its natural orientation is such that $\boldsymbol{\beta}$ points from higher to lower momentum scales $k$, from the UV to the IR. (This is the direction of increasing `coarse-graining' in which the microscopic
dynamics is `averaged' over increasingly large spacetime volumes.)
The integral curves of this vector field, $k\mapsto \Gamma_k$, are
the {\it RG trajectories},  \index{RG flow} and the pair
$(\mathcal{T},\boldsymbol{\beta})$ is called the {\it RG flow}. It
constitutes the dynamical system alluded to earlier.
%

One usually assumes that every $A\in \mathcal{T}$ can be expanded as
$A[\Phi,\bar{\Phi}]= \sum_{\alpha=1}^{\infty} \bar{u}_{\alpha}
P_{\alpha} [\Phi,\bar{\Phi}]$ where the set $\{ P_{\alpha}\}$ forms
a basis of invariant functionals. Writing the RG trajectory
correspondingly, $\Gamma_k[\Phi,\bar{\Phi}]=
\sum_{\alpha=1}^{\infty} \bar{u}_{\alpha}(k) P_\alpha [\Phi
,\bar{\Phi}]$, one encounters infinitely many {\it running coupling
constants}, $\bar{u}_{\alpha}(k)$, whose $k$-dependence is governed
by an infinite coupled system of differential equations: $k
\partial_k \bar{u}_{\alpha}(k)= \bar{\beta}_{\alpha}
(\bar{u}_1,\bar{u}_2,\cdots; k)$. The dimensionful {\it beta
functions} $\bar{\beta}_{\alpha}$ arise by expanding the RHS of the
FRGE: $\frac{1}{2}\text{STr}[\cdots]=
\sum_{\alpha=1}^{\infty}\bar{\beta}_{\alpha}
P_{\alpha}[\Phi,\bar{\Phi}]$. The coefficients
$\bar{\beta}_{\alpha}$ are similar to the familiar beta functions of
perturbative quantum field theory (where, however, only the finitely
many beta functions of the relevant couplings are considered.)

Reexpressing the RG equations in terms of dimensionless couplings
$u_{\alpha}\equiv k^{-d_{\alpha}}\bar{u}_{\alpha}$ with $d_{\alpha}$
the canonical mass dimension of $\bar{u}_{\alpha}$, the resulting
{\it FRGE in component form} is autonomous, i.e. its
$\beta$-functions have no explicit $k$-dependence: $ k \partial_k
u_{\alpha}(k)= \beta_{\alpha}(u_1(k),u_2(k),\cdots) $. The coupling
constants $(u_{\alpha})\equiv u$ serve as local coordinates on
$\mathcal{T}$, and the $\beta_{\alpha}$'s are the components of the
vector field $\boldsymbol{\beta}\equiv (\beta_{\alpha}(u))$.

\index{UV and IR fixed points|(} \index{RG flow}
Later on {\it fixed points} of the RG flow will be of special
interest. At a fixed point, $\boldsymbol{\beta} = 0$, so its
coordinates $(u_{\alpha}^{\ast})=u^{\ast}$ satisfy the  infinitely
many conditions $\beta_{\alpha}(u^{\ast})=0$. The fixed point's {\it
UV critical hypersurface}, $\mathcal{S}_{\text{UV} }$, or
synonymously its {\it unstable manifold} is defined to consist of
all points in $\mathcal{T}$ which are pulled into the fixed point
under the inverse RG flow, i.e. for increasing scale $k$.
Linearizing the flow about $u^{\ast}$ one has
$k \partial_k u_{\alpha}(k) = \sum_{\gamma} B_{\alpha\gamma}
\big(u_{\gamma}(k)-u_{\gamma}^{\ast}\big)$
with the {\it stability matrix} $\boldsymbol{B}=
(B_{\alpha\gamma})$, $ B_{\alpha\gamma}\equiv \partial_{\gamma}
\beta_{\alpha}(u^{\ast})$. If the eigenvectors of $\boldsymbol{B}$
form a basis, its solution reads $u_{\alpha}(k)=
u_{\alpha}^{\ast}+\sum_{\alpha}C_I V^I_{\alpha} \big(k_0 \slash
k\big)^{\theta_I}$. Here the $C_I$'s are constants of integration
and the $V^I$'s denote the right-eigenvectors of $\boldsymbol{B}$
with eigenvalues $-\theta_I$, i. e. $\sum_{\gamma} B_{\alpha\gamma}
V^{I}_{\gamma}= -\theta_I V_{\alpha} ^I$. In general
$\boldsymbol{B}$ is not symmetric and the {\it critical exponents}
$\theta_I$ are complex. Along eigendirections with
$\text{Re}\,\theta_I >0$ ($\text{Re}\, \theta_I <0$ ) deviations
from $u_{\alpha}^{\ast}$ grow (shrink) when $k$ is lowered from the
UV towards the IR; they are termed relevant (irrelevant).

A trajectory $u_{\alpha}(k)$ within $\mathcal{S}_{\text{UV}}$, by
definition, approaches $u_{\alpha}(k\rightarrow \infty)=
u_{\alpha}^{\ast}$ in the UV. For the constants $C_I$ in its
linearization this implies that $C_I= 0$ for all $I$ with
$\text{Re}\, \theta_I<0$. Hence the trajectories in $
\mathcal{S}_{\text{UV}}$ are labeled by the remaining $C_I$'s
related to the critical exponents with $\text{Re}\, \theta_I >0$.
(For simplicity we assume all $\text{Re}\, \theta_I$ nonzero.) As a
consequence, {\it the dimensionality of the critical hypersurface,
$s\equiv\text{dim}\big(\mathcal{S}_{\text{UV}}\big)$, equals the
number of critical exponents with $\text{Re}\, \theta_I >0$}, i.e.,
the number of relevant directions.

A fixed point is called {\it Gaussian} if it corresponds to a free
field theory. Its critical exponents agree with the canonical mass
dimension of the corresponding operators. A fixed point whose
critical exponents differ from the canonical ones is referred to as
nontrivial or as a {\it non-Gaussian fixed point} (NGFP).
\index{functional renormalization group FRG|)}

\subsubsection{Asymptotic Safety \label{as.secC}}

The construction of a quantum field theory involves finding an RG
trajectory which is infinitely extended in the sense that it is a
curve, entirely within theory space, with well defined limits
$k\rightarrow 0$ and $k\rightarrow \infty$, respectively.
\emph{Asymptotic Safety is a proposal for ensuring the existence of
the second limit.} Its crucial prerequisite is a nontrivial RG fixed
point $\Gamma_{\ast}$ on $\mathcal{T}$. Let us assume there is such
a fixed point. Then it is sufficient to simply pick any of the
trajectories within its hypersurface $\mathcal{S}_{\text{UV}}$ to be
sure that the trajectory has a singularity free ultraviolet behavior
since it will always hit the fixed point for $k\rightarrow \infty$.
There exists a $\text{dim}\big(\mathcal{S}_{\text{UV}}
\big)$-parameter family of such trajectories. 

Most probably an UV fixed point is not only sufficient but also
necessary for an acceptable theory without divergences. Therefore,
in the simplest case when there exists only one, the physically
inequivalent asymptotically safe quantum theories one can construct
are labeled by the $\text{dim}\big(\mathcal{S}_{\text{UV}}\big)$
parameters characterizing trajectories inside
$\mathcal{S}_{\text{UV}}$. Thus the degree of predictivity of
asymptotically safe theories is essentially determined by the number
of {\it relevant} eigendirections at $\Gamma_{\ast}$. If this is a
finite number $s \equiv\text{dim}\big(\mathcal{S}_{\text{UV}}\big)$,
it is sufficient to measure only $s$ of the couplings
$\{u_{\alpha}(k)\}$ characterizing $\Gamma_k$ in order to predict
the infinitely many others. In particular, at $k =0$ the standard
effective action $\Gamma\equiv\Gamma_0$ is obtained which `knows'
all possible predictions.

The only input required for this construction is the theory space
$\mathcal{T}$, 
that is the field contents and the symmetries. It
fully determines the FRGE and its fixed point properties.  Since
$\Gamma_{k\rightarrow\infty}$ is closely related to the bare action
$S$, the Asymptotic Safety program essentially consists in {\it
computing} $S\sim \lim_{k\rightarrow \infty}\Gamma_k =
\Gamma_{\ast}$ from the fixed point condition. In this sense the
approach amounts to a selection process among quantum theories
rather than the quantization of a classical system known beforehand.
It has become customary to call \index{QEG quantum Einstein gravity|(}
{\it Quantum Einstein Gravity}, or
QEG, any quantum field theory of metric-based gravity, regardless of
its bare action, which is defined by a trajectory on the theory
space $\mathcal{T}_ {\text{QEG}}$ of diffeomorphism invariant
functionals $A[g_{\mu\nu},\bar{ g}_{\mu\nu},C^{\mu},\bar{C}_{\mu}]$.

A priori the functional integral over `all' metrics is only formal
and plagued by mathematical problems. Knowing $\Gamma_{*}$ and the
RG flow in its vicinity one can give a well defined meaning to it.
The only extra ingredient that needs to be selected is an UV
regularization for the integral. It is then possible to use the
information encoded in the flow of $\Gamma_k$ near $\Gamma_*$ in
order to determine how the (`bare') parameters, on which the integral 
depends, must be tuned in order to obtain a meaningful limit when the
UV regulator is removed \cite{elisa1}. Thus the mathematical subtleties
of the functional integral are overcome if the long time-behavior of
the associated dynamical system on $\mathcal{T}_{\text{QEG}}$ can be
controlled, e.g. by means of a fixed point. For an evolution
equation as complicated as the FRGE, on an infinite dimensional
theory space, it is by no means clear from the outset that this is
possible, i.e. that there exist RG trajectories that extend to
infinite values of the evolution parameter. An essential part of the
Asymptotic Safety program consists in demonstrating that this is
indeed the case, for the concrete reason that the trajectory
hits a fixed point in the long time-limit.

Practical computations require a nonperturbative approximation
scheme. The method of choice consists in a {\it truncation of theory
space}. One sets all but a certain subset of couplings $u_{\alpha}$
to zero, and expands $\Gamma_k$ in terms of the appropriately chosen
reduced set $\{P_{\alpha},\, \alpha=1,\cdots,N\}$ where, as before,
$P_{\alpha}$ is a basis of invariant functionals in terms of which
now only the actions in the truncated theory space can be expanded.
Hence the FRGE boils down to a system of $N$ coupled differential
equations. This amounts to a severe restriction, of course, which
needs to be justified a posteriori by systematically changing and
enlarging the subset chosen. This difficulty is not specific to
gravity; the same strategy is followed in FRGE-based investigations
of matter field theories on flat space and in statistical physics.

\begin{figure}[t]
 \begin{center}
   \includegraphics[width=0.6\textwidth]{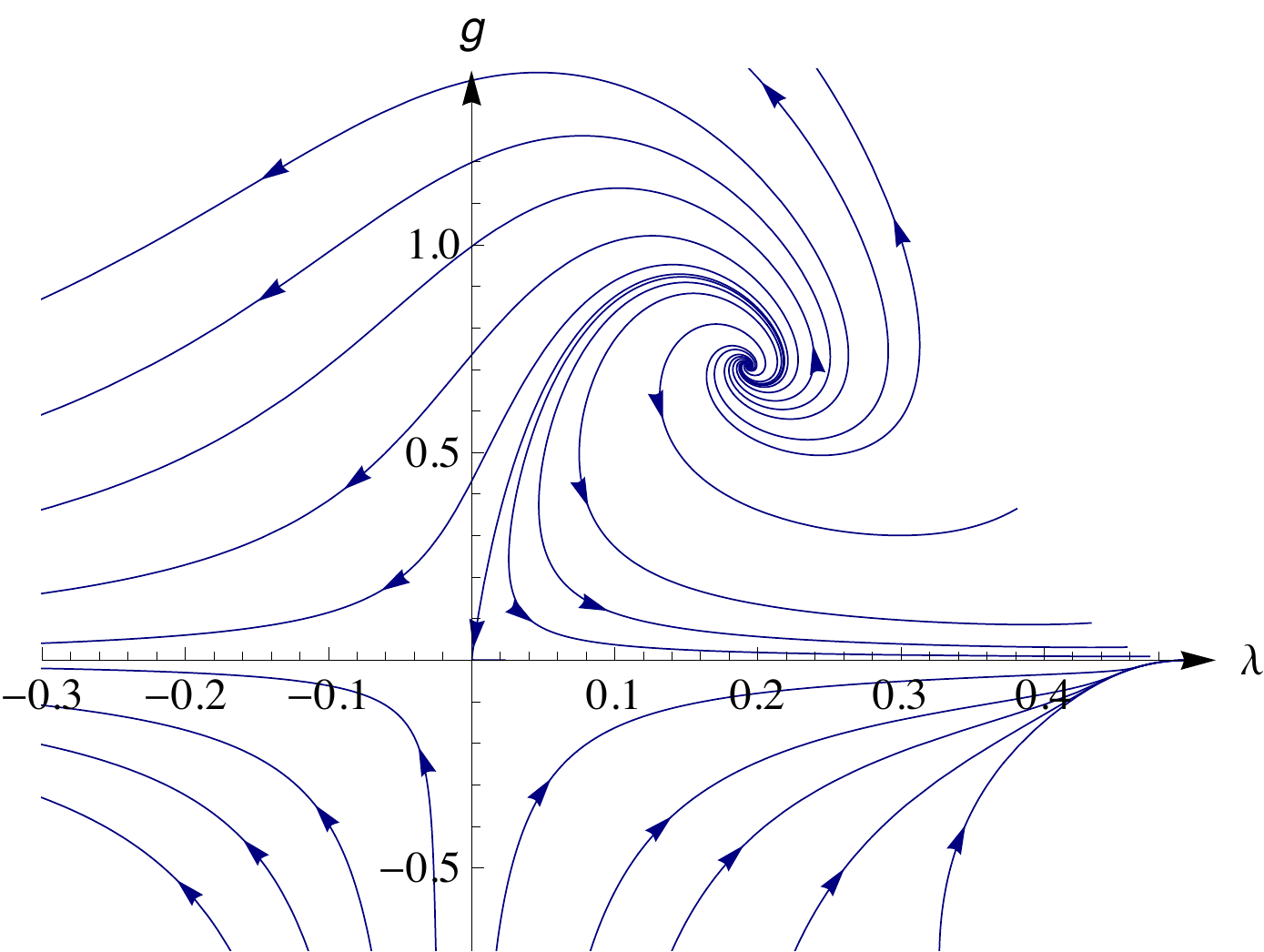}
 \end{center}
 \caption{RG flow of the Einstein-Hilbert truncation on the
 ($g$,$\lambda$)-plane. The arrows point towards decreasing scales
 $k$. (First obtained in \cite{frank1}).} \label{asFig.1}
\end{figure}
%

At the time Weinberg conjectured the possibility of Asymptotic
Safety, due to the lack of nonperturbative computational techniques,
a NGFP was known to exist only for a single coupling, Newton's
constant, and only in $d=2+\epsilon$ spacetime dimensions
\cite{wein}. The situation changed when the EAA-based methods became
available \cite{mr}. Starting from early work on the
`Einstein-Hilbert truncation' \cite{mr,oliver1} and a generalization
with an additional $R^2$-term \cite{oliver2}, a considerable number
of truncations with increasingly large subsets $\{P_{\alpha}\}$ were
analyzed in the following decade \cite{codello}. Quite remarkably,
they all agree in that \emph{QEG indeed seems to possess a NGFP
suitable for the Asymptotic Safety construction.} \index{QEG quantum Einstein gravity} Although a
complete proof is not within reach, by now there is highly
nontrivial evidence for a NGFP on the full (un-truncated) theory
space, rendering QEG nonperturbatively renormalizable
\cite{reviews1,reviews2,reviews3}. As a representative example, Fig.
\ref{asFig.1} shows the phase portrait of the Einstein-Hilbert
truncation \cite{mr} based upon the running action $\Gamma_k =
\frac{1}{16\pi G_k} \int \text{d}^d x \sqrt{g}\,
(-R(g)+2\Lambda_k)+S_{\text{gf}}+S_{\text{gh}}$ which has $N=2$. It
involves the approximation of neglecting the $k$-dependence in the
gauge fixing and ghost sectors which can be justified by BRST
methods \cite{mr}. This ansatz contains a running Newton constant
$G_k$ and cosmological constant $\Lambda_k$, \index{cosmological constant} their dimensionless
analogs being, in $d$ spacetime dimensions, $g(k)\equiv k^{d-2}G_k$
and $\lambda(k)\equiv \Lambda_k \slash k^2$, respectively. Their
beta functions $(\beta_g,\beta_{\lambda})\equiv \boldsymbol{\beta}$
have been computed for any $d$ \cite{mr}. The first steps of the
calculation are reminiscent of those in perturbatively quantized
general relativity \cite{perturbGR} but this is a coincidence
due to the specific ansatz for $\Gamma_k$.
Moreover,  $\beta_{g}$ and $\beta_{\lambda}$ are quite different
from  beta-functions in perturbation theory. They sum up
contributions from arbitrary orders of perturbation theory and, what
is more important, they contain also information about the strong
power law-type renormalization effects which are not seen usually in
perturbative calculations employing dimensional regularization. This
is important however, for instance in order to `tame' the notorious
quadratic (and higher) divergences due to the non-zero mass
dimension of $G$.

Fig. \ref{asFig.1} shows the flow diagram obtained by solving the
coupled equations $k\partial_k g(k)=\beta_g (g,\lambda)$ and
$k\partial_k \lambda(k) =\beta_{\lambda}(g,\lambda)$ for $d=4$
\cite{frank1}. Besides a Gaussian fixed point at
$g_{\ast}=0=\lambda_{\ast}$ there is indeed a second, {\it
non-Gaussian} fixed point at $g_{\ast}>0$, $\lambda_{\ast}>0$. Both
of its critical exponents have a positive real part. Hence $s
\equiv{\rm dim}\, S_{\rm UV}\,=\, 2$. In sufficiently general
truncations one finds that $s<N$, and the reduced dimensionality
then allows us to predict $N-s$ couplings after $s$ couplings have
been measured. The predictions are encoded in the way the
$s$-dimensional hypersurface $\mathcal{S}_{\text{UV}}$ is immersed
into the (truncated or complete) theory space. There are general
arguments suggesting that $s$ should saturate at a small finite
value when $N$ is increased \cite{wein}. Concrete calculations
confirmed this picture, first in $d=2+\epsilon$  where the
`$\Lambda+R+R ^2$ truncation', having $N=3$, yields a NGFP with
$s=2$ \cite{oliver2}. So, given two input parameters, the third one
is a prediction. One might for instance express the coefficient of
the term $\int \sqrt{g}R^2$ added to the Einstein-Hilbert action in
terms of $g$ and $\lambda$. In $d=4$, all known truncations
confirmed that the projection of the flow onto the
$g$-$\lambda$-plane has the same structure as in Fig. \ref{asFig.1},
with `perpendicular' directions added.
By now, there exist very impressive analyses of $f(R)$ truncations,
with $f$ a polynomial of high degree. They do indeed display the
expected stabilization of $s$ at a small finite value when $N$ is
made large \cite{dimSurface}. Furthermore, first explorations of
infinite dimensional truncated theory spaces were performed
\cite{creh2} and truly functional flows in non-polynomial $f(R)$
truncations are within reach now \cite{frank-fR}. Trying to make the
truncations more accurate it is not sufficient to generalize their
$g_{\mu\nu}$-dependence only; at the same time we must also allow
for a more general dependence of $\Gamma_k$ on the background
metric. The first results on such `bi-metric truncations' which
treat the $g_{\mu\nu}$- and $\bar{g}_{\mu\nu}$-dependence on a
similar footing further support the viability of the Asymptotic
Safety program \cite{bimetric}. The same is true for a different
type of generalization, the inclusion of scale dependent surface
terms into $\Gamma_k$ for spacetimes with boundaries \cite{daniel1}.
There is yet another important, but technically difficult
generalization, namely  {\it non-local} terms. They are particularly
important in the infrared where they are expected to cure a problem 
of the Einstein-Hilbert truncation not visible in Fig. \ref{asFig.1}:
singularities of the beta-functions at $\lambda=1\slash2$ which 
indicate that the truncation becomes insufficient in the IR.
In \cite{frank2} a simple but genuinely `functional' flow of a
non-local EAA was analyzed which turned out to possess an infrared
fixed point, i.e. a non-local `fixed functional'.

Besides a better understanding of the RG flow in QEG, 
\index{QEG quantum Einstein gravity|)} future work
will also have to address the question of observables. The running
couplings parameterizing the EAA have no direct physical
significance in general. While under very special circumstances it
might be possible to deduce observable effects directly from the
$k$-dependence of certain couplings (by some kind of `RG
improvement'), the general strategy is to first construct the
functional integral, then find interesting observables  in terms of
the fundamental fields, and finally compute their expectation
values. In this respect the status of observables within the
Asymptotic Safety program is not different from any other functional
integral based approach.

\subsubsection{Causal Dynamical Triangulations \label{as.secD}}

\index{causal dynamical triangulations|(}%
\index{CDT|see{causal dynamical triangulations}}
The partition functions of standard model-like quantum field
theories, analytically continued to Euclidean space and discretized,
have been extensively studied by Monte Carlo techniques. It is
therefore natural to apply similar ideas to gravity and to attempt a
definition of the formal functional integral \eqref{as.2} as the
$a\rightarrow 0$ limit of  the partition function belonging to a
suitably chosen statistical mechanics model, specified by a choice
of dynamical variables, bare action $S[g]$ and measure
$\mathcal{D}g$. Here the discretization scale $a$ is analogous to a
lattice spacing. A priori the `lattice units' defined by $a$ are
unphysical; they can be converted to physical lengths or masses only
later when it comes to computing observables.

The limit $a\rightarrow0 $ is to be taken indirectly, as follows.
The statistical system has a chance of describing physics in the
continuum if $a$ can be made much smaller than any relevant physical
length scale $\ell$, or more adequately, if all lengths $\ell$ are
much larger than $ a$. In fact, in numerical simulations where $a$
is necessarily nonzero ($ a=1$, say) the requirement $(\ell\slash a)
\gg 1$ is met if the free parameters of the statistical model (bare
couplings) are tuned such that its correlation length diverges and
$\ell$, in lattice units, becomes very large. 
\index{continuum limit} Thus the {\it
continuum limit} $a\slash \ell \rightarrow 0$ amounts to $\ell
\rightarrow \infty$ with $a$ fixed (rather than $a\rightarrow 0$ and
$\ell$ fixed). As it is well known from the statistical physics  of
critical phenomena, for instance, the correlation length does indeed
diverge at second order phase transition points. So the strategy
will be to propose a plausible statistical model, compute
numerically its partition function in dependence of the bare
parameters, and search for points in parameter space where the
correlation length diverges. If such a critical point exists one
would use it to define a continuum theory and explore its
properties.

The statistical systems underlying critical phenomena are
conveniently analyzed in terms of their RG flow under successive
`coarse graining'. While there is considerable ambiguity in how this
is done concretely, it typically boils down to a space averaging of
the degrees of freedom (block spin transformation, etc.) which, in a
continuum language, amounts to a step-by-step integrating out of
field modes with increasing wavelengths. In this setting systems at
second order phase transition points, displaying no preferred length
scale, are described by fixed points of the RG flow.

This observation brings us back to Asymptotic Safety: The discrete
system describes a continuum theory when its bare parameters are
tuned to their fixed point values. Then the partition function $Z$
is a sum over contributions from fluctuations whose wavelengths, in
physical units, range from zero to infinity. In EAA language this
amounts to specifying a {\it complete} trajectory $\Gamma_k$, well
behaved in particular in the UV since $\lim_{k\rightarrow
\infty}\Gamma_k = \Gamma_{\ast}$. One can show that $\Gamma_{\ast}$
is indeed very closely related to the RG fixed point of the
statistical model, and that the large $k$ behavior of $\Gamma_k$,
with minimal additional input, can be mapped onto the RG flow of the
model near the second order phase transition point \cite{elisa1}.

The CDT approach \cite{ajl,agjl,agjl-rev} is a specific proposal for
a statistical system representing gravity. It sums over the class of
piecewise linear 4-geometries which can be assembled from
\index{simplicial decomposition}
4-dimensional simplicial building blocks (with link length $a$) in
such a way that the resulting spacetime is `causal' in a certain
technical sense. A priori the spacetimes $\mathcal{M}$ summed over
have Lorentzian signature. However, to ensure the existence of a
generalized Wick rotation \index{Wick rotation}
they are restricted to be globally
hyperbolic which allows introducing a global proper-time foliation,
$\mathcal{M}=I \times \Sigma$, where $I$ denotes a `time' interval
and space is represented by 3-dimensional leaves $\Sigma$ whose
topology is not allowed to change in time. A choice extensively
studied is $\Sigma = S^3$ so that at each proper-time step $t_n\in
I$ the spatial geometry is represented by a triangulation of $S^3$.
It is made up of equilateral spatial tetrahedra with positive
squared side-length $\ell_s^2 \equiv a^2>0$. The number $N_3(t)$ of
tetrahedra, and the way they are glued together to form a piecewise
flat 3-dimensional manifold will change in general when  we go from
$t=t_n$ to the next time slice at $t_{n+1}$. In order to constitute
a 4-dimensional triangulation, the 3-dimensional slices must be
connected in a `causal' way, preserving the $S^3$-topology at all
intermediate times. (This ensures that a branching of the spatial
universe into several disconnected pieces (baby universes) does not
occur.) For the gluing of two consecutive time slices $S^3(t_n)$ and
$S^3(t_{n+1} )$ it is sufficient to introduce four types of
4-simplices, namely the so-called (4,1)-simplices, which have 4 of
its vertices on $S^3(t_n)$ and 1 on $S^3(t_{n+1})$, the
(3,2)-simplices with 3 vertices on $S^3(t_n)$ and 2 on
$S^3(t_{n+1})$, as well as (1,4)- and (2,3)- simplices defined the
other way around. The integration over spacetimes $\mathcal{M}$
boils down to a sum over all possible ways to connect given
triangulations of $S^3(t_n)$ and $S^3(t_ {n+1})$ compatible with the
topology $I\times S^3$, along with a summation over all
3-dimensional triangulations of $S^3(t)$, at all times $t$.

Denoting by $\ell_t$ and $\ell_s$ the length of the time-like and
the space-like links, respectively, one has $\ell_t^2=-\alpha
\ell_s^2$ where the constant $\alpha$ is positive in the Lorentzian
case, whence $\ell_t^ 2<0$. It was shown \cite{ajl} that there
exists a well defined rotation in the complex $\alpha$ plane
$(\alpha \rightarrow -\alpha)$ which, thanks to the restriction to a
given foliation in the simplicial decomposition, connects the
Lorentzian to the Euclidean signature, with $\ell_t^2 =
|\alpha|\ell_s^2>0$. \index{simplicial decomposition}
This turns oscillating exponentials
$e^{\mathrm{i} S}$ into Boltzmann factors $e^{-S}$, so that the
resulting partition function can be computed with Monte Carlo
integration methods. It reads \cite{ajl}: \index{Regge calculus}
\begin{align}
Z(\kappa_0,\kappa_4,\Delta)&= \sum_{T}\frac{1}{C_{T}} \exp \big( -
S_{\text{Regge}}[T]\big) \label{as.5}
\end{align}
The symmetry factor $C_{T}$ equals the order of the automorphism
group of the triangulation $T$, and $S_{\text{Regge}}$ is the
Regge-discretized Einstein-Hilbert action:
$ S_{\text{Regge}}= - (\kappa_0 + 6\Delta) N_0 + \kappa_4
\big(N_4^{(4, 1)}+N_4^{(3,2)}\big) + \Delta
\big(2N_4^{(4,1)}+N_4^{(3,2)}\big)$.
Here $N_4^{(4,1)}$ and $N_4^{(3,2)}$ denote the number of (4,1)- and
(3,2)-simplices in $T$, respectively, and $N_0$ is the total number
of vertices. The couplings $\kappa_0$ and $\kappa_4$ correspond to
$1\slash G$ and $\Lambda\slash G$, respectively, and $\Delta$
parameterizes a possible asymmetry between $\ell_t$ and $\ell_s$; it
is nonzero if $|\alpha|\neq 1$.

Extensive Monte Carlo simulations of the partition function
\eqref{as.5} have been performed at a number of points in the space
of bare couplings ($\kappa_0$, $\kappa_4$, $\Delta$).  Three
different phases were discovered, and one of them seems indeed
capable of representing continuum physics.
A surface in parameter space, $\kappa_4=\kappa_4(\kappa_0,\Delta)$,
has been identified on which the 4-volume becomes large. This
`infinite' volume limit should however not be confused with the
continuum limit. \index{continuum limit}
The crucial question is whether the latter can
actually be realized \`{a} la Asymptotic Safety by tuning the
remaining two parameters to a second order phase transition point.
The answer is not known yet, but this is the topic of very active
current research.

The most important result of the CDT model is that it is able to
describe the emergence of a classical 4-dimensional de Sitter
universe with small superimposed quantum fluctuations. The
calculation is carried out in the Euclidean signature, but thanks to
the above $\alpha$-rotation it admits a Lorentzian interpretation.
The reason this result is interesting is that it resolves a difficulty of previous attempts to address quantum gravity with dynamical triangulations: the 4-d Euclidean triangulation models without the `causality' constraint produced only states with Hausdorff dimensions $d_{\text{H}}=2$ and $d_{\text{H}}=\infty$, respectively, contradicting the classical limit.

In these CDT simulations the link length $a$ is still as large as
about 2 Planck lengths so they do not yet probe the physics on
sub-Planckian length scales \cite{agjl}. Once simulations well
beyond the Planck scale become feasible they should be able to make
contact with the RG fixed point predicted by the EAA based
calculations in the continuum. Indeed, it has been shown already
\cite{oliver-frac,frank-frac} that the CDT and EAA predictions for
the running spectral dimension agree quite precisely in the
semiclassical regime. It is also known how, at least in principle,
the information about the $k$-dependence of the  EAA can be used to
predict the expected RG running of a statistical model near the
continuum limit \cite{elisa1}.  \index{continuum limit}
In this respect it should also be
mentioned that while most EAA studies have been performed for {\it
Euclidean signature}, they also apply to the {\it Lorentzian} case
almost unchanged \cite{frank-sig}.
It will be very interesting to see whether future Monte Carlo
results lead to the same picture of physics near the fixed point as
the FRGE studies. 
\index{UV and IR fixed points|)}
\index{asymptotic safety|)} \index{causal dynamical triangulations|)}

Finally, CDT breaks Lorentz invariance because the spatial and temporal cut-offs are independent. There is no general argument that           
Lorentz invariance must be restored in the continuum limit. If it is not restored, the classical limit of the continuum theory 
might not be general relativity, but rather something akin to the Ho\v{r}ava--Lifshitz theory, which is non-Lorentz invariant \cite{Horava}. Ho\v{r}ava--Lifshitz theory is renormalizable, and if it were physically viable, it would represent a possible solution to the quantum gravity problem.

\subsection{Hamiltonian Theory and Quantum Geometry} \label{s2.2}
\index{loop quantum gravity!Hamiltonian theory|(}

The \emph{Asymptotic Safety} program generalizes the procedures that
have been successful in Minkowskian quantum field theories (MQFTs)
by going beyond traditional perturbative treatments. An avenue that
is even older is canonical quantization, pioneered by Dirac,
Bergmann, Arnowitt, Deser, Misner and others. Over the past 2-3
decades, these ideas have inspired a new approach, known as 
\emph{Loop Quantum Gravity} (LQG).

While the point of departure is again a Hamiltonian framework, as
explained in section \ref{s1}, there is an important conceptual
shift: the idea now is to construct a \emph{quantum} theory of
geometry and then use it to formulate quantum gravity
systematically. This theory was constructed in detail in the 1990s.
Since then, research in LQG has progressed along two parallel 
avenues. In the first, discussed in this sub-section,
one continues the development of the canonical quantization program,
now using quantum geometry to properly handle the field theoretical
issues. In the second, discussed in the next sub-section, one
develops a path integral framework and defines dynamics via
transition amplitudes between quantum 3-geometries. In the final
picture, the fundamental degrees of freedom are \emph{quite
different} from those that would result in a `direct' quantization
of GR --they are not metrics and extrinsic curvatures but chunks, or
atoms, of space with quantum attributes. Classical geometries emerge
only upon coarse graining of their coherent superpositions.

This subsection is divided into three parts. The first summarizes
the Hamiltonian framework that provides the point of departure, the
second explains the basic structure of quantum geometry and the
third sketches the status of quantum dynamics in canonical LQG.

\index{effective average action|)}

\subsubsection{Connection Dynamics}
\label{s2.2.1}

The key idea underlying the Hamiltonian framework used in LQG
is to cast GR in the language of gauge theories that successfully
describe the electroweak and strong interactions. This requires a
shift from metrics to connections; Wheeler's `geometrodynamics'
\cite{jw} is replaced by a dynamical theory of spin-connections
\cite{aa-newvar}. \index{geometrodynamics}
Once this is achieved, the phase space of general
relativity becomes \emph{the same} as that of gauge theories: All
four fundamental forces of Nature are unified at a kinematical
level. However, dynamics of GR has two distinguishing features.
First, whereas the Hamiltonian of QED or QCD \index{QCD} \index{QED}
uses the flat
background metric, the Hamiltonian constraints that generate
dynamics of GR are built entirely from the spin connection and its
conjugate momentum; the theory is manifestly \emph{background
independent.} Second, the gauge group now refers to rotations in the
physical space rather than in an abstract, internal space. This is
why in contrast to, say, QCD, \emph{spacetime} geometry can now
emerge from this gauge theory. As we will see, these two features
have a powerful consequence: one is led to a \emph{unique} quantum
Riemannian geometry.

Fix a 3-manifold $M$ which is to represent a Cauchy surface in
spacetime. The gravitational phase space $\gps$ is coordinatized by
pairs $(\LA_a^j, E^a_j)$ of an $\SU(2)$ connection $\LA_a^j$ and its
conjugate `electric field' $E^a_j$ on $M$, where $j$ refers to the
Lie algebra $\su(2)$ of $\SU(2)$ and $a$ to the tangent space of
$M$. Thus, the fundamental Poisson brackets are:
\be \label{pbs1} \{\LA_a^j(x), \, E^a_k\} = - i \kappa_{\rm N}\,
\delta_a^b\, \delta^j_k\, \delta^3(x,y) \ee
where $\kappa_{\rm N} = 8\pi\, G_{\rm N}$ is the gravitational
coupling constant. As remarked above, although the phase space
variables have the familiar Yang-Mills form, they also admit a
natural interpretation in terms of spacetime geometry. To spell it
out, let us first recall from Chapter 8 that the standard Cauchy
data of GR consists of a pair, $(q_{ab},\, K_{ab})$, representing
the intrinsic positive definite metric $q_{ab}$ and the extrinsic
curvature $K_{ab}$ on $\Sigma$. If we denote by $e^a_j$ an
orthonormal triad
---a `square root' of $q^{ab}$--- then in the Lorentzian signature
we have:
\be \label{ADM}  E^a_j = \sqrt{q}\, e^a_j \quad\quad {\rm and}
\quad\quad 
\LA_a^j = \Gamma_a^j - \iota K_a^j\ee
where $q$ denotes the determinant of the metric $q_{ab}$,
$\Gamma_a^j$ is the intrinsic spin connection on $M$ defined by
$e^a_j$,  $K_a^j = K_{ab}\,e^b_j$ and $\iota= 1$ in the Euclidean
signature and $\iota = i (\equiv \sqrt{-1})$ in the Lorentzian
signature (used in most of this Chapter). The connection $\LA_a^j$
parallel transports left handed (or unprimed) spacetime spinors.
In the
final solution, its curvature ${\underbar{F}}_{ab}^j := 2\partial_{[a}\, \LA_{b]}^j + \epsilon^{jkl}\,\LA_{ak}\, \LA_{bl}$ represents the
(pull-back to $M$ of the) self-dual part of the spacetime Weyl curvature.%
\footnote{For simplicity we assume that $M$ is compact; in the
asymptotically flat case, one has to specify appropriate boundary
conditions at infinity and keep track of boundary terms. See. e.g.,
\cite{aa-newvar,alrev,ttbook}. Since the electric field $E^a_i$ is a
density of weight 1, mathematically, it is often simpler to work
with its dual $\Sigma_{ab}^j := \eta_{abc} E^c_j$, which is just a
2-form on $M$. Finally, as is standard in Yang-Mills theories, the
internal indices $j,k,\ldots$ are raised and lowered using the
Cartan-Killing metric on $\su(2)$.}

As is well-known, dynamics of GR is generated by a set of
constraints. While they are rather complicated and non-polynomial
functionals of the geometrodynamical ADM variables\index{ADM formulation}, they become low
order polynomials in the connection variables. In absence of matter
sources, they are \cite{aa-newvar}:
\be \label{constraints} \mathcal{G}_j := \underbar{D}_aE^a_j =0, \,\,
\mathcal{D}_b := E^b_j\,\underbar{F}_{ab}^j =0, \,\, {\rm and} \,\,
\mathcal{H}:= \epsilon^{jkl}\, \big(\underbar{F}_{abl}\, - \Lambda
\eta_{abc}E^c_l \big)\, E^a_jE^b_k =0\ee
The first constraint is just the familiar Gauss law of Yang Mills
theory, the second is the Diffeomorphism constraint of GR, and the
third the Hamiltonian constraint. Interestingly, these are the
simplest gauge invariant, local expressions one can construct from a
connection and its conjugate electric field \emph{without} reference
to a background metric. Indeed, these are the \emph{only} such
expressions that are at most quartic in the canonical variables
$A_a^i,\,E^a_j$. At first one might expect that it would be
difficult to couple matter to gravity using these connection
variables since they refer only to the self dual part of spacetime
curvature. But this is not the case; one can couple spin zero, half
and one fields keeping the simplicity \cite{art,aabook} and recently
the framework has also been extended to include higher dimensions
\cite{ttetal1}, and supersymmetry \cite{ttetal2}.

Since the constraints are polynomial in the connection variables, so
are the equations of motion. Furthermore, the framework represents a
small extension of GR: Since, in contrast to the ADM variables\index{ADM formulation}, none
of the equations require us to invert $E^a_j$, they remain viable
even when $E^a_j$ become degenerate. At these phase space points one
no longer has a (non-degenerate) spacetime metric but connection
dynamics continues to remain meaningful. The standard causal
structures have been extended to such configurations \cite{causal}.
Finally, the connection dynamics \index{connection dynamics}
framework provides a natural
setting for proofs of the positive energy theorems a la Witten
\cite{witten}; one can establish the positivity of the gravitational
Hamiltonian not only on the constraint surface as in the original
theorems but also in a neighborhood of the constraint surface, i.e.,
even `off-shell' \cite{aa-gth}.

As we noted after Eq. (\ref{ADM}), in the Lorentzian signature the
connection $\LA_a^i$ is complex valued, or, equivalently, it a
1-form that takes values in the Lie algebra of $\mathbb{C}{\rm
SU(2)}$, the complexification of $\SU(2)$. While this feature does
not create any obstacle at the classical level, a key mathematical
difficulty arises in the passage to quantum theory: Because
$\mathbb{C}{\rm SU(2)}$ is \emph{non-compact}, the space of
connections $\LA_a^j$ is not known to carry diffeomorphism invariant
measures that are necessary to construct a satisfactory Hilbert
space of square integrable functions of connections. To bypass this
difficulty, the main-stream strategy has been to replace the
complex, left handed connections $\LA_a^j$ with \emph{real} $\SU(2)$
connections $A_a^j$, obtained by replacing $i$ in (\ref{pbs1}) and
(\ref{ADM}) by a real, non-zero parameter $\gamma$ \cite{fb}. Then,
both the phase space variables are real and the fundamental
Poisson-brackets become
\be \label{pbs2} \{\LA_a^j(x), \, E^a_k\} = \gamma \kappa_{\rm N}\,
\delta_a^b\, \delta^j_k\, \delta^3(x,y)\, . \ee
$\gamma$ is known as the Barbero-Immirzi parameter \cite{gi} and
taken to be positive for definiteness. As we will see in section
\ref{s2.2.2}, one can introduce well-defined measures on the space
$\mathcal{A}$ of these real connections $A_a^j$ and develop rigorous
functional analysis to introduce the quantum Hilbert space and
operators without any reference to a background geometry. This
passage from left handed to real connections represents a systematic
generalization of the Wick rotation \index{Wick rotation}
one routinely performs to obtain
well-defined measures in MQFTs. However, the rotation is now
performed in the `internal space' rather than spacetime. Indeed, the
\emph{spacetime} Wick rotation does not naturally extend to general
curved spacetimes while this internal Wick rotation does and serves
the desired purpose of taming the functional integrals.

However, the strategy has two limitations. First, the form of the
constraints (and evolution equations) is now considerably more
complicated. But thanks to several astute techniques introduced by
Thiemann \cite{tt,ttbook}, these complications can be handled in the
canonical approach, and they are not directly relevant to spin
foams. The second limitation is that, while the connection $A_a^j$
is well-defined on $M$ and continues to have a simple relation to
the ADM variables\index{ADM formulation}, it does not have a natural 4-dimensional
geometrical interpretation in solutions to the field equations
\cite{samuel}. Nonetheless, one \emph{can} arrive at the canonical
pair $(A_a^j, E^a_j)$ by performing a Legendre transform of a
4-dimensionally covariant action $S(e,\, \omega)$ that depends on a
space-time co-tetrad $e^I_\mu$ and a Lorentz connection
$\omega_\mu^{IJ}$ \cite{holst}.
%
%


\subsubsection{Quantum Riemannian Geometry}
\label{s2.2.2}
\index{quantum Riemannian geometry|(}
The first step in the passage to quantum theory is to select a
preferred class of \emph{elementary phase space functions} which are
to be directly promoted to operators in the quantum theory without
factor ordering ambiguities. In geometrodynamics, these are taken to
be the positive definite 3-metric $q_{ab}$ on $M$ and its conjugate
momentum, $P^{ab}= \sqrt{q}\,(K^{ab} - K q^{ab})$ (integrated
against suitable test fields). In connection dynamics the choice is
motivated by structures that naturally arise in gauge theories.
Thus, the configuration variables are now the \emph{Wilson lines, or
holonomies $h_{\ell}$} \index{holonomies}
which enable one to parallel transport left
handed spinors along 1-dimensional (curves or) \emph{links} $\ell$
in $M$, and the conjugate momenta are the \emph{`electric field
fluxes'} $E_{f,S}$ across 2-dimensional \emph{surfaces} $S$ (smeared
with test fields $f^i$ that take values in $\su(2)$)
\cite{rs1,ai,al4,alrev,crbook,ttbook}: \index{connection dynamics}
\label{hol-flux} \be h_\ell := \mathcal{P}\, \exp \int_\ell A, \quad
\quad {\rm and}\quad\quad E_{f,S} := \int_S\, d^2S_a\, f^i(x)
E^a_i(x)\, .\ee
Note that the definitions do not require a background geometry;
since $A$ is an $\su(2)$-valued 1-form, it can be naturally
integrated along 1-dimensional links to yield $h_\ell \in \SU(2)$,
and since $E$ is the Hodge-dual of a ($\su(2)$-valued) 2-form the
second integral is also well-defined without any background fields.
However, the Poisson brackets between these variables fail to be
well-defined if $\ell$ and $S$ are allowed to have an infinite
number of intersections. Therefore they have to satisfy certain
regularity conditions. Two natural strategies are to use piecewise
linear links and 2-surfaces or piecewise analytic ones (more
precisely, `semi-analytic' in the sense of \cite{lost,alrev,gs}).
The first choice is well-adapted to the simplicial decompositions
often used in Spinfoam models while the second is commonly used in
canonical LQG. \index{simplicial decomposition}

Formal sums of products of these elementary operators $\hat{h}_\ell$
and $\hat{E}_{n,S}$ generate an abstract algebra $\Al$. This is the
analog of the familiar Heisenberg algebra in quantum mechanics and
one's first task is to find its representations. The Hilbert space
$\Hkg$ underlying the chosen representation would then serve as the
space of kinematical quantum states, the quantum analog of the
gravitational phase $\gps$ of GR, the arena to formulate dynamics.

In quantum mechanics, von-Neumann's theorem guarantees that the
Heisenberg algebra admits a unique representation satisfying certain
regularity conditions (see, e.g., \cite{emch}). 
However, in MQFTs, because of the infinite number of degrees of
freedom, this is not the case in general: The standard result on the
uniqueness of the Fock vacuum assumes \emph{free field dynamics}
\cite{nonunique,segal}.
What is the situation with the algebra $\Al$ of LQG? Now, in
addition to the standard regularity condition, we can and \emph{have
to} impose the strong requirement of background independence.
\index{background independence!\& quantum geometry|(} A
fundamental and surprising result due to Lewandowski, Okolow,
Sahlmann, and Thiemann \cite{lost} and Fleishhack \cite{cf} is that
the requirement is in fact so strong that it suffices to single out
a unique representation of $\Al$, without having to fix dynamics.
Thus, \emph{thanks to background independence, quantum kinematics is
unique} in LQG.

This powerful result lies at the foundation of much of LQG because
the unique representation it selects leads to the fundamental
discreteness in quantum geometry. Therefore let us discuss the key
features of this representation and compare and contrast it with
representations used in MQFTs. The underlying Hilbert space $\Hkg$
is the space $L^2(\Ab,\, d\mu_o)$ of square integrable functionals
of (generalized) connections with respect to a regular, Borel
measure $\mu_o$. As one would expect, the holonomy operators
$\hat{h}_\ell$ act by multiplication while their `momenta'
$\hat{E}_{f,S}$ act by differentiation. There is
a state $\Psi_o$ in $\Hkg$ 
which is cyclic in the sense that $\Hkg$ is generated by repeated
actions of $\hat{h}_\ell$ on $\Psi_o$. These properties are shared by
MQFTs
%
%
where the Fock space can also be represented as the space of
square-integrable functionals over the space of
(distribution-valued) fields on $\mathbb{R}^3$ and the vacuum plays
the role of $\Psi_o$. In these theories, the vacuum state is
Poincar\'e invariant and this invariance implies that the Poincar\'e
group is unitarily implemented in the quantum theory. In LQG, the
state $\Psi_o$ is invariant under the kinematical symmetry group
${\rm SU(2)}_{\rm loc} \ltimes \Diff(M)$ of connection dynamics
---the semi-direct product of the local $\SU(2)$ gauge
transformations and diffeomorphisms of $M$--- and this group is
unitarily represented on $\Hkg$. This fact provides a natural point
of departure in the imposition of quantum constraints, discussed
below. \index{connection dynamics}

However, the representation also has two unfamiliar features: i)
$\Hkg$ is non-separable, and, ii) while the holonomies $\hat{h}_\ell$
\index{holonomies}
are well-defined operators on $\Hkg$, the connection operators
themselves do not exist (because $\hat{h}_\ell$ fail to be continuous
with respect to the links $\ell$). These aspects of LQG kinematics
have caused some unease among researchers outside LQG (see, e.g.
\cite{hnrev}) because it is not widely appreciated that they are
\emph{not} peculiarities of LQG but follow, in essence, just from
background independence. 
In particular, if one seeks a
representation of the properly constructed kinematical algebra of
geometrodynamics  \index{geometrodynamics}
in which the cyclic state is invariant under the
kinematical symmetry group $\Diff(M)$, \emph{one again finds that
the representation inherits these two features} \cite{aa-ehlers}.
%
%
Intuition derived from the $\Diff(S^1)$ group used, e.g., in string
theory does not carry over to higher dimensions in tnis respect.

Let us now discuss quantum states and operators in some detail.
Recall that in MQFTs, while the characterization of the Fock space
as the space of square integrable functionals of (generalized)
fields is succinct, detailed calculations are most efficiently
performed in a convenient basis that diagonalizes the number
operators. The situation with $\Hkg$ is analogous. More precisely,
it is convenient to decompose $\Hkg$ into orthogonal subspaces,
$\Hkg = \bigoplus\, \H_{\alpha}$, associated with graphs $\alpha$ in
$M$ with a finite number of oriented links $\ell$. Next, if one
labels each link $\ell$ of $\alpha$ with a non-trivial, irreducible
representation $j_\ell \not=0$ of $\SU(2)$, one obtains a further
decomposition \cite{rs2,jb2}
\be \label{decomposition} \Hkg = \bigoplus_{\alpha}\, \H_{\alpha} =
\bigoplus_{\alpha,\, j_\ell}\, \H_{\alpha,\, j_\ell}\, .  \ee
If $\alpha$ has $L$ links, $\H_{\alpha,\, j_\ell}$ is a
\emph{finite} dimensional Hilbert space which can be identified with
the space of quantum states of a system of $L$ spins. Therefore
(\ref{decomposition}) is called a \emph{spin-network decomposition}
of $\Hkg$. To make this relation explicit, note first that a
(generalized) connection $A$ assigns to each link $\ell$ a holonomy
$h_\ell$ and elements $\Psi$ of $\Hkg$ are functions of these
(generalized) connections. States $\Psi$ in $\H_{\alpha,\, j_\ell}$
are of the form
\be \label{state} \Psi(A) = \psi (h_{\ell_1},\, \ldots,\,
h_{\ell_L}) \ee
where $\psi$ is a function of the $L$ $\SU(2)$ group-elements in its
argument, which is square integrable with respect to the Haar
measure on $[\SU(2)]^L$. They know only about the action of the
connection $A$ pulled back to the $L$ links of $\alpha$. Thus, by
restricting attention to a single graph $\alpha$, one truncates the
theory and focuses only on a finite number of degrees of freedom.
The spirit is the same as in MQFTs. In any calculation with Feynman
diagrams of a weakly coupled theory (such as low energy QED) one
truncates the theory by allowing only a \emph{finite} number of
virtual particles. Similarly, in strongly coupled theories (such as
low energy QCD) \index{QCD} \index{QED}
one truncates the theory by making a lattice
approximation. In both cases, the full Hilbert space is recovered in
the limit in which the degrees of freedom are allowed to go to
infinity. In LQG this is achieved by taking a well-defined
(projective) limit in the space of graphs \cite{al4}. Finally, the
second equality in (\ref{decomposition}) is obtained by carrying out
Fourier transforms (using the Peter-Weyl theorem) on $[\SU(2)]^L$.

Such truncations are useful if the operators of interest leave the
truncated Hilbert spaces invariant. This is indeed the case with
geometric operators of LQG. 
As one would expect from the phase space description, these
operators are constructed from $\hat{E}_{f,S}$ since the electric
field $E^a_i$ also serves as the orthonormal triad in the classical
theory. The action of $\hat{E}_{f,S}$ on a state $\Psi \in
\H_{\alpha}$ is non-trivial only if the surface $S$ intersects one
or more links of the graph $\alpha$ and then the action involves
only group theory at the intersection \cite{alrev,ttbook,gs}. This
is just the structure one would expect from background independence!
To construct geometric operators such as those corresponding to
areas of 2-surfaces and volumes of 3-dimensional regions, one first
expresses their classical expressions in terms of the `elementary'
phase space functions $\hat{E}_{f,S}$ and then promotes the classical
expression to a quantum operator. In the intermediate stages one has
to introduce auxiliary structure but the procedure ensures that the
final expressions are background independent
\cite{alrev,crbook,ttbook}.

Let us now consider the operator $\widehat{\Ar}_{S,\, \alpha}$ on
$\H_\alpha$, representing the area of a 2-surface $S$ (without
boundary) \cite{rs4,almmt1,al5} which has played a particularly
important role in LQG. Let us first suppose that the surface $S$
intersects $\alpha$ only at a node $n$. Then, one can naturally
define a \emph{node-Laplacian operator} $\Delta_{\alpha,\,S,\,n}$
whose action on $\Psi$ of (\ref{state}) is an appropriate sum of the
Laplacians on the copies of $\SU(2)$ associated with links $\ell_i$
that intersect $S$ at $n$ \cite{al4,al5,alrev}. As one might expect,
$\Delta_{\alpha,\,S,\,n}$ is a negative definite self-adjoint
operator on $\H_\alpha$. The final expression of the area operator
$\widehat{\Ar}_{S,\,\alpha}$ is given by \vskip-0.1cm
\be \label{area} \widehat{\Ar}_{S,\,\alpha} = 4\pi\, \gamma\, \lp^2\,
\sqrt{\Delta_{\alpha,\,S,\,n}}\ee
If there are multiple intersections $n_i$ between $\alpha$ and $S$,
$\widehat{\Ar}_{S,\,\alpha}$ is just the sum of these operators for each
$n_i$. The non-trivial result is that operators defined on various
$\H_{\alpha}$ can be naturally \emph{glued together} to obtain a
self-adjoint operator $\widehat{\Ar}_S$ on the entire $\Hkg$.

Properties of $\widehat{\Ar}_{S,\,\alpha}$ have been analyzed in detail.
Its spectrum is discrete in the sense that all its eigenvectors are
normalizable. 
%
%
%
%
%
In the special case when all intersections between $\alpha$ and $S$
are at bi-valent nodes at which `straight' links pierce $S$,
the expression of eigenvalues simplifies to a from that is useful in
many applications \cite{al5,flr}:
\be \label{simplified} a_S = 8\pi\,\gamma\, \lp^2\, \sum_n
\sqrt{j(j+1)}\, . \ee
There is a smallest non-zero eigenvalue among these:
\be \Delta a_S\ =\ 4\pi\gamma\, \lp^2 \,\, \sqrt{3}.\label{gap} \ee
This \emph{area gap} pays an important role in the theory. The level
spacing between consecutive eigenvalues is \emph{not} uniform but
decreases \emph{exponentially} for large eigenvalues \cite{al5}.
This implies that, although the eigenvalues are fundamentally
discrete, the continuum approximation becomes excellent \emph{very
rapidly}. \index{area gap}

For the volume and length operators, the strategy is the same and
the background independence of LQG again fixes the precise form of
the final expressions \cite{rs4,al6,alrev,ttbook}. 
\index{background independence!\& quantum geometry|)}
However, the detailed procedure is technically more complicated.
The length operator has not had significant applications. The volume
operator has been investigated in greater detail because features
prominently in the dynamical considerations of the canonical theory
\cite{tt,kg-tt,ttbook}. The problem of finding its spectrum has been
cast in a form that makes it accessible to numerical studies \cite{br}.
Although the eigenvalues are discrete, there are indications that,
in contrast to the area operator, the spectrum of the volume
operator may not have an volume gap. This is but one indication that
the quantum geometry has qualitatively different features from what
one may naively expect from the classical Riemannian geometry or a
naive discretization thereof.

Let us summarize. The kinematical framework of LQG is well
developed, with full control on functional analysis. In particular,
the infinite dimensional integrals are not formal symbols but
performed with well defined measures \cite{al2,jb1}. There are two
key results that simplify the analysis: the uniqueness theorem
\cite{lost} and the spin-network decomposition of the full
Hilbert space \cite{rs2,jb2}. 
The natural truncation of the theory is achieved by restricting
oneself to 
the Hilbert space $\H_\alpha$ defined by a graph $\alpha$. Elements
of these $\H_\alpha$ describe elementary quanta of geometry; to
obtain classical geometries one needs to coherently superpose a
large number of them. \index{quanta of geometry}

Perhaps the simplest way to visualize the elementary quanta is to
introduce a simplicial decomposition \index{simplicial decomposition}
 $\S$ of the 3-manifold $M$ and
consider a graph $\alpha$ which is dual $\S$: Each cell in $\S$ is a
topological tetrahedron $T_n$, dual to a node $n$ of $\alpha$; each
face $F_\ell$ of $\S$, is dual to a link $\ell$. 
In Regge calculus, \index{Regge calculus}
every $T_n$ has the geometry of a tetrahedron in
flat space and the curvature is encoded in the holonomies of the
connection around `bones' that lie at the intersection of any two
faces of $T_n$. What is the situation in LQG? \index{holonomies}
To bring out the
similarities and contrasts, it is convenient to consider a basis
$\Psi_{\alpha,v_n, a_{\ell}}$ in $\H_{\alpha, j_\ell}$ that
simultaneously diagonalizes the volume operator associated with the
tetrahedron $T_n$, and the area operators associated with the faces
$F_{\ell}$, for all $n, \ell$. Each of these spin-network states
describes a specific \emph{elementary} quantum geometry. One can
think of the node $n$ as a `grain' or a `quantum' of space captured
in the (topological) tetrahedron $T_n$. As in Regge calculus each
$T_n$ has a well defined volume $v_n$ and each of its faces $F_\ell$
has a well-defined area $a_\ell$. But now the $v_n, a_\ell$ are
\emph{discrete.} More importantly, because the operators $J^i_\ell$
do not commute, $T_n$ no longer has the sharp geometry of a
geometrical tetrahedron in the Euclidean space. In particular,
operators describing angles between any two distinct faces $F_\ell,
F_{\ell^\prime}$ of a $T_n$ are \emph{not} diagonal in the basis.
Furthermore, although the area of any common face $F$ of two
adjacent tetrahedrons is unambiguous, in contrast with the Regge
geometry, \index{Regge calculus} curvature now resides not just at the bones of tetrahedra
but also along the faces; the geometry is `twisted' in a precise
sense \cite{twisted}. These properties of the quantum geometry
associated with the basis $\Psi_{\alpha,v_n, a_{\ell_n}}$ are
closely analogous to the properties of angular momentum captured by
the basis $|j,m\rangle$ in quantum mechanics: it too diagonalizes
only some of the angular momentum operators, leaving values of other
angular momentum observables fuzzy. Thus, each of the elementary
cells in the simplicial decomposition is now a `tetrahedron' in the
same heuristic sense that the a spinning particle in quantum
mechanics is a `rotating body'. \index{simplicial decomposition}

To conclude, we note that tri-valent spin-networks were introduced
by Roger Penrose already in 1971 in a completely different approach
to quantum gravity \cite{rp}. He expressed his general view of that
construction as follows: \emph{``I certainly do not want to suggest
that the universe `is' this picture \ldots But it is not unlikely
that essential features of the model I am describing could still
have relevance in a more complete theory applicable to more
realistic situations''}. In LQG one finds that the trivalent graphs
$\alpha_{\rm tri}$ are indeed `too simple' because all states in the
$\H_{\alpha_{\rm tri}}$ have zero volume \cite{loll}. Also, we now
have detailed geometric operators and find that the angles cannot be
sharply specified. Nonetheless, Penrose's overall vision is realized
in a specific and precise way in the LQG quantum geometry.

\index{quantum Riemannian geometry|)} 
\subsubsection{Quantum Einstein's equations}
\label{s2.2.3}

Recall from (\ref{constraints}) that we have three sets of
constraints. In the classical theory, the Gauss and the
Diffeomorphism constraints generate kinematical symmetries while
dynamics is encoded in the Hamiltonian constraint. In the quantum
theory the physical Hilbert space $\Hp$ is to be constructed by
imposing the quantum constraints $\hat{C}\Psi_{\phy} = 0$ a la
Dirac. This requires one to solve two non-trivial technical
problems: i) Introduce well-defined constraint operators $\hat{C}$
on $\Hk$ starting from the classical constraint functions $C$; and
ii) Introduce the appropriate scalar product on the solutions
$\Psi_{\phy}$ to obtain $\Hp$. The second step is non-trivial
already for systems with a finite number of degrees of freedom if
the constraint operator $\hat{C}$ has a continuous spectrum because
then the kinematical norm of physical states $\Psi_{\phy}$ diverges.
In geometrodynamics, \index{geometrodynamics}
the operators $\hat{C}$ have been defined only
formally and generally the issue of scalar product is not addressed.
In LQG by contrast, the availability of a rigorous kinematical
framework provides the necessary tools to address both these issues
systematically.

For the kinematical constraints, both these steps have been carried
out \cite{almmt1}. Since these constraints $C_{\rm kin}$ have a
natural geometrical interpretation, the quantum operators
$\hat{C}_{\rm kin}$ simply implement those geometrical
transformations on the kinematical (spin-network) states $\Psi_{\rm
kin}$ in $\Hk$. The second task, that of introducing the appropriate
scalar product, is carried out using a general strategy called
\emph{group averaging} \cite{almmt1,dm}. The detailed
implementations of these ideas is straightforward for the Gauss
constraint but there are important subtleties in the case of the
diffeomorphism constraint \cite{almmt1,alrev,gs,ttbook}. In
particular,
%
%
the strategy described here allows only the exponentiated version of
the diffeomorphism constraint, i.e., \emph{finite} diffeomorphisms,
and one has to specify the precise class of diffeomorphisms that are
allowed.%
\footnote{With a natural choice of this class, $\Hdiff$ is
\emph{separable}, although $\Hk$ is not. This may seem surprising at
first. But the situation is completely analogous to what happens
already in the quantum theory of free Maxwell theory with the Gauss
constraint if one does not wish to work with an indefinite metric
\cite{thirring, aa-ehlers}.}
The Hilbert space $\Hdiff$ on which both the kinematical constraints
are satisfied provides a completion of the Dirac quantization
program.

For the Hamiltonian constraint $C_H$, on the other hand, the
situation is still in flux. There is a non-trivial result due to
Thiemann that one \emph{can} regulate this constraint systematically
on $\Hdiff$ \cite{tt}. By contrast, no such regularization is
available for the WDW equation of geometrodynamics. \index{geometrodynamics} But
the procedure involves introduction of additional structures in the
intermediate steps, whence the final result is ambiguous.
Furthermore, the \emph{physical} meaning of the additional
structures has remained unclear. Finally, recall that in GR, the
Poisson bracket between Hamiltonian constraints smeared with lapse
functions $N$ and $M$ is the diffeomorphism constraint smeared with
a `q-number' shift field $K^a = q^{ab}(ND_aM - MD_aN)$. An important
question is whether this Poisson bracket structure is reflected in
the quantum theory. In these regularizations, in the quantum theory
the commutator of the two Hamiltonian constraints vanishes and so
does the diffeomorphism constraint on right hand side on the
`kinematical habitat' on which the calculation is carried out
\cite{habitat}. While this establishes consistency, one would hope
for a better scheme in which neither side vanishes and the
commutator structure captures the non-trivial, off-shell relation
between the constraints.

Recently, a promising approach to this problem has been introduced
by Laddha, Varadarajan and others \cite{mv1,alok}. The first
underlying idea is to take hints from earlier work on 1+1
dimensional parameterized field theories where well-understood,
unconstrained field theories are recast in an extended setting with
constraints. The constraints mimic those of GR
in that they are again related to space-time diffeomorphisms \cite{mv2}. 
One finds that the techniques used in LQG provide a natural avenue
to implement quantum constraints in the parameterized form, leading
to the correct final quantum theory. The second and deeper
observation is motivated by the fact that in connection dynamics the
diffeomorphism and the Hamiltonian constraints of
(\ref{constraints}) can be naturally combined in the spinorial
setting as $E^{aA}{}_B\, E^{bB}{}_C\, F^C_{ab\,D} =0$ where $A
\ldots D$ are spinorial indices \cite{aabook}. (The trace over $A$
and $D$ yields the Hamiltonian constraint while the trace-free part
yields the diffeomorphism constraint). \index{connection dynamics}
This unity suggests that
although Hamiltonian constraint generates time evolution, this
action could be recast in terms of geometric operations
\emph{within the 3-manifold} $M$. This has been shown to be the case
\cite{alok}: in the classical theory, time evolution can be
re-expressed as the action of \emph{diffeomorphism and Gauss
constraints} smeared with certain `q-number' smearing fields on $M$.
As a consequence, an entirely new perspective emerges. 
These features do not carry over to geometrodynamics since there the
two constraints cannot be naturally combined into a single one. In a
simplified theory, where the gauge group $\SU(2)$ is replaced by the
Abelian group ${\rm U(1)}^3$, the program has been carried out and
it has been shown that the algebra of constraints \emph{closes
off-shell non-trivially.} There is ongoing research to extend these
results to the full theory using $\SU(2)$. \index{geometrodynamics}

Finally there has been considerable research on coupling matter
fields to gravity, particularly those that can serve as physical
clocks and rods \cite{clocks1,clocks2,gs}. The idea, as in
geometrodynamics, is to use the matter fields to `deparameterize'
the constraints and study the ensuing \emph{relational} dynamics. On
the conceptual side, these ideas will play a key role in the
physical interpretation of canonical LQG. On the technical side, it
is rather surprising that certain matter fields do make the quantum
constraints manageable enabling one to extract the notion of
`evolution' from the solution to quantum constraints. Once there is
a fully satisfactory implementation of the Hamiltonian constraint,
these ideas will play a key role in extracting physics from
canonical LQG. A qualitative understanding has already begun to
emerge because the strategy of \cite{mv1} to better regularize the
constraints and that of \cite{clocks1,clocks2} to deparameterize the
theory using matter can be seen as generalizations to the full
theory of the successful strategies used in loop quantum cosmology
to first obtain and then interpret the quantum theory.

A complementary approach to dynamics is provided by Spinfoams,
discussed in the next subsection.
\index{loop quantum gravity!Hamiltonian theory|)}


\subsection{Covariant Loop Gravity: Spinfoams}
\label{s2.3}

\index{loop quantum gravity!spin foams|(}

The covariant or \emph{Spinfoam} formulation of LQG is built again
on the quantum theory of geometry discussed in section \ref{s2.2},
but the now dynamics is specified by defining the transition
amplitudes, order by order in a suitable expansion. This is akin to
spirit used by Feynman to build QED directly in terms of the Feynman
rules, which streamlined and simplified the theory.\index{QED}

As with Feynman diagrams, the amplitudes defined in this manner can
also be seen as given by a sum over histories. The relevant
histories, however, describe spacetime as the \emph{evolution of
individual quanta of geometry}, \index{quanta of geometry}
rather than of classical
configurations. Thus, a 3-geometry is still represented by a
spin-network, and a 4-geometry, by a history of spin-networks. These
histories are called \emph{Spinfoams} \cite{rr,jb3,crbook}. The sum
over Spinfoams define transition amplitudes between quantum
3-geometries and, as discussed in section \ref{s3.3}, the $n$-point
functions of non-perturbative quantum gravity.

The main results of covariant LQG to date are the following:

i) The amplitudes are \emph{finite} at every order.

ii) At each order, the amplitudes have a well-defined classical
limit, related to a truncation of classical general relativity.

iii) The theory has been extended to include fermions and Yang-Mills
fields \cite{fermions}.

Regarding point (i), there are two potential sources of infinities
in the theory. The ultraviolet (UV) divergence which corresponds to
the conventional infinities of perturbative Feynman diagrams, and
infrared (IR) diverges that can arise from the contributions of
intermediate states with large-scale geometries. The UV divergences
are naturally cured by the discreteness of the underlying quantum
geometry itself. The IR divergences are cured by the presence of a
positive cosmological constant ${\underline{\Lambda}}$. \index{cosmological constant}
Therefore, interestingly, the
structure of the theory is such that a cosmological constant
\emph{with a positive sign} naturally acts as a physical IR
regulator. Regarding point (ii), recall that the classical limit of
lattice QCD \index{QCD}
on a fixed triangulation is just the classical lattice
theory. Similarly, the classical limit of covariant loop quantum
gravity at a fixed order is related to Regge calculus \index{Regge calculus} on a finite
triangulation in a precise sense. We will now discuss these issues
in detail.

\subsubsection{Transition Amplitudes}

For simplicity, we describe the case without fermions and Yang-Mills
fields and with ${\underline{\Lambda}} =0$ (thus ignoring IR problems).  We later
discuss the necessary modifications to incorporate ${\underline{\Lambda}} >0$.

In quantum theory, one calculates the transition amplitudes between
initial and final states. In LQG these are states of quantum
geometry and therefore belong to the Hilbert spaces ${\H}^{\rm
diff}_\alpha$ labeled by (abstract) graphs $\alpha$. In the absence
of an external time parameter, there is no distinction between the
initial and the final states. Therefore, it is convenient to combine
the two graphs that refer to the boundary states; we denote this
total graph by $\Gamma$. Then, if $\Gamma$ has $N$ nodes $n$ and $L$
links $l$, ${\H}^{\rm diff}_\Gamma$ is spanned by states $\psi(U_l)$
which are in $L^2(\SU(2))^L$ and invariant under $\SU(2)$ gauge
transformations at the nodes. Thus, ${\H}^{\rm diff}_\Gamma$ is the
same as the Hilbert space of a $\SU(2)$ lattice Yang-Mills theory.
The theory defines a transition amplitude for each of these states
$\Psi(U_l)$.

\begin{figure}[t] 
{\includegraphics[width=.40\textwidth]{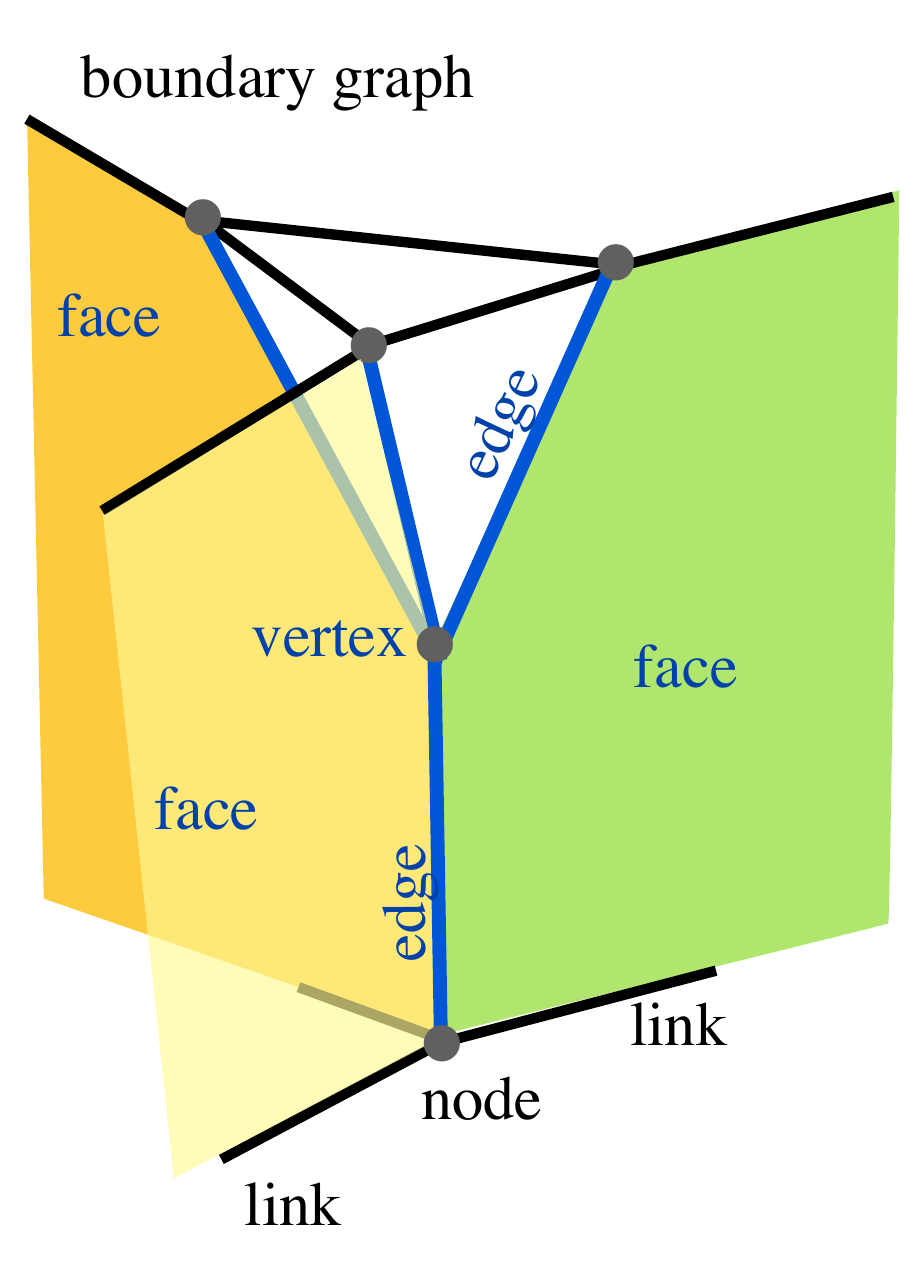} }
\caption{2-complex terminology.}
\label{twocomplex}
\end{figure}

To any given order, the transition amplitudes are labeled by a
\emph{2-complex} $\mathcal{C}$, a higher-dimensional analog of a graph: it is
defined as a (combinatorial) set of \emph{faces} $f$ meeting at
\emph{edges} $e$, which in turn meet at \emph{vertices} $v$ (see
Fig. \ref{twocomplex}). It can be regarded as a history of a
spin-network in which each link $l$ of the spin-network `evolves' to
form a face $f$, each node $n$ `evolves' to an edge $e$, and
non-trivial dynamics occurs when a new vertex $v$ appears. The
number of vertices in $\mathcal{C}$ defines the order in the expansion of the
transition amplitude.

To grasp the interplay with spacetime geometry, it is useful to note
that a triangulation $\Delta$ of the four dimensional spacetime
defines a dual 2-complex: each 4-simplex of $\Delta$ corresponds to
a vertex in $\mathcal{C}$, each tetrahedron of $\Delta$ to an edge 
in $\mathcal{C}$, and each triangle in $\Delta$ to a face of $\mathcal{C}$. 
For simplicity, we will only consider 2-complexes dual to triangulations. 
The boundary $\Gamma=\partial{\cal C}$ of a 2-complex is a graph that the state
$\psi(U_l)$ refers to (see Fig. \ref{twocomplex}).

Let us fix a 2-complex with boundary graph $\Gamma$ and $v$
vertices. Then, the $v$th order term in the expansion of the
transition amplitude associated with a state $\psi\in {\H}^{\rm
diff}_{\Gamma}$ is defined as the scalar product in ${\H}^{\rm
diff}_{\Gamma}$ of the state $\psi$ with the function $W_{\mathcal{C}}(U_l)$
defined as follows:
\begin{equation}
W_{\cal C}(U_l)=\int_{\SU(2)} dh_{vf} \ \prod_f \delta(h_f)\
\prod_v A_v(h_{vf}).
\label{ta}
\end{equation}
This $W$ is the key object in Spinfoams because, to order $v$ in our
expansion, quantum dynamics is encoded in $W$. The integral in
(\ref{ta}) is over one $\SU(2)$ variable $h_{vf}$ associated to each
vertex-face pair, the delta distribution is over $\SU(2)$ and its
argument $h_f$ is the (oriented) product of the variables $h_{vf}$
around a face.   In the 4d Lorentzian theory, the ``vertex
amplitude''  $A_v$ has the form
\begin{equation}
A_{v}(h_{vf})=\int_{\SL(2,C)} dg_{ve} \
\prod_e \sum_j (2j+1)\ {\rm Tr}_j[Y^\dagger g_eg_{e'}Yh_f]\, .
\label{va}
\end{equation}
Here the $\SL(2,C)$ integration variables $g_{ve}$ are associated to
each edge emerging from $v$ and the two edges $e$ and $e'$ in the
trace are those bounding the face $f$. $Y$ is a map from the spin
$j$ representation of $\SU(2)$ to the unitary representation 
of $\SL(2,C)$ with continuous quantum number $\gamma(j+1)$ and discrete
quantum number $j$, defined by%
\footnote{The maps extends easily from functions on $\SU(2)$ to
functions of $\SL(2,C)$.}
\begin{equation}
Y:\,\,\, |j;m\rangle\quad \mapsto \quad |\gamma(j+1),\gamma;j,m\rangle
\label{Y}
\end{equation}
where, as before, $\gamma$ is the Barbero-Immirzi parameter. These
three equations completely define the theory. Quite remarkably, to
order $v$ in the vertex expansion, they encode the entire quantum
dynamics.

This form of the amplitude is variously denoted EPRL, EPRL-FK, or
EPRL-FK-KKL amplitude. It was derived in \cite{eprl} building on
results in \cite{fk,ls} and extended to arbitrary 2-complexes in
\cite{kkl}, and forms the basis of the 4d Lorentzian theory so far.
Variants could be interesting; some of these have been considered
for the Euclidean theory \cite{bo}.

Let us examine the structures of this amplitude. The appearance of
$\SL(2,C)$ is not surprising: it reflects the local Lorentz
invariance of GR. The appearance of the \emph{unitary} (infinite
dimensional) representations of this group should not be too
surprising either, given that unitary representations of symmetry
groups are ubiquitous in quantum gravity. Indeed, one may wonder why
the mathematics of the infinite dimensional unitary representations
of $\SL(2,C)$ has played such a small role in the attempts to
construct a quantum theory of gravity so far. The map $Y$ on the
other hand is a new ingredient that constitutes the technical core
of the Spinfoam model and deserves explanation. For this, let us
first return to the classical theory discussed in section
\ref{s2.2}. In the space-time picture, the momentum conjugate to the
$\SL(2,C)$ connection $\omega$ is \cite{holst} 
\begin{equation}
\pi_{IJ} =\frac{1}{4\kappa_{\rm N}}(\epsilon_{IJKL} e^K\wedge e^L+
\frac{1}{2\gamma\kappa_{\rm N}} e_I\wedge e_J)
\label{B}
\end{equation}
On a boundary of a spacetime region, the one-form normal to the
boundary contracted with the tetrad gives a vector in the internal
Minkowski space, which determine a preferred Lorentz frame. We can
decompose $\pi_{IJ}$ in this frame in the same manner in which the
Maxwell field $F_{IJ}$ decomposes in the electric and magnetic
field. Simple algebra then shows that the electric and magnetic
parts of  $\pi_{IJ}$, denoted respectively $\vec K$ and $\vec L$
satisfy the algebraic equation
\be
      \vec K\, =\, \gamma\ \vec L
\label{sc}
\ee
This is a key equation in covariant loop quantum gravity, called the
\emph{simplicity constraint}. To ensure a correct classical limit,
this constraint has to be implemented in the quantum theory in an
appropriate fashion. This is precisely what the map $Y$ does: in the
quantum theory \eqref{sc} holds on the image of this map as a weak
operator equation (i.e., for all matrix elements of the operators)
\cite{ding}.

\subsubsection{Classical Limit}

Spinfoam dynamics presented in the last subsection was arrived at
from several independent considerations: the Hamiltonian LQG
\cite{rr}, the fact that GR can be regarded as a constrained BF
theory \cite{jb-BF}, the Ponzano-Regge and Turiev-Viro models
\cite{pr,regge1,regge2,regge3} \index{Regge calculus} for quantum gravity in 3-dimensions \index{group field theory}
and group field theory \cite{crbook,gft1}. Furthermore, the overall
paradigm underlying Spinfoams is borne out in symmetry reduced,
cosmological models, where the transition amplitudes obtained by
summing over quantum geometries have been shown to be finite and in
agreement with the Hamiltonian theory \cite{ach2}. While these
considerations provide a reasonably strong motivation, one still
needs direct evidence in favor of the specific proposal (\ref{ta}).
Analysis of the classical limit provides a natural avenue to test
its viability.

In any quantum theory, the classical limit is obtained in a regime
where quantum numbers are large. Then the relevant actions are large
compared to the Planck constant and the limit can be interpreted as
$\hbar \to 0$. For example, for a particle with a Hamiltonian $H$,
in the  $\hbar \to 0$ limit we have:
\be W(x,t;x',t')\sim \int [Dx]\ e^{\frac{i}{\hbar}S[x]} \sim  A\
e^{\frac{i}{\hbar} S(x,t;x',t')}\, . \ee
where the integration is over the paths from $(x,t)$ to $(x',t')$
and $S(x,t;x',t')$ is the Hamilton function, namely the value of the
action on the solution of the classical equations of motion that
start at $(x,t)$ and ends at $(x',t')$. In gravity the analogous
procedure requires us to consider areas and volumes that are large
compared to the Planck scale. Thus, to study the classical limit of
Spinfoam dynamics, one can compare the large $j$ limit of the
transition amplitude \eqref{ta} with the classical action. The
asymptotic analysis of the vertex amplitude \eqref{va} is
nontrivial, and has been carried out mainly by the Nottingham group
\cite{barrett}. For the simplest case where the 2-complex $\mathcal{C}$ has
only one vertex $v$, and the results can be summarized as follows.
Recall that the amplitude is a function of the boundary quantum
state $\psi$ and quantum geometries are more general than Regge
geometries. \index{Regge calculus} If $\psi$ does not endow the 4-simplex $\Delta$ dual to
$\mathcal{C}$ with a consistent classical geometry, the transition amplitude
is suppressed exponentially. If it does, then the asymptotic form of
the amplitude is given by
\be A_v \sim A \left(e^{\frac{i}{\hbar}
(S_R+\frac\pi{4})}+e^{-\frac{i}{\hbar} (S_R+\frac\pi{4})}\right).
\label{asym} \ee
where $S_R$ is the Regge action of $\Delta$.
\begin{table}[t]
  \centering
\multirow{2}{10mm}{\begin{sideways}\parbox{20mm}{$\xleftarrow{
\hspace*{2em}{\rm Continuum\ limit} \hspace*{2em}}$}\end{sideways}}
\begin{tabular}{@{} ccc @{}}
   \parbox{3.5cm}{\center Loop-gravity  \\  transition~amplitudes} \hspace*{.5em}& $\xrightarrow{j\to\infty}$ &\hspace*{.5em} \parbox{3cm}{ \center Regge theory} \\[1cm]
{\begin{sideways}\parbox{10mm}{ $\xleftarrow{{\cal C}\to\infty}$}\end{sideways}} \hspace*{1em}&  &\hspace*{1em} {\begin{sideways}\parbox{10mm}{ $\xleftarrow{{\Delta}\to\infty}$}\end{sideways}}  \\[2mm]
   \parbox{4cm}{ \center  Exact \\ transition amplitudes}\hspace*{1em} & $\xrightarrow{j\to\infty}$ & \hspace*{1em}\parbox{3cm}{ \center  General relativity }  \\
  \end{tabular}\\[1cm]
\parbox{35mm}{$\xrightarrow{ \hspace*{4em}{\rm Classical\ limit} \hspace*{3em}}$}
  \caption[Continuum versus classical limit]{Relation between continuum limit and classical limit of the transition amplitudes.}
  \label{tab:tavola} \index{continuum limit}
\end{table}
The presence of two terms in (\ref{asym}) is a consequence of the
fact that, as we saw in section \ref{s2.2}, the starting point of
the analysis is tetrad gravity and, when the tetrad changes
orientation, the first order LQG action changes sign, while the
Einstein-Hilbert action does not. Consequently, for each classical
metric solution we have two tetrad solutions whose action is equal
in magnitude but with opposite signs. The $\frac\pi{4}$ is also well
understood: it is the Maslov index that always appear in the
semiclassical limit when the two classical solutions sit on
different branches of the solution space \cite{hal}. Therefore the
result \ref{asym} has the following simple interpretation: the
classical limit of the transition amplitude defined by a 2-complex
$\mathcal{C}$ dual to a spacetime triangulation $\Delta$ is the Regge
amplitude associated with that triangulation. This is precisely what
one would have hoped. In this sense the proposal passes the
viability criterion and it is reasonable to regard equations
\eqref{ta}, \eqref{va} and \eqref{Y} as providing a tentative
definition of the dynamics of LQG. \index{Regge calculus}

This covariant formulation of LQG has some similarities with the
path-integral approach based on Regge Calculus \cite{hamber1,
hamber2} where one sums over configurations representing a Regge
discretization of general relativity. This approach was introduced
already in the 1980s and has evolved considerably since then
\cite{hamber3}. 
In spite of the formal structural similarity, there is an important
conceptual difference between the two approaches. In Regge calculus,
the lengths of the individual links can be arbitrarily small. By
\index{Regge calculus}
contrast, the geometries that are summed over in Spinfoams represent
histories of \emph{quanta} of space, whence the areas of plaquettes
cannot be arbitrarily small; they are bounded below by the area-gap
\index{area gap}
of LQG. This fundamental discreteness naturally removes the UV
divergences and introduces the Planck scale already in the
permissible histories that are summed over. Consequently, the
scaling structure of the theory with respect to the Regge Calculus
is quite different.

We conclude this discussion by noting that the classical limit we
have discussed here should not be confused with the continuum limit
\index{continuum limit}
of the theory. The first is the standard $\hbar\to 0$ limit while
the second refers to refinement, i.e., adding more and more degrees
of freedom. Recall that the classical limit of lattice QCD on a
fixed lattice is of course a classical lattice theory. \index{QCD}
In LQG, the
lattice is replaced by a triangulations, but with the crucial
difference that its geometry is not pre-specified but constitutes
the dynamical variable. Nonetheless, situation with respect to the
classical limit is similar: in covariant loop quantum gravity, this
limit is related in a precise way to Regge calculus on a finite
triangulation. \index{Regge calculus}
As well known, classical Regge calculus converges to
full GR in the limit in which the triangulation is refined. The
structure of the theory is therefore as in the Table
\ref{tab:tavola}: To arrive at GR from Spinfoams, one can start from
the upper left corner of the diagram, move first to the right and
then down.

\subsubsection{Cosmological Constant and IR Finiteness}

As we noted above, the reason behind the UV finiteness of the
Spinfoam amplitude (\ref{ta}) is intuitively simple: because of the
discreteness of space at the Planck scale, there is an in-built and
natural physical cut off preventing the standard quantum field
theory divergences. In other words, there are no degrees of freedom
at arbitrary small scales. 
Therefore the sum over intermediate states in a perturbation
expansion does not include field configurations of arbitrary high
momentum.
From this perspective, the UV divergences of standard quantum field
theory can be interpreted as pathologies introduced by the fact of
neglecting the discrete nature of space.

However, the amplitude \eqref{ta} can have IR divergences. This can
happen every time the 2-complex has a \emph{bubble}, i.e.,  a set of
continuous faces with the topology of a two-sphere. These bubbles
are the Spinfoam analog of loops in Feynman diagrams: The quantity
circulating around a Feynman diagram loop is the momentum, and high
momentum means UV; while the quantity circulating around a Spinfoam
bubble is the area, and high area means IR. On a bubble, the sum
over spins $j$ in \eqref{ta} can lead to divergent terms because $j$
is unbounded above. Geometrically, these divergences correspond to
`spikes, representing large regions of spacetime bounded by small
hypersurfaces.'

Remarkably, these divergences disappear naturally if there is a
positive cosmological constant ${\underline{\Lambda}}$ in the theory. 
\index{cosmological constant} Technically, the
effect of a positive ${\underline{\Lambda}}$ is to replace $\SL(2,C)$  with a quantum
deformation of $\SL(2,C)$. The mathematics for implementing this
deformation has been developed \cite{E.Buffenoir:kx,Noui:2002ag} and
the Spinfoam amplitude with the cosmological constant has also been
defined \cite{Fairbairn:2010cp,Han:2011aa}. The transition
amplitudes of the theory with a quantum deformation of $\SL(2,C)$
are finite, and the classical limit of their vertex amplitude is
still given by Eq. \eqref{asym}. But now the Regge action that
appears in the classical limit has a cosmological constant ${\underline{\Lambda}}$,
related to the deformation parameter $q$ of the quantum group via
\cite{Ding:2011hp,Han:2011aa}.
\be
    q\, = \, \exp{{\underline{\Lambda}}\hbar G}\, .
\ee
Therefore the full theory now depends on \emph{two} dimensionless
parameters: $q$ or ${\underline{\Lambda}}\hbar G$, and the Barbero-Immirzi parameter
$\gamma$. The bare cosmological constant enters the theory as a free
parameter, therefore the theory does not prescribes its value. To
explore various limiting regimes, one has to calculate the behavior
of physical observables, keeping appropriate combinations of these
constants fixed and let a complementary combination tend to the
desired value. \index{cosmological constant}

%
%

\subsubsection{QED, QCD and LQG}

Similarities between the Spinfoam model defined in the last three
subsections and QCD\index{QCD} \index{QED}
on a fixed lattice are evident: In
both cases, we have a discretization of the classical theory where
the connection is replaced by group elements, and a quantum theory
defined by an integral over configurations of an amplitude which is
a product of local quantities. The use of a triangulation in
Spinfoams instead of a square lattice simply reflects the fact that
a square lattice is unnatural in absence of a flat metric. However,
there is also a crucial difference. The Wilson QCD action depends on
an external parameter, the lattice spacing $a$, while appropriate
discretizations of the Einstein-Hilbert action, like the Regge
action, do not. \index{Regge calculus}
To recover the continuum theory, in QCD it is not
sufficient to increase the total size of the lattice; it is also
necessary to send $a$ to zero. Equivalently, the lattice spacing $a$
can be absorbed in the coupling constant $\beta$ in front of the
action and, in order to recover the continuum limit, it is necessary
to tune $\beta$ to its critical value, $\beta=0$. \index{continuum limit}   
In gravity, instead, the Regge action (or any other admissible
discretization) does \emph{not} include a lattice spacing $a$ (nor,
therefore, a coupling constant that needs to be tuned to a critical
value as $a\to 0$). The reason is simply that the lattice spacing
$a$ refers to a background geometry ---the Yang-Mills theory depends
on a fixed, externally given spacetime metric--- while in gravity
the geometry is included in the dynamical variables. It can be shown
in general \cite{ditt} that the discretization of a
reparametrization invariant theory can be defined \emph{without} a
parameter that needs to be tuned to a critical value in the
continuum limit. \index{continuum limit} Accordingly, in a suitable 
\index{continuum limit} discretization of general relativity the 
continuum limit can be defined just by making the triangulation (or the 
two complex) increasingly finer.%
\footnote{In concrete physical calculations, however, only finite
triangulations suffice, as is generally the case in QCD. Similarly,
in QED a finite number of Feynman graphs suffice. \index{QCD}
\index{QED}}
%
An alternative approach to the continuum limit is discussed in
\cite{bianca}. \index{continuum limit}


Interestingly, there are also similarities between Spinfoams and
perturbative QED\index{QED}. The nodes of the graph can be seen as
quanta of space and the 2-complex can be read as a history of these
quanta, showing where these quanta interact, join and split, just as
real and virtual particles do in the Feynman graphs. The analogy is
reinforced by the fact that the Spinfoam amplitude can actually be
concretely obtained as a term in a Feynman expansion of a `group
field theory' (see for instance Chapter 9 of \cite{crbook} and
\cite{gft1,bo}). The specific group field theory that gives the
gravitational amplitude (\ref{ta}) has been derived (in the
Euclidean context) in \cite{Krajewski:2010yq}. \index{group field theory}

Thus, the Spinfoam paradigm shares some key features with QCD as
\index{QCD} \index{QED}
well as QED, our two most successful, fundamental quantum theories.
In addition, Spinfoams bring out a novel interplay between these
theories and quantum gravity. A Feynman graph of QED is a history of
quanta of a field while the lattice used in QCD is a collection of
discrete chunks of spacetime. They are distinct and unrelated. But
general relativity taught us that spacetime itself is a field
---the gravitational field--- and in LQG its discrete chunks are the
quanta of this field. Therefore, once we recognize that the
gravitational field is both dynamical and quantum, the quantum
gravity analog of the lattice used in QCD can be seen as a Feynman
graph of a quantum theory, representing the history of gravitational
quanta. In this sense, the Feynman graphs of QED and lattices of QCD
merge in LQG via Spinfoams. 
\index{loop quantum gravity!spin foams|)}
%

\section{Applications}
\label{s3}

Exploration of the physical consequences of Asymptotic Safety is
still at its beginning. First investigations on both cosmological
\cite{cosmo1,entropy} and black-hole spacetimes \cite{bh1,evap} have
been performed within asymptotically safe QEG. \index{QEG quantum Einstein gravity} The main idea is to
employ a method often used in particle physics that goes under the
name \emph{RG improvement.} Here, it amounts to replacing $G$,
$\Lambda$ with $G_k$, $\Lambda_k$ and identifying $k$ with an
appropriately chosen dynamical or geometrical scale. Since this
identification suffers from a certain degree of ambiguity,
ultimately the method will have to be to be replaced by a more
precise one. Nonetheless, these investigations have already provided
a first idea of the QEG effects to be expected. Because the subject
is still evolving, we will discuss these ideas in the `Outlook'
section \ref{s4.2}.

On the other hand, three applications of LQG have been investigated
in detail over the last 10-15 years, resulting in thousands of
publications whose results have been summarized in several detailed
reviews (see, e.g.,\cite{asrev,mbrev,barraurev,blvrev,perez,PoS}).
In this section we will present some highlights of those
developments. Even though LQG is still far from being a complete
theory, advances could be made by using a \emph{truncation
strategy}: One first chooses the physical problem of interest,
focuses just on that sector of the full theory which is relevant to
the problem, and then uses LQG techniques to analyze it, making full
use of the quantum geometry summarized in sections \ref{s2.2}.

The section is divided into three parts. In the first we discuss the
very early universe; in the second, quantum aspects of black holes,
and in the third, the issue of defining $n$-point functions in a
manifestly background independent theory.

\subsection{The Very Early Universe}
\label{s3.1}

It is evident from Chapter 3 that there has been a huge leap in our
understanding of the early universe over the past two decades.
However, on the conceptual front a number of issues have remained in
the Planck era of the very early universe. Over the last decade
these issues have been systematically addressed in Loop Quantum
Cosmology (LQC). \index{loop quantum gravity!loop quantum cosmology|(}%
\index{LQC|see{loop quantum gravity, loop quantum cosmology}}  

In particular, the big bang singularity was resolved and cosmological perturbations are being analyzed following several approaches \cite{asrev,barraurev,aan,madrid,french}. For
brevity, we will focus on one of these which provides an internally
consistent paradigm starting from the Planck regime, with detailed
predictions that are compatible with the WMAP and Planck data. 
\index{WMAP} In
the first two parts of this subsection we summarize the main results
and in the third we present a critical analysis of adequacy of the
truncation strategy that underlies the discussion of quantum
cosmology in \emph{any} approach.

\subsubsection{Singularity Resolution}
\index{singularity resolution in LQG|(}
Every expanding Friedman, Lemaitre, Robertson, Walker (FLRW)
solution of GR, has a big bang singularity if matter satisfies the
standard energy conditions. But scalar fields with potentials that
feature in the inflationary scenarios violate these energy
conditions. Therefore, initially there was a hope that the standard
singularity theorems of GR \cite{hebook} could be avoided in the
inflationary context. However, this turned out not to be the case:
Borde, Guth and Vilenkin \cite{bgv} showed, \emph{without any
reference to energy conditions}, that if the expansion of a
congruence of past directed time-like or null geodesics is negative
(on an average), then they are necessarily past incomplete; the
finite beginning represented by the big bang in GR is not avoided.
But these arguments assume a smooth, classical geometry all the way
back to the big bang which has no physical basis since quantum
effects cannot be ignored in the Planck regime. Thus, although it is
often heralded as reality, big bang is a prediction of classical
gravity theories in a domain in which they are \emph{not}
applicable. A key result of LQC is that the quantum geometry effects
in the Planck regime lead to a natural resolution of the big bang in
a wide variety of cosmological models \cite{asrev,mbrev}.

\begin{figure}[]
  \begin{center}
\includegraphics[width=3.2in,height=2.5in,angle=0]{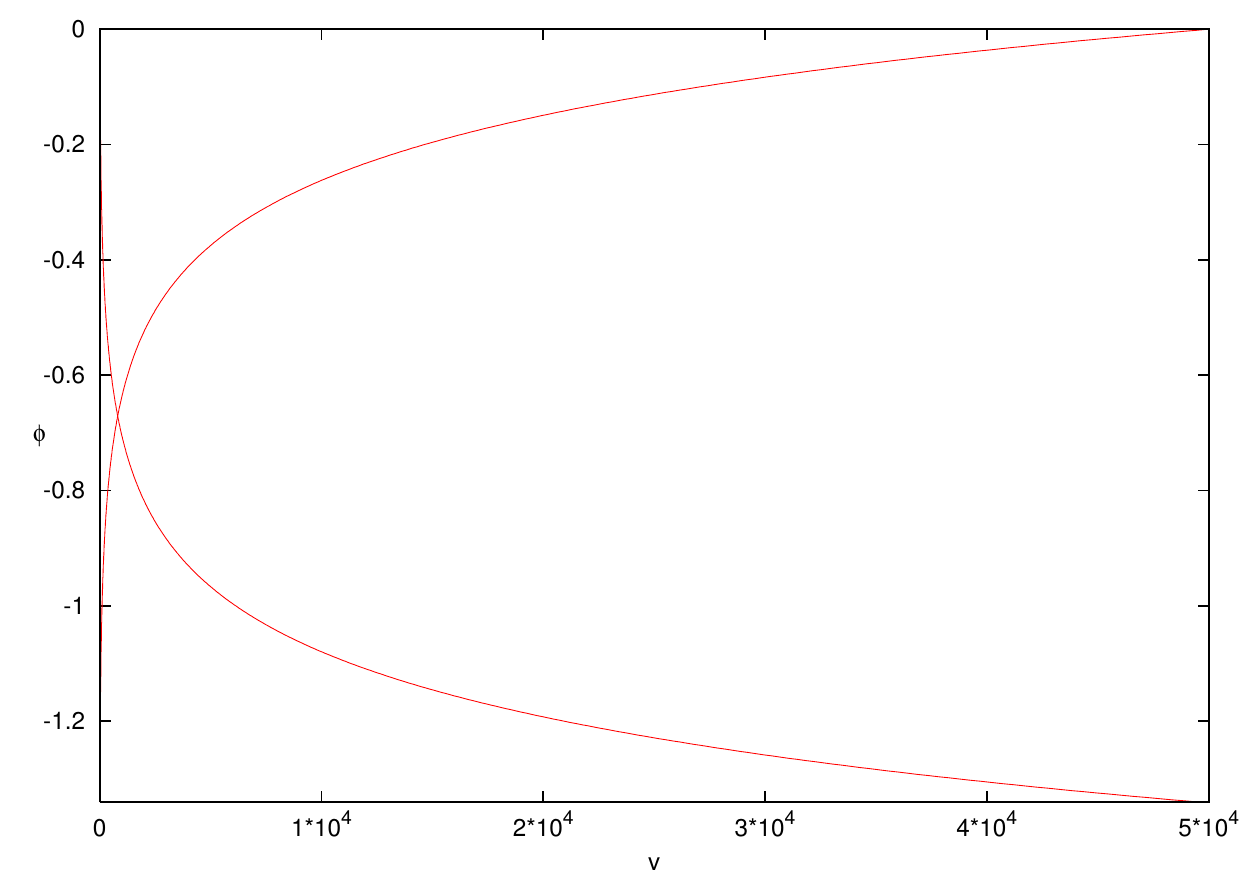}
\includegraphics[width=3.2in,height=2.5in,angle=0]{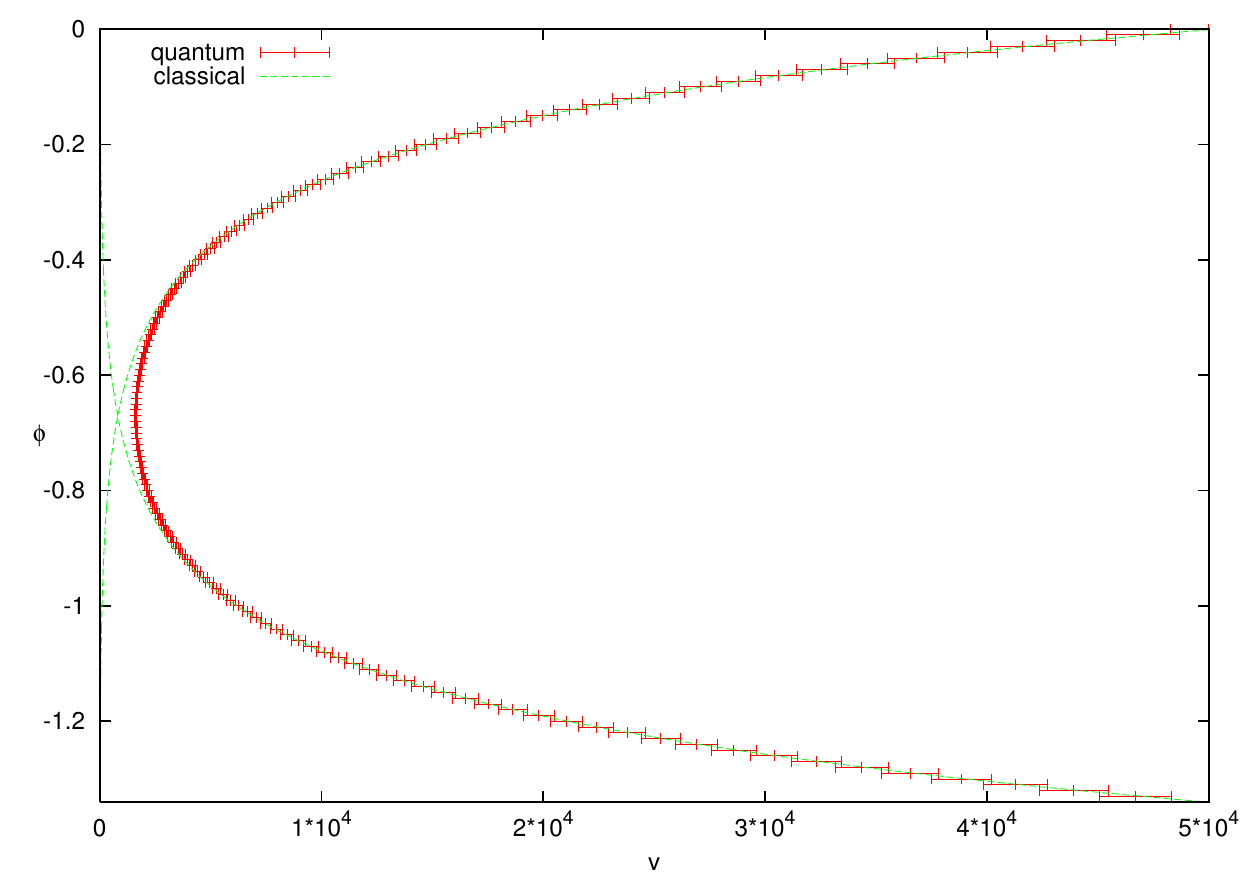}
\caption{The FLRW model with a massless scalar field. \textit{Left Panel:}
In classical GR, there are two types of solutions; those which begin with
a big bang expand forever and those that start out with zero energy density
and contract to a big crunch singularity. \textit{Right Panel:} In LQG,
quantum geometry effects create a novel repulsive force which starts becoming
significant when the energy density and curvature are $\sim\,\,10^{-3}$
in Planck units. The force grows with curvature and the big bang is replaced
by a quantum bounce. From A.~Ashtekar, Gen. Rel. Grav. 41, 707-741 (2006). With kind permission from Springer Science and Business Media. }
\label{fig:lqc}
\end{center}
\end{figure}

To illustrate how this comes about, consider the simple example of
the $k=0$ FLRW spacetimes with a massless scalar field $\phi$ as a
source. It is convenient to fix a fiducial cell $\mathcal{C}$ in co-moving
coordinates and plot these solutions directly in terms of physical
variables of the problem, the scalar field and the volume $v$ of
$\mathcal{C}$. As the left panel of Fig. \ref{fig:lqc} suggests, one can
regard the scalar field as a \emph{relational} time variable, in
term of which the volume $v$ ---and hence the curvature---
`evolves'. In Bianchi models, \index{Bianchi models}
the `evolving' quantities would also
include anisotropies and in, say, the Gowdy model, the
inhomogeneities that encapsulating gravitational waves. Since the
massless scalar field \emph{does} satisfy all the energy conditions,
all these solutions are singular. In the k=0 FLRW case 
the universe either expands starting with the big bang or contracts
into the big crunch singularity. Quantum cosmology was introduced in
the 1970s in the hope that these classical singularities would be
tamed by quantum effects \cite{jw}. However, in the WDW quantum
geometrodynamics \index{geometrodynamics}
of the simple model under consideration
unfortunately this hope is not realized \cite{aps,consistent2}. The
ideas was revived some three decades later in a pioneering paper by
Bojowald who showed that the situation is quite different in LQC
\cite{mb}. Subsequent conceptual completions and technical
improvements of this reasoning have provided a systematic
understanding of how this comes about.

First, the uniqueness theorem of LQG kinematics descends to LQC
\cite{ac} making LQC inequivalent to the WDW theory already at the
kinematical level. The WDW differential equation turns out not to be
well-defined on the LQC Hilbert space and one has to return to the
classical Hamiltonian constraint and systematically construct the
corresponding quantum operator making full use of the underlying
quantum geometry of LQG \cite{aa-badhonef}. This construction is
subtle and brings out a delicate interplay between the discreteness
of \emph{physical} areas and mathematics underlying the definition
of the Hamiltonian constraint in the connection dynamics framework
\cite{aps,apsv}. \index{connection dynamics} A detailed analytical argument \cite{acs} has
established that the density operator is well defined on the
physical Hilbert space with an upper bound $\rcr = {3}/({8 \pi
\gamma^2 G (\Delta a_S)})$ \index{area gap}
which is \emph{directly} controlled by
the area gap $\Delta a_S$ of LQG. Thus, density and curvature cannot
diverge in \emph{any} physical state. Numerical simulations showed
that this upper bound is in fact reached in states which are sharply
peaked at late times and, not surprisingly, it is reached precisely
at the bounce \index{quantum bounce}. Finally, the LQC singularity resolution has also been
established in the consistent histories approach \cite{consistent3}.

Details of LQC dynamics can be summarized as follows. Let us first
consider classical solution depicted by an expanding trajectory
(Fig. \ref{fig:lqc}, left panel). Fix a point at a late time,
consider a quantum state sharply peaked at that point and evolve it
using the LQC Hamiltonian constraint. One then finds that the wave
packet remains sharply peaked on the classical trajectory so long as
the matter density or curvature is less than $\sim\, 10^{-3}$ of the
Planck scale. Thus, in this regime there is good agreement with GR.
However, if we evolve the quantum state backwards towards the
singularity, instead of following the classical trajectory into the
singularity ---as is the case in the WDW theory--- the wave packet
bounces. \index{quantum bounce}
The expectation value of the energy density now starts
decreasing and once it reaches a few thousandths of $\rho_{\rm Pl}$,
the peak of the wave packet again follows a classical trajectory
along which the universe expands as we continue to move
\emph{backward} in time (Fig. \ref{fig:lqc}, right panel). An
important feature of LQC dynamics is that while the quantum geometry
effects are strong enough to resolve the big bang, agreement with GR
is recovered quickly, already when the curvature has fallen by a
factor only of $\sim 10^{-3}$ from the Planck scale. One can modify
Einstein's equations by introducing some quantum gravity effects by
hand and resolve the singularity. But such modifications generally
lead to departures from GR already at the density of water! LQC
naturally achieves the delicate balance: the UV pathology is tamed
leaving GR in tact rather close to the Planck regime.

Although in the Planck regime the peak of the wave function deviates
very substantially from the general relativistic trajectory, rather
surprisingly it follows an \emph{effective trajectory} with very
small fluctuations (see Fig. \ref{fig:lqc}). This effective
trajectory was derived \cite{vt} using techniques from geometric
quantum mechanics. The effective equations incorporate the leading
corrections from quantum geometry. They modify the left hand side of
Einstein's equations. However, to facilitate comparison with the
standard form of Einstein's equations, one moves this correction to
the right side through an algebraic manipulation. Then, one finds
that the Friedmann equation $(\dot{a}/a)^2 = (8\pi G\, \rho/3)$ is
replaced by
\be \left(\frac{\dot{a}}{a} \right)^2 = (8\pi G\,\rho /3)\, \left(1 -
\frac{\rho}{\rcr}\right) \, . \ee
At $\rho =\rcr$, the right side vanishes, whence $\dot{a}$ vanishes
and the universe bounces. \index{quantum bounce}
This can occur because the LQC correction
$\rho/\rcr$ \emph{naturally} comes with a \emph{negative} sign which
gives rise to an effective `repulsive force'. The occurrence of a
negative sign is non-trivial: in the standard brane world scenario,
for example, Friedmann equation is also receives a $\rho/\rcr$
correction but it comes with a positive sign (unless one makes the
brane tension negative by hand) whence the singularity is not
resolved. Finally, there is an excellent match between analytical
results within the quantum theory, numerical simulations and
effective equations. In particular, the effective equations capture
the leading LQC corrections to Einstein's equations very
efficiently.

This analysis has been extended to include the cosmological constant
\index{cosmological constant} \index{cosmological constant}
of either sign \cite{lambda}, the k=1 FLRW models \cite{apsv}, the
Bianchi I, II and IX models \index{Bianchi models}
which include anisotropies \cite{awe,we}
and the Gowdy models which include inhomogeneities \cite{gowdy}.
Furthermore the effective equations have been used to show that in
LQC \emph{all} curvature singularities ---including, e.g., the big
rip--- are resolved in all FLRW models \cite{ps}. These results
suggest that the quantum geometry effects of full LQG may well lead
to a resolution of all space-like, strong curvature singularities of
GR.

Finally, note that in all the models that have been studied in
detail, singularity resolution occurs \emph{generically} without any
exotic matter or need to fine-tune initial conditions. Furthermore,
one does not have to introduce a new boundary condition such as in
the Hartle-Hawking proposal. Why then does the LQC singularity
resolution not contradict the standard singularity theorems of
Penrose, Hawking and others? These theorems are inapplicable because
\emph{the left hand side} of the classical Einstein's equations is
modified by the quantum geometry corrections of LQC. What about the
more recent singularity theorems that Borde, Guth and Vilenkin
\cite{bgv} proved in the context of inflation? They do not refer to
Einstein's equations. But, motivated by the eternal inflationary
scenario, they assume that the expansion is positive along any past
geodesic. Because of the pre-big-bang contracting phase, this
assumption is violated in the LQC effective theory.
\index{singularity resolution in LQG|)}

 \subsubsection{Phenomenology: Implications of the Pre-inflationary Dynamics}

The inflationary scenario has had an impressive success in
accounting for the observed $1$ part in $10^5$ anisotropies in the
CMB. Therefore, although many of the LQC results hold in a broad
class of early universe paradigms (see, e.g., \cite{other}), for
brevity and concreteness we will restrict ourselves to inflation
here.

The resolution of the big bang singularity opens a natural avenue to
extend this scenario to the Planck regime by systematically
investigating the pre-inflationary dynamics. It is often argued that
while this phase is conceptually important, it can not be relevant
for observations because the near-exponential expansion during
inflation would wash away all memory of prior dynamics. The
reasoning is that modes seen in the CMB cannot be excited by the
pre-inflationary dynamics because, when evolved back in time
starting from the onset of the slow roll, their physical wave
lengths $\lambda_{\rm phy}$ continue to remain within the Hubble
radius $\mathfrak{R}_{\rm H}$ all the way to the big bang. However,
this argument is flawed on two accounts. First, what matters to the
dynamics of these modes is the curvature radius $\mathfrak{R}_{\rm
curv} = \sqrt{6/R}$ determined by the Ricci scalar $R$, and not
$\mathfrak{R}_{\rm H}$, and the two scales are equal only during
slow roll. Thus we should compare $\lambda_{\rm phy}$ with
$\mathfrak{R}_{\rm curv}$ in the pre-inflationary epochs. The second
and more important point is that the pre-inflationary evolution
should not be computed using general relativity, as is done in the
argument given above. One has to use an appropriate quantum gravity
theory since the two evolutions can well be very different in the
Planck epoch. Therefore, modes that are seen in the CMB could have
$\lambda_{\rm phy} \gtrsim \mathfrak{R}_{\rm curv}$ in the
pre-inflationary phase. If this happens, these modes \emph{would be}
excited and the quantum state at the onset of the slow roll could be
quite different from the Bunch Davies (BD) vacuum used at the onset
of the slow roll.

Now, another common assumption was that even if there are such
excitations over the BD vacuum at the onset of inflation, they would
have no effect because they would be diluted away during inflation.
However, this is not the case: stimulated emission compensates for
expansion so the excitations persist at the end of inflation
\cite{parker69,agullo-parker}. Indeed, the difference from the
standard prediction could well be so large that the resulting power
spectrum is incompatible with the amplitude and the spectral index
observed by WMAP. \index{WMAP}
In this case, that particular quantum gravity
scenario would be ruled out. On the other hand, the differences
could be more subtle: the new power spectrum for scalar modes could
be the same but there may be departures from the standard
predictions that involve tensor modes or higher order correlation
functions of scalar modes, changing the conclusions on
non-Gaussianities. In this case, the quantum gravity theory would
have interesting predictions for future observational missions
\cite{agullo-parker}. Thus, pre-inflationary dynamics can provide an
avenue to confront quantum gravity theories with observations.

To analyze what happens during the pre-inflationary phase, in LQC
one proceeds as follows. Since in the inflationary paradigm it is
adequate to consider just the FLRW geometries and first order scalar
(or curvature) and tensor perturbations $\mathcal{R},\,
\mathcal{T}$, one first truncates the full phase of GR to this
sector, replaces the FLRW metrics with the quantum wave functions
$\Psi_o$ provided by LQC and investigates the dynamics of first
order quantum perturbations $\hat{\mathcal{R}},\, \hat{\mathcal{T}}$
on these \emph{quantum} FLRW geometries \cite{akl,aan}. Since
quantum perturbations now propagate on quantum geometries which are
all regular, free of singularities, \emph{the framework
automatically encompasses the Planck regime.} What is then the
status of the `trans-Planckian issues' discussed in the context of
inflation? A careful examination shows that they boil down to the
following question: Is the LQC truncation scheme self-consistent?
That is, is it consistent to ignore the back reaction and work just
with first order quantum perturbations on quantum FLRW backgrounds?
This central issue is extremely difficult to analyze in any approach
to quantum gravity because it requires a careful treatment of
regularization and renormalization of the stress energy tensor of
quantum perturbations on FLRW \emph{quantum} geometries.

The LQC analysis was carried out in detail using the simplest
$\frac{1}{2}\,m^2\phi^2$ potential that is compatible with the
current observations. It has revealed three interesting features
\cite{aan,barraurev,asrev}. First, there exist quantum states of the
background FLRW geometry that remain sharply peaked on solutions to
effective equations all the way from the bounce till the curvature
has fallen by several orders of magnitude, when general relativity
is an excellent approximation. \index{WMAP}
Therefore, one can focus on effective
dynamics and ask if these solutions would generically encounter the
phase of slow roll inflation that is compatible with observations.
It turns out that these solutions are completely determined by the
value $\phi_{\rm B}$ of the inflaton at the bounce and it is 
constrained to lie in a finite interval, $|\phi_{\rm B}| \in  [0, 
7.47\times 10^5]$.\emph{This is the parameter space of LQC.} For 
definiteness, let us suppose that the inflaton and its time derivative 
have the same sign at the bounce. Then, the detailed analysis shows 
that the dynamical trajectory \emph{will} encounter an inflationary 
phase compatible with observations (within the WMAP error bars) 
provided $\phi_{\rm B}>0.93$, i.e., in almost the entire parameter 
space \cite{as3}. \index{quantum bounce}

So we can choose an effective trajectory with $\phi_{\rm B} >0.93$, 
select a quantum state $\Psi_o$ which is sharply peaked on it and consider
quantum fields $\hat{\mathcal{R}},\,\hat{\mathcal{T}}$ representing
scalar and tensor perturbations on the quantum geometry $\Psi_o$. At
first the problem of studying their dynamics seems intractable.
However, the detailed investigation has brought out a second
non-trivial and completely unforeseen feature: assuming that the
back reaction can be neglected, dynamics of
$\hat{\mathcal{R}},\,\hat{\mathcal{T}}$ is on quantum geometry
$\Psi_o$ is \emph{identical} to that of quantum fields
$\hat{\mathcal{R}},\,\hat{\mathcal{T}}$ propagating on a smooth,
classical FLRW metric $\bar{g}$ constructed from $\Psi_o$. This
construction is quite subtle and involves rather complicated
combinations of the expectation values of various operators in the
state $\Psi_o$. Thus, although the scalar and tensor modes
$\hat{\mathcal{R}},\,\hat{\mathcal{T}}$ propagate on the quantum
geometry $\Psi_o$, their dynamics is sensitive to only those
features of $\Phi_o$ that are captured in $\bar{g}$. This $\bar{g}$
is a `dressed' effective metric: While the metric determined by the
effective equations discussed above knows only about the expectation
values, $\bar{g}$ knows also about certain fluctuations, i.e., a
finite number of `higher moments' of $\Psi_o$. Physics behind this
result can be intuitively understood in terms of a simple analogy:
As light propagates in a medium, while there are many interactions
between the Maxwell field and the atoms of the medium, the net
effect can be neatly coded in just a few parameters such as the
refractive index. In LQC, the result provides a powerful technical
simplification because it enables one `lift' various well-developed
mathematical techniques from QFT on classical FLRW spacetimes to
$\hat{\mathcal{R}},\,\hat{\mathcal{T}}$ propagating on quantum
geometries $\Psi_o$.

However, this analysis assumes that the back reaction can be
neglected. One can always start by restricting oneself to states
$\psi$ for which this assumption holds at the bounce.
\index{quantum bounce} But there is
no guarantee that the condition will continue to be satisfied under
evolution especially in the Planck regime immediately after the
bounce. Does the energy density of the fields
$\hat{\mathcal{R}},\,\hat{\mathcal{T}}$ remain negligible all the
way from the deep Planck regime of the bounce to the onset of slow
roll, removed from the bounce by some 11 orders of magnitude in
curvature? This issue can be settled only numerically. These
simulations require great care because: i) the renormalization
procedure subtracts two diverging terms whence even a tiny loss of
precision can result in a significant error; ii) the simulation has
to be carried over a very large number of time steps; and, iii)
since the background density falls rapidly, even extremely small
numerical errors (of the order of one part in $10^{15}$) can be
comparable to the background energy density. Simulations with all
the due care have been performed to establish firm \emph{upper
bounds} on the energy density in perturbations. They showed that if
$\phi_{{\rm B}} >1.23$, there is a natural choice of initial conditions
for $\psi$ at the bounce such that the back reaction can indeed be
ignored from the bounce to the onset of inflation. 
\index{quantum bounce} Furthermore,
there are analytical arguments to show that if a state $\psi$
satisfies this condition, then all states in an open neighborhood do
so. Any of these states provide a \emph{self consistent} solution in
which the initial truncation hypothesis is seen to be satisfied in
the final solution. This is the third non-trivial result. Together,
the three results establish that, LQC does provide a self-consistent
extension of the standard inflationary scenario to the Planck regime
for almost all of the LQC parameter space.

What are then the phenological predictions of these self-consistent
solutions? The power spectrum and the spectral index have been
calculated and, as in the standard inflationary calculations, they
agree with observations within error bars. However, there is a small
window in the LQC parameter space where certain LQC predictions
differ from those of standard inflation. For example, the standard
`consistency relation' $r = -8n_t$ relating the ration $r$ of the
tensor to scalar power spectra to the tensor spectral index is
modified \cite{aan}. These deviations arise precisely by the
mechanism we discussed above: the LQC effective dynamics of the FLRW
background is qualitatively different from that of GR so that
certain modes \emph{can} have wave lengths $\lambda$ larger than the
curvature radius. Therefore, at the onset of inflation the LQC
quantum state $\psi$ of perturbations has excitations over the BD
vacuum in these modes. This departure from the BD vacuum also has
implications for the CMB and galaxy distribution
\cite{agullo-parker} and observational tests for such effects have
already been proposed \cite{observations}. A careful analysis of
this window in the LQC parameter space is a focus of current
research.

To summarize, LQC has led to a natural resolution of the initial
singularity in cosmological models of direct physical interest via
quantum geometry effects that replace the big bang with a big bounce
\cite{mb,aps,apsv}. \index{quantum bounce}
Cosmological perturbations on these quantum
geometries have been studied in detail \cite{aan,madrid,french}.
There are natural choices of states at the bounce for which one
obtains self consistent extensions of the inflationary scenario all
the way to the Planck regime of the bounce. By combining these
results with the very rich set of results on inflationary and
post-inflationary dynamics, one obtains is a coherent paradigm to
account for large scale structure, starting right at the quantum
bounce. Furthermore, in a small window of the parameter space, this
analysis provides results that differ from standard inflation,
thereby opening an avenue to extend the reach of observational
cosmology to the Planck scale.

\subsubsection{Is Quantum Cosmology Justified?}

As emphasized in section \ref{s2.3}, in our most successful
theories, such as QED and QCD, the \emph{actual} calculations of
physical effects have always involved truncations. 
\index{QCD} \index{QED}
The mini and midi superspace were introduced in the 1960s in the hope that this
truncation would be sufficient to capture the salient quantum
effects that tame cosmological singularities. Now that this hope is
borne out, it is appropriate to reexamine the strategy and ask: Is
this truncation where one ignores an infinite number of degrees of
freedom not too severe?

The LQC strategy is guided by the following considerations. First,
there is an analogy with Dirac's solution to the Hydrogen atom
problem. From the perspective of full QED, Dirac's restriction to
spherical symmetry is a drastic truncation because it removes all
physical photons and ignores all but a finite number of degrees of
freedom. \index{QED} 
But the results of this truncated theory are in excellent
agreement with observations and we need quantum corrections from QED
only when the accuracy of experiments is at the level of the Lamb
shift when the vacuum fluctuations of the photon field cannot be
ignored. The viewpoint is that the situation is similar in
cosmology: An analysis of the problem in the mini-superspace
approximation \emph{that appropriately takes into account quantum
geometry effects from the full theory} should provide a good
approximation to the predictions of the full theory. The second
source of intuition is provided by the Belinskii Khalatnikov
Lifshitz (BKL) conjecture in GR discussed in Chapter 9. It suggests
that as one approaches a generic space-like singularity in GR, the
local evolution is well approximated by the Bianchi I, II and IX
models. \index{Bianchi models} \index{quantum bounce}
Therefore the fate of singularities in Bianchi models is of
special interest. A common concern is that even if the big bang is
replaced by a big bounce in the isotropic case, typically this
singularity resolution would not survive in Bianchi models
(primarily because the anisotropic shear terms diverge as $1/a^6$
where $a$ is the scale factor). In LQC, by contrast, the big bang
singularity is again resolved once the quantum geometry effects from
full LQG are correctly incorporated \cite{awe,we}. Furthermore, if
one traces the Hamiltonian constraint of the Bianchi I model over
anisotropies, one is led precisely to the FLRW hamiltonian
constraint, bringing out robustness of the scheme. 
Finally, in the
CDT simulations one finds that even when one allows all fluctuations
in geometry keeping only the scale factor fixed, the behavior of the
scale factor, including quantum fluctuations, is described
accurately by a mini superspace model which assumes homogeneity and
isotropy from the outset \cite{agjl-rev}. Putting together these
diverse results, it is not unreasonable to hope that these models
adequately capture the behavior of global observables (such as the
scale factor and average matter density) that would be predicted by
the full theory.

What about the truncation used in treating cosmological
perturbations $\hat{\mathcal{R}},\,\hat{\mathcal{T}}$? Full LQG
\emph{will} admit states in which there are huge quantum
fluctuations in the Planck regime whose physics cannot be captured
by states of the type $\Psi_o\otimes\psi$ where $\Psi_o$ is a state
of the quantum FLRW geometry and $\psi$ is the state of linear
quantum perturbations $\hat{\mathcal{R}},\,\hat{\mathcal{T}}$. It is
often implicitly assumed that \emph{all} states of the full quantum
gravity will have huge fluctuations. LQC has provided concrete
evidence that this need not be the case: there do exist states of
the type $\Psi_o\otimes\psi$ for which truncation is \emph{self
consistent}. These states lead to an unforeseen, tame behavior in
which $\hat{\mathcal{R}},\,\hat{\mathcal{T}}$ evolve as linear
perturbations on a background quantum geometry $\Psi_o$ carrying
energy densities that are negligible compared to that in the
background. The non-triviality lies in the fact that these self
consistent, truncated solutions lead to the power spectrum and
spectral index that are consistent with observations. Thus, the
situation is similar to that in the standard $\Lambda$CDM\index{$\Lambda$CDM} model
where it suffices to restrict oneself to the simplest cosmological
solutions. The early universe appears to be simpler than what one
would have a priori imagined!

\index{loop quantum gravity!loop quantum cosmology|)}

\subsection{Black Holes}
\label{s3.2}

\index{quantum black holes|(}

As is clear from Chapter 4, black holes (BHs) serve as powerful
engines that drive the most energetic astrophysical phenomena. But,
as discussed in Chapter 10, they have also driven developments in
fundamental physics, particularly quantum gravity, raising deep
conceptual questions about the statistical mechanical origin of the
Bekenstein-Hawking entropy \cite{jdb,swh} and a quantum gravity
description of the BH evaporation process \cite{swh}. In this
subsection we will provide a brief summary of developments in LQG in
this area.

\subsubsection{Quantum Horizon Geometry and Micro-canonical Entropy}

\index{quantum horizons|(} \index{black hole!entropy|(}
In statistical mechanics, entropy is generally associated with
systems in equilibrium. BHs in equilibrium were first modeled using
event horizons of stationary space-times in GR. However, in
statistical mechanics equilibrium refers only to the system under
consideration, and not the entire universe. Therefore, about 15
years ago, a quasi-local framework was introduced through the notion
of \emph{isolated horizons} (IHs) to better model BHs which are
themselves in equilibrium, allowing for dynamical processes in the
exterior \cite{akrev}. Event horizons of stationary space-times as
well as the cosmological horizons in de Sitter space-time are
special cases of IHs. Interestingly, the first law of BH
thermodynamics naturally extends to IHs, with a further advantage
that mass and angular momentum in the law now refer to the BH
itself, defined at the IH, rather than to the ADM quantities\index{ADM formulation} defined
at infinity which receive contributions also from the exterior
region \cite{abl}. Its form is again similar to the first law of
thermodynamics, suggesting that a multiple of the area $a_\Delta$ of
the IH $\Delta$ should be interpreted as entropy $S_\Delta$.
Hawking's analysis of quantum radiance provides the numerical value
of the multiple, yielding the Bekenstein-Hawking formula, now for
IHs: $S_\Delta = a_\Delta/4G_{\rm N}\hbar$.

In LQG, one investigates the statistical mechanical origin of this
entropy, $S_\Delta$ \cite{blvrev}. The shift of focus to IHs has two
advantages. First, one can consider realistic, astrophysical black
holes: Not only does one not have to invoke `charges' to make BHs
near-extremal, but one can even allow for distortions in the horizon
geometry that may be caused by matter rings or other black holes.
Second, the cosmological horizons (for which thermodynamic
considerations are known to hold) are automatically incorporated.
The idea is to first investigate the quantum geometry of these IHs
\cite{abck,aev} and then calculate the number of quantum microstates
in the specified ensemble \cite{counting,blvrev}. This procedure
provides a statistical mechanical derivation of entropy in terms of
quantum
geometry. 
We will now summarize these developments.

As before, one carries out a truncation of the theory that is
motivated by the physical problem of interest. Thus, one begins with
the phase space of GR in connection dynamics, \index{connection dynamics} now with a spatial
3-manifold $M$ that is asymptotically flat \emph{and} has an
internal boundary $S$, the intersection of $M$ with an IH 3-manifold
$\Delta$.
Detailed analysis shows that the total phase space can now be
written as $\mathbf{\Gamma} = \mathbf{\Gamma}_{\rm bulk} \times
\mathbf{\Gamma}_{\rm S}$ where $\mathbf{\Gamma}_S$ turns out to be
the phase space of an $\U(1)$ Chern-Simons theory. The IH boundary
condition relates the curvature $F$ of the $\U(1)$ Chern-Simons
connection to the pull-back $\underline{\Sigma}$ of the 2-form
$\eta_{abc}E^c_i r_i$ where $r^i$ is the unit internal vector normal
to $S$: $F = -(2\pi/a_{\Delta})\, 8\pi G_{\rm N} \gamma
\underline{\Sigma}$.

\begin{figure}[h]
\includegraphics[height=6cm,width=3.7in]{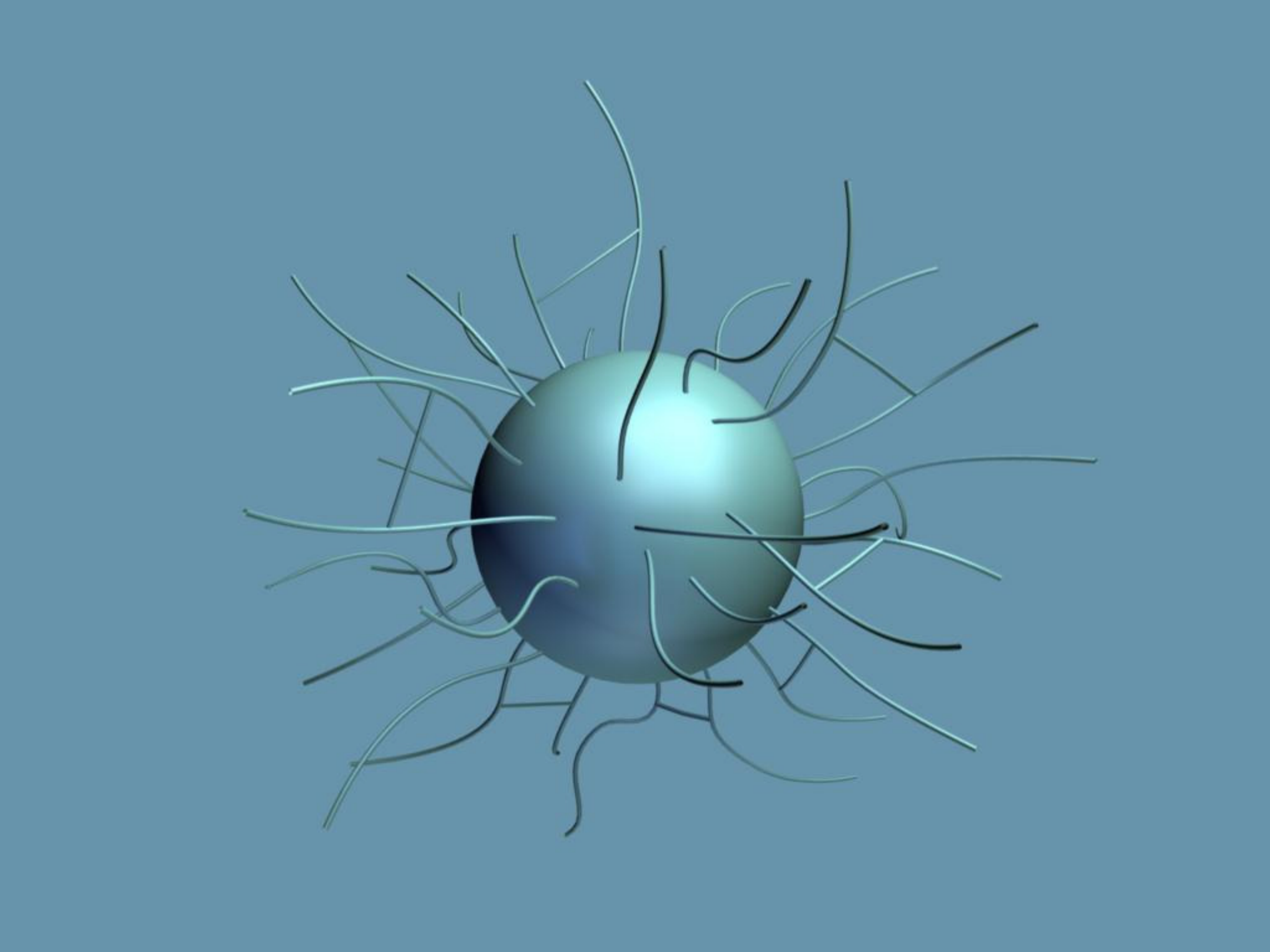}
\caption{An artist's rendering of an isolated horizon punctured by 
spin-network links. Image credit: Alejandro Corichi.}
\label{bh}
\end{figure}

For this sector of GR, one has to extend the quantum geometry
framework of section \ref{s2.2} to allow for an inner boundary $S$
corresponding to an IH. The bulk Hilbert space $\H_{\rm bulk}$ is
again spanned by spin networks. \index{spin network}
However, the links of these
spin-networks can now end on the boundary $S$, piercing it on a node
(see Fig. \ref{bh}). The surface Hilbert space $\H_{\rm CS}$ is now
the Hilbert space of an $\U(1)$ Chern-Simons theory on the resulting
punctured sphere, with the level (or, dimensionless coupling
constant) $k = a_{\Delta}/4\pi \gamma\lp^2$. The total kinematical
Hilbert space is now a tensor product $\Hk = \H_{\rm bulk} \otimes
\H_{\rm CS}$. States in $\Hk$ are now subject to the \emph{quantum
horizon boundary condition} which is an operator equation:
\be \label{qhbc} (1\otimes \hat{F}) \Psi = -\frac{2\pi}{a_\Delta}\,
8\pi G_{\rm N}\gamma (\underline\Sigma \otimes 1) \Psi \ee
Note that solutions to (\ref{qhbc}) can exist only if $\hat{F}$ on
the surface Hilbert space$\H_{CS}$ has the same eigenvalues as the
triad operator $\underline\Sigma$ on $\H_{\rm bulk}$. This is a
highly non-trivial condition since the two operators have been
defined \emph{completely independently} on two \emph{distinct}
Hilbert spaces. However, the framework passes this severe test
because the two operators share an infinite number of eigenvalues.
Finally, the physical meaning of this condition is as follows: The
intrinsic curvature of the IH can fluctuate and so can the bulk
geometry in its neighborhood, but they have to fluctuate in tandem,
satisfying (\ref{qhbc}).

Next, one has to impose quantum constraints. There are some
interesting subtleties which lead to mapping class groups and
quantum deformation of $\U(1)$ on $S$ \cite{abck}. The final result
is that what matters is only the number of punctures, not their
location on $S$, \emph{and} that each puncture has to be treated as
`distinguishable.' We emphasize that these are systematic
implications of the constraint equations, and \emph{not} additional
inputs, as is sometimes thought. 
The net result is that, assuming the Hamiltonian constraint does
admit a sufficient number solutions in the bulk, LQG provides a
coherent description of quantum space-times with IHs \cite{abck}.
This quantum geometry is depicted in Fig. \ref{bh}. (For more
detailed summaries, see e.g. \cite{alrev,blvrev}).

To calculate entropy, one has to fix an ensemble and count the
number of states compatible with the macroscopic parameters
characterizing the ensemble. This is done via the notion of
multipole moments that characterize the geometry of $\Delta$ in a
diffeomorphism invariant manner \cite{aev}. (In the simplest case
when all moments except the mass monopole vanish, the IH is
spherically symmetric with respect to \emph{some} ${\rm SO(3)}$
action.) The ensemble is specified by requiring that all multipoles
lie in a small interval around some pre-specified values. The idea
is to calculate the number $\mathcal{N}$ of quantum microstates of
the horizon geometry that satisfy this constraint. Its logarithm
gives the microcanonical entropy of the ensemble.

To determine $\mathcal{N}$ one has to count specific types of finite
sequences of half integers subject to certain constraints
\cite{blvrev}. This problem has been investigated in detail in a
series of mathematical papers by Barbero, Villase\~{n}or and others
that are of interest in their own right \cite{counting}. They
combine known types of Diophantine equations
with techniques involving generating functions and Laplace transforms. 
The final result is that the micro-canonical entropy is given by
\be S_{\rm micro} = \frac{\gamma_o}{\gamma} \frac{a_{\Delta}}{4 \lp^2} 
+ O(\ln \frac{a_{\Delta}}{\lp^2}) \ee
where $\gamma_o \sim 0.2$ is a root of an algebraic equation%
\footnote{Because the surface states on $\Delta$ are intertwined
with the bulk spin network states, \index{background independence}
a priori one can give assign two
meanings to the term `pure surface terms' that are to be counted.
They lead to two values $\approx 0.27$ and $0.24$ of $\gamma$. (See,
e.g., \cite{blvrev}.) In detailed LQG calculations this difference
only changes numerical values by small amounts. But conceptually it
is important to better understand and resolve this ambiguity.}
\cite{blvrev}. Thus, in the sector of the theory where the BI
parameter is set to $\gamma_o$, one recovers the Bekenstein-Hawking
result to the leading order with a logarithmic correction. The
Barbero-Immirzi parameter is a quantization ambiguity in LQG, rather
similar to the $\theta$-ambiguity in QCD \cite{holst}. In QCD, the
value of this parameter is determined experimentally. \index{QCD}
In LQG, an experimental measurement of, say, the area gap \index{area gap} would similarly
determine $\gamma$. The LQG viewpoint is that while such
measurements are completely out of reach of current technology, the
Bekenstein-Hawking formula can be used as a theoretical constraint
to determine $\gamma$. Note that once we set $\gamma=\gamma_o$ to
get agreement with this formula for one type of IH (say spherical
ones), the agreement extends to all IHs.

\subsubsection{Semiclassical Considerations and Dynamical Processes}

The description of quantum horizons we just summarized has the
advantage that it is fully background independent. But that very
feature makes it difficult to relate it to the rich body of
semi-classical results that have been derived in Kerr and Rindler
space-times. Therefore, over the last three years, two independent
avenues have been introduced to make closer contact with
semi-classical results and study quantum dynamical processes. In
this subsection we will briefly describe their current status.

In the first approach, developed by Ghosh, Perez and others,
\cite{perez-bhs} one considers the near horizon geometry of Kerr
space-times and asks the question: How would near horizon,
stationary observers describe physics \emph{within} LQG? Denote by
$\chi^a$ the Killing vector which is the null normal to the Kerr
horizon $\Delta$ and consider observers $\mathcal{O}$ with
4-velocity $u^a = \chi^a/\sqrt{\chi\cdot\chi}$, at a fixed distance
$d \ll R_{\Delta} = \sqrt{(a_{\Delta}/4\pi)}$ from $\Delta$. Note
that 
the observers $\mathcal{O}$ are approximately at rest with respect
to $\Delta$ since their angular momentum is $O(d/R_\Delta)$. If one
were to consider the Hamiltonian framework with a boundary at the
location of the observers $\mathcal{0}$, one would find that the
Hamiltonian acquires, in addition to the ADM surface
integral\index{ADM mass}\index{ADM momentum} at
infinity, a 2-surface integral $H_{\mathcal{O}}$ at the inner
boundary which, one argues, is given by $H_{\mathcal{O}} =
a_{\Delta}/(8\pi G d)$. In LQG, the corresponding operator is
$\hat{H}_{\mathcal{O}} = \widehat{\rm Ar}_{S}/(8\pi G d)$, where
$\widehat{\rm Ar}_{S}$ is the area operator of section \ref{s2.2},
now associated with the intersection $S$ of $\Delta$ with a partial
Cauchy surface $M$ used in the Hamiltonian framework. Next, since
the acceleration of $u^a$ is given to the leading order by $1/d$,
one assumes that observers $\mathcal{O}$ would experience the Unruh
temperature $T_{\rm U} = 1/(2\pi d)$ \cite{wu}. This is supported by
two independent considerations: i) if one red-shifts $T_{\rm U}$ to
infinity, one obtains the Hawking temperature $T_{\rm H}$, and, ii)
detectors carried by the observers $\mathcal{O}$ coupled to
$\hat{H}_{\mathcal{O}}$ would read local temperature $T_{\rm U}$
\cite{perez-bhs}.

Using these ingredients, one arrives at the following physical
picture: the observers $\mathcal{O}$ would describe the punctured
quantum horizon $\Delta$ as a grand canonical ensemble
$\tilde{\rho}_{\mathcal{O}}[\beta, \mu; \gamma]$ of \emph{punctures}
$p$ endowed with spin labels $j_p$, at an inverse temperature
$\beta_{\rm U} = (2\pi d)/\hbar$ and a chemical potential $\mu$.
(The dependence on the Barbero-Immirzi parameter $\gamma$ comes from
$\hat{H}_{\mathcal{O}}$). As usual, this is equivalent to a
canonical ensemble $\rho_{\mathcal{O}}[\beta_{\rm U}; \gamma]$ in
which $-T \frac{\partial S}{\partial N}\mid_E$ equals the chemical
potential $\mu$ of the grand canonical ensemble. An explicit
calculation of $-T \frac{\partial S}{\partial N}\mid_E$ provides $\mu$
as a function $\mu(\gamma)$ of $\gamma$.
%
Finally, recall from section \ref{s2.2} that the level spacing
between eigenvalues of the area operator goes to zero exponentially
for large areas. Hence the energy required to create a new puncture
is arbitrarily small for large black holes. One therefore makes the
final assumption that, as for photons, the physical value of the
chemical potential $\mu$ should be zero. This condition determines
$\gamma$ uniquely and the value is precisely the  $\gamma_o$ arrived
at by state counting in the micro-canonical ensemble irrespective of
the choice of $d$ (which enters only in the local temperature that
$\mathcal{O}$ attribute to the BH). In the resulting canonical
ensemble $\rho_{\mathcal{O}}[\beta_{\rm U}; \gamma_o]$ that the
observers $\mathcal{O}$ would use to describe the BH, the entropy is
given by $S = a_{\Delta}/4\lp^2$ to leading order, exactly as in the
micro-canonical ensemble. Thus, these semi-classical considerations
provide the same final result but with a novel description of the
quantum horizon as a gas of punctures carrying spins. Therefore,
this approach opens new avenues to describe dynamical processes,
including the BH evaporation. \index{black hole!evaporation}\medskip

The second and complementary development is due to Gambini and
Pullin \cite{gp} and follows a strategy that is analogous to the one
used in LQC. It considers a different truncation of GR, that of
spherically symmetric space-times. While this truncation was
discussed in the LQG literature already in the 90s, the
\emph{global} structure of the quantum space-times ---including both
the asymptotic part and the portion that is classically inside the
horizon--- was analyzed relatively recently.

In this symmetry reduced model, it suffices to consider spin
networks with graphs along just the radial line. However, the nodes
now carry additional labels that encode information about the
connection and geometry in the two transverse directions. While
there are close similarities with LQC, there is a major difference:
Since the 3-geometry is now inhomogeneous, we have infinitely many
Hamiltonian as well as diffeomorphism constraints, smeared with
radial lapse and shift fields \cite{bs}. Remarkably, it is possible
to express solutions to the Hamiltonian constraints in a closed form
as a linear combination of the spin networks \cite{gp}. 
\index{background independence} The
diffeomorphism constraint can then be solved as in section
\ref{s2.3} by group averaging \cite{almmt1}. In the resulting
physical Hilbert space, the ADM mass\index{ADM mass} is a Dirac observable as in the
classical theory. Furthermore, as in LQC, by appropriately
deparameterizing the theory, one can also express the metric as a
parameterized Dirac observable. As one would expect from quantum
geometry, the metric is an operator valued distribution concentrated
at the nodes of the spin networks. There are semi-classical states
which upon coarse graining on an appropriate scale ---say, a
thousand times the Planck length--- yield smooth classical
geometries. However, as in LQC, the quantum space-time is
singularity-free and, as was anticipated by calculations within
effective LQG equations for this model, the quantum space-time is
`larger' than that of classical GR. At a technical level, the fact
that one can solve the infinite set of both Hamiltonian and
diffeomorphism constraints is highly non-trivial.

As in LQC, it is now natural to investigate the behavior of test
quantum fields on the quantum geometry of the symmetry reduced
model. For this, one now truncates the theory allowing linear scalar
fields on spherically symmetric space-times, again ignoring the back
reaction in the first step. Then, as in LQC, the scalar field
$\hat{\Phi}$ now propagates on a quantum state $\Psi_o$ of the
background geometry that, on coarse graining, yields the classical
Schwarzschild geometry of a large black hole. In the interaction
picture, in the approximation in which the back reaction is ignored,
the field $\hat\Phi$ again propagates on an effective dressed
quantum geometry. The main effect of the background quantum space
time on quantum field theory is to replace the partial differential
equation governing  $\hat\Phi$ with a difference equations. However,
for frequencies (at infinity) which are significantly smaller than
the Planck frequency, there is negligible difference from the
thermal spectrum at infinity. This is not surprising because the
Hawking radiation is robust with respect to the near horizon
microstructure of space-time \cite{robust}. But conceptually the
underlying discreteness of quantum geometry does have one important
effect: it removes the UV divergences encountered in the Boulware
and Unruh vacua at the horizon \cite{gp}.

This recent development has provided a coherent framework to
describe Hawking radiation from first principles using the strategy
of truncating LQG to the physical problem of interest. At a
technical level, as we indicated, there is a close similarity with
the framework used in LQC. 
On the physical side, on the other hand, there is a difference.
Since the issue of the back reaction of \emph{quantum} perturbations
is significant only in the very early universe, in LQC one could
analyze this issue systematically and show that the truncation used
is physically self-consistent. For black holes, on the other hand,
it is the back reaction that drives the evaporation process.
Therefore, the truncation used so far for black holes is not
adequate to systematically analyze the issue of information loss.
\index{quantum horizons|)} \index{black hole!entropy|)}
\index{black hole!evaporation} \index{quantum black holes|)}

\subsection{$n$-point Functions in a Diffeomorphism Invariant Theory}
\label{s3.3}

\index{background independence!\& n-point functions|(}
As Wightman emphasized already in the 1950s,
in MQFTs the $n$ point functions
\be W(x_1,\ldots,x_n) = \langle{0|{\phi(x_n)\,\ldots\, \phi(x_1)}|0}\rangle,  \label{npoint} \ee
completely determine the theory \cite{Wightman:1959fk}. In
particular, one can calculate the scattering amplitudes from these
distributions. However, since they make an explicit reference to the
Minkowski metric, it is far from being a priori clear that these
ideas can be extended in a meaningful manner to non-perturbative
quantum gravity. Indeed, at first it may appear that, because
manifolds do not admit non-trivial, diffeomorphism invariant n-point
distributions, a background independent framework cannot lead to
non-trivial n-point functions either. However, as we will see, this
argument is too naive. The $n$-point functions refer to a state and
in gravity that state can naturally encode information about a
specific geometry which can then appear in the expressions of these
distributions. In particular, LQG does lead to non-trivial $n$-point
functions. Furthermore, to the leading order they have been shown
\cite{Bianchi:2011hp,Bianchi:2009ri,Rovelli:2011kf,Han:2013tap} to
agree in the appropriate sense with the $n$-point functions
calculated in the effective low energy quantum general relativity
\cite{donoghue} referred to in section \ref{s2.1}. These
calculations have created a bridge from the rather abstract and
unfamiliar background independent framework of LQG to notions and
techniques used in concrete calculations in familiar MQFTs.

\begin{figure}
\begin{center}
\includegraphics[height=35mm]{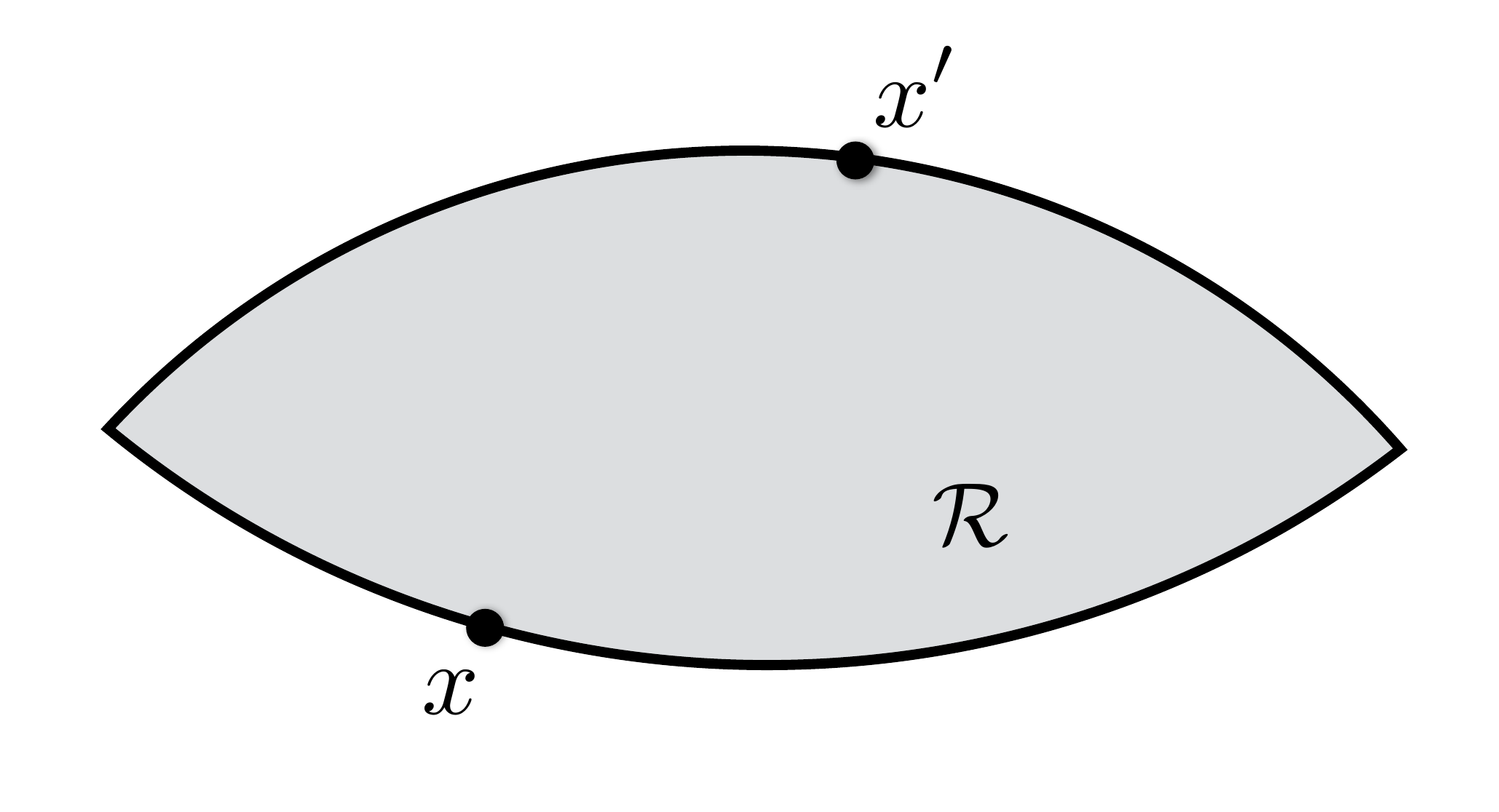}
\caption{Geometrical interpretation of the transition amplitude.}
\label{box}
\end{center}
\end{figure}

To spell out the construction, let us first return to MQFTs and
recall that an $n$-point function can be written as a path integral.
 \be
 W(x_1,\ldots,x_n)=\langle{0|{\phi(x_n),\,\ldots,\,\phi(x_1)}|0}
 \rangle = \int D\phi \ \phi(x_n)...\phi(x_1)\, e^{i{\mathcal S}[\phi]}
 \label{pathintegralfield}
 \ee
For simplicity, consider the two-point function. We can organize the
integration in \eqref{pathintegralfield} as follows. Select an
arbitrary \emph{compact} region $\R$, as in the Fig. \ref{box}, such
that the points $x,x^\prime$ of interest lie on its boundary $b$.
Denote by $W(\varphi)$ the integral over fields $\phi$ defined only
on $\R$ with the boundary value $\varphi$ on $b$, and by
$\Psi_b[\varphi]$ the integral over fields defined only on the
exterior region $M - \R$, again with the boundary value $\varphi$ on
$b$ (and appropriate fall-off at infinity). Then, we have:
\be 
W(x,x^\prime) =\int D\varphi\,\,  W[\varphi]\,
\varphi(x)\varphi(x^\prime)\, \Psi_b[\varphi_b] \label{carl} \ee
From the perspective of the region $\R$, this expression can be
interpreted as providing the 2-point function for the boundary state
$\Psi_b[\varphi]$, the transition amplitude ($\sim e^{iS_\R}$) being
given by $W[\varphi]$.

In the form (\ref{carl}), the functional integral can be taken over
to quantum gravity using Spinfoam transition amplitudes $W_\mathcal{C}$
introduced in section \ref{s2.2}. But there are crucial differences
in the underlying structures that are needed in MQFTs versus LQG. In
MQFTs, in addition to the value $\varphi$ of the field on the
boundary $b$, we must \emph{also} use the background metric to fix
the shape and geometry of the boundary $b$, and the spacetime
distance between $x^\prime$ and $x$. Only then can we calculate the
transition amplitude $W[\varphi]$ associated with the region $\R$,
and the boundary state $\Psi_b[\varphi]$:
\be  W[\varphi] = \int D\phi \ e^{i\int_{\cal R}L[\phi]},\quad {\rm
and} \quad \Psi_b[\varphi] = \int D\phi \ e^{i\int_{M-\cal
R}\,L[\phi]},   \ee
But in LQG, these expressions cannot and do not make reference to a
background metric. As we saw in section \ref{s2.2}, a signature of
the non-triviality of the LQG construction is that the transition
amplitude $W[\varphi]$ is a function only of the field $\varphi$. It
does not depend on any background structures; it refers only to the
dynamical fields and the physical process of interest.

How can one then make contact with the effective field theory
calculations which do refer to a background geometry? This is
achieved through the state $\Psi_b[\varphi]$. \index{effective field theory}
Because of the functional integration involved in its definition,
$\Psi_b[\varphi]$ depends on the dynamics of the theory as well as
the boundary conditions at infinity. But as in MQFTs, these
ingredients are invisible to the calculation of the 2-point function
(\ref{carl}); what matters directly is the state $\Psi_b[\phi]$
itself. But the key difference from MQFTs is that, because we are
considering gravitational fields, \emph{this state now encodes
information about the geometry.} For appropriate boundary conditions
at infinity, $\Psi_b[\varphi]$ would be peaked at a certain
classical (intrinsic and extrinsic) 3-geometry of $b$. Therefore, it
assigns to $b$ a certain shape and size and also selects a classical
4-geometry in $\R$ that extremizes the classical action for the
given boundary geometry $\varphi$ at $b$. The relative position of
the $n$ arguments of the $n$-point function is well defined with
respect to that 4-geometry. In particular, for comparison with the
effective theory, one picks a region $\R$ of Minkowski spacetime and
approximates it with a triangulation that is sufficiently fine to
capture the relevant dynamical scale of the phenomenon of interest.
One then determines the intrinsic and extrinsic geometry of the
boundary of this region, and picks a quantum state
$\Psi\in\H_\Gamma$ of gravity peaked on these data. Then the two
points $\vec x$ and $\vec x'$ sit on nodes of the boundary graph.
The operators $E_\ell$, associated to these nodes and links, provide
the geometric interpretation of the quanta in the boundary state.

This construction resolves a longstanding confusion in quantum
gravity, referred to in the very beginning of our discussion.
Formally, if $S[g]$ is a generally covariant action and $D[g]$ a
generally covariant measure, the distribution
\be
 W(x_1,...,x_n)=\int D[g]\ g(x_1)... g(x_n)\ e^{iS[g]}
\label{wrong} \ee
would be trivial, i.e., independent of the position of the points
$(x_1,...,x_n)$ (as long as they do not overlap). Therefore
\eqref{wrong} are not the physically interesting $n$-point functions
in a generally covariant theory. In particular, they are completely
unrelated to the $n$ point functions used in the effective field
theory: while the \eqref{wrong} refer only to the $n$ points $x_1
\ldots x_n$ in the spacetime manifold, those in the effective theory
depend on \emph{physical distances} between these points computed in
the background metric. In the LQG construction, the $n$ point
functions \emph{do know} about the physical distances between $x_n$
in the background geometry determined by the boundary state
$\Psi_b[\varphi]$. However, under a spacetime diffeomorphism that
sends $\R \to \R^\prime$,\, $b \to b^\prime$ and $ (x_1, \ldots x_n)
\to (x_1^\prime, \ldots, x_n^\prime)$ and $\varphi \to
\varphi^\prime\, $ the boundary state also transforms covariantly,
$\Psi_b(\phi) \to \Psi_{b^\prime}(\phi^\prime)$, whence the geodesic
distances between any two points $x_i, x_j$ on the boundary and
their images $x_i^\prime, x_j^\prime$ are preserved. Therefore, the
final results are diffeomorphism invariant.

Calculation of a $n$-point functions have been performed in this
framework, using states and operators of the canonical theory and
the transition amplitude \eqref{ta} provided by Spinfoams
\cite{Bianchi:2011hp,Bianchi:2009ri,Rovelli:2011kf}. The result is
that, in a suitable semiclassical limit (to terms
$\mathcal{O}(\hbar))$ the two-point function \emph{exactly matches}
\cite{Bianchi:2009ri} with the one obtained from Lorentzian Regge
calculus \cite{regge1}. \index{Regge calculus}
In turn, this limit is consistent with the
effective field theory \cite{donoghue}. \index{effective field theory}
Thus, although the basic notions and techniques appear to be very different from those used in perturbative treatments the final results show that, as in the Asymptotic Safety scenario, a peaceful co-existence with low energy results is possible. A program to make a systematic connection with effective field theory has been initiated recently \cite{Han:2013tap}. Some radiative corrections have also computed using more refined 2-complexes, containing bubbles
\cite{Riello:2013bzw}. Ideally, one would hope that the low energy
behavior of the $n$ point functions computed in the non-perturbative
theory would agree with the effective theory, while at high energies
the LQG calculations would provide an UV completion of the
non-renormalizable perturbation theory. Whether this is the case is
still an open question.

The conceptual non-triviality of results to date lies in the fact
that they provide a streamlined approach to compare the
non-perturbative theory with background dependent effective
theories, without sacrificing the underlying diffeomorphism
invariance. Reconciling the two had been a long standing open issue
in quantum gravity.
\index{background independence!\& n-point functions|)}

\section{Discussion}
\label{s4}

In this section we will first summarize the main ideas and results,
putting them in a broader context, and then discuss open issues that
remain.

\subsection{Summary}
\label{s4.1}

Approaches discussed in this Chapter are rooted in well-established
physics: principles of GR and QFT. The viewpoint is that ideas that
have no observational support should not constitute an integral part
of the foundation of quantum gravity, even when they can lead to
rich mathematical structures. In particular, these approaches do not rely on a negative cosmological constant, \index{cosmological constant}
or extended objects or specific matter content involving towers of fields and particles.%
%
\footnote{In the same spirit, they do not \emph{demand}
supersymmetry nor higher dimensions, but the methods used in these approaches have been extended to incorporate these possibilities \cite{percacci,ttetal1,ttetal2}.}
The primary goal is not unification with other fundamental forces.
Rather, the emphasis is on qualitatively new insights into
\emph{quantum} spacetimes that can emerge from non-perturbative
techniques. In classical GR, the dynamical nature of geometry led to
new phenomena ---such as gravitational waves, black hole horizons
and the big bang--- that could not even be imagined before. As
earlier Chapters in this volume vividly bring out, these notions
have had deep impact on the subsequent developments in astrophysics,
cosmology, computational physics, and geometric analysis. In these
developments one can often use perturbative techniques but they have
to be built around the novel non-linear configurations and use
qualitatively new boundary conditions, dictated by full general
relativity. The expectation is that the situation will be similar in
the quantum domain with the new, unforeseen features of quantum
geometry. Results to date, e.g. on UV finiteness and physics of the
very early universe discussed in sections \ref{s2} and \ref{s3} provide 
concrete evidence in favor of this expectation.

We saw in section \ref{s2.1} that the new notion of 
Effective Average Action (EAA) in the continuum can be used to
give a meaning to the basic functional integral by reformulating the
problem as a question about solutions of a functional flow equation
\cite{mr}. These renormalization group trajectories possess a well
defined ultraviolet limit, allowing one to reconstruct the
functional integral from them if they hit a non-Gaussian fixed point
in this limit. We summarized the evidence for the existence of such
a non-trivial fixed point \cite{mr,oliver1,frank1,oliver2}. We then
discussed CDT, a lattice approach based on statistical mechanics
ideas \cite{agjl-rev,ajl,agjl,SDS}. \index{UV and IR fixed points}


Over the last two decades, 
concrete progress has occurred by carrying out suitable, finite
dimensional truncations of the infinite dimensional theory space 
$\mathcal{T}$. The initial 
truncation had only two coupling constants, $G_{\rm N}$ and $\Lambda$,
corresponding to the Einstein-Hilbert and the cosmological constant
terms. \index{cosmological constant}
By now, the truncations have reached a mature level, allowing
for \emph{nine} different coupling constants
in the gravitational sector. 
Not only has the non-trivial fixed point persisted but
there is consistency with the two dimensional truncation in a
precise sense. The analysis has also been extended beyond pure
gravity and several matter couplings have been investigated in
detail \cite{percacci}. These results have provided highly
non-trivial evidence in support of Asymptotic Safety. 
\index{UV and IR fixed points}

The CDT approach has led to an unforeseen result that had eluded
earlier lattice simulations: In the Euclidean signature, de Sitter
spacetime with small fluctuations was shown to emerge from Monte
Carlo simulations using the discretized Einstein-Hilbert action. The
primary importance of this demonstration is not so much that it is
the de Sitter spacetime that resulted but rather that the result has
an interpretation as a 4-dimensional classical geometry in the first
place. Indeed, previous dynamical triangulation simulations had led only to `crumpled' or `polymer-like' phases rather than the one corresponding to a smooth macroscopic geometry. 

In LQG, the emphasis is again on non-perturbative methods. But while
in the EAA framework one introduces a background metric
$\bar{g}_{ab}$ in the intermediate stages, splits the physical
metric $\tilde{g}_{\mu\nu}$ as $\tilde{g}_{\mu\nu}= \bar{g}_{\mu\nu} +
h_{\mu\nu}$ and interprets $\mathcal{D}{\tilde{g}}_{\mu\nu}$ as an
integration over the nonlinear fluctuations,
$\mathcal{D}\tilde{h}_{\mu\nu}$, the LQG framework is \emph{manifestly}
background independent. Quantum geometry, developed in the canonical
framework \cite{alrev,crbook,ttbook,PoS} provides well-defined
techniques to carry out path integrals directly in terms of
Spinfoams \cite{crbook,PoS,perez} which represent physical, quantum
spacetimes, without any split. 

In section \ref{s2.3} we discussed LQG dynamics via a Spinfoam model  
\cite{eprl,fk,ls,kkl}. This model has drawn a great deal of
attention because it represents a notable confluence of ideas from
apparently distinct directions: Canonical LQG \cite{alrev,ttbook},
Regge calculus \cite{regge1,regge2,regge3}, \index{Regge calculus} topological field \index{group field theory}
theories and group field theory \cite{crbook,gft1,bo}. The number of
simplices that feature in the underlying 4-geometries provides a
mathematically natural expansion parameter to calculate transition
amplitudes. These amplitudes are UV finite to any order in this
expansion \cite{eprl,fk,ls,kkl} and, in presence of a positive
cosmological constant, \index{cosmological constant}
there is an elegant procedure involving a
quantum deformation of $\SL(2,C)$ (the double cover of the local
Lorentz group) that provides a natural infrared regulator
\cite{Fairbairn:2010cp,Han:2011aa}. Finally, 
in \ref{s3.3} we 
summarized the construction of $n$-point functions in the semi-
classical limit of this model. 
The leading term in the
2-point function reproduces the low energy graviton propagator in a
precise sense \cite{Bianchi:2011hp,Bianchi:2009ri,Rovelli:2011kf}.
These developments have begun to create a bridge \cite{Han:2013tap}
from the background independent, non-perturbative framework of LQG
to effective field theories that encompass low energy scattering
processes in quantum gravity. 
Thus, LQG offers a well-defined set of fundamental equations  
describing quantum spacetime, free of UV and IR divergences in a 
natural expansion, with substantial evidence for GR to emerge in a 
suitable limit.
 
Over the past two decades, LQG has also been used to analyze
long-standing issues which originally constituted the main
motivation for quantum gravity. As in Asymptotic Safety, progress
has occurred by truncating the full theory appropriately and
analyzing the truncated sectors in detail. But now truncations are
motivated directly by each physical problem under consideration.
In the cosmological truncation, discussed in section \ref{s3.1}, not only are the strong curvature singularities naturally resolved by the quantum geometry effects \cite{mb,aps,apsv,awe,gowdy,asrev,ps}
but the standard paradigms have been extended all the way to the bounce by facing the Planck regime squarely, using quantum field theory on \emph{quantum} cosmological space-times \cite{akl,aan,madrid,french}. Furthermore there is a small window in the parameter space where the theory can be confronted with future observations. In section \ref{s3.2}, we summarized the current status of quantum black holes in LQG. 
\index{quantum black holes}
%
Using the notion of isolated horizons and techniques from quantum geometry one can treat all black hole and cosmological
horizons in one go, without having to restrict oneself to 
extremality \cite{abck,aev}. More recently, semi-classical
considerations have brought the LQG description closer to the more
familiar treatments in terms of energy and temperatures measured by
suitable families of near-horizon observers \cite{perez-bhs}.
Finally, 
there is now a novel approach to investigate the quantum evaporation process, using quantum field theory on \emph{quantum} space-times 
describing black holes in LQG \cite{gp}.
%
%
\subsection{Outlook}
\label{s4.2}

Every quantum gravity program faces two types of issues: i) those
which are \emph{internal} to any given program which must be
resolved before one has a conceptually complete, coherent theory
with the correct low energy limit in 4 spacetime dimensions; and,
ii) those which are \emph{common} to all programs, addressing the
long standing physical questions. As our summary illustrates,
concrete advances have occurred on both these fronts. However, a
number of important challenges remain. In particular, so far
\emph{none} of the approaches to quantum gravity satisfies the
`internal' criterion of completeness.

We will now illustrate these challenges and ensuing opportunities
through examples. Strategies summarized here are necessarily
provisional; our primary intent is only to provide a general idea of
the directions that are being currently pursued.

\textbf{$\bullet$ Infrared Issues:} 
In the EAA approach of the Asymptotic Safety program there exist trajectories admitting the non-trivial UV fixed points which are known 
to reduce to GR in the low energy limit. 
\index{UV and IR fixed points}
In the CDT approach, because of the usual limitations on the size of
lattices that can be handled in simulations, the smallest physical
length $a$ of the link is still about $2\lp$, and the infrared
regime corresponds to $\sim 20\lp$ \cite{agjl-rev}. 
The IR behavior is already illuminating in that not only does a 
classical de Sitter geometry with small quantum fluctuations arise in 
the Euclidean signature, but this occurs even for universes of radius
$\sim 20 \lp$!%
\footnote{Interestingly, the same scale arose completely
independently in LQC: in the $k=1$ Lorentzian FLRW cosmology, for
example, dynamics of the quantum wave functions is accurately
described by GR once the radius of the universe exceeds $8\lp$ even
in the case when the universe grows to a radius only of $23\lp$
before undergoing a recollapse a la GR \cite{apsv}.}
An important open issue is whether other physically interesting
spacetimes can be recovered from CDT.

In full LQG, as we saw in sections \ref{s2.3} and \ref{s3.3}, low energy limit is recovered in a certain well-defined sense. 
However, there is considerable room for improvement. In particular,
although the boundary states currently used are well-motivated,
being peaked on the metric as well as the extrinsic curvature of the
boundary induced by the Minkowski metric, 
there is still considerable ambiguity in their choice which descends to 
the transition amplitude and $n$-point functions. Conceptually, this is
not problem because both these quantities are, by definition,
functions of the boundary state.
But different boundary states would give rise to different
sub-leading terms, making comparison with the effective theory
ambiguous. A principle to select \emph{canonical} states
corresponding to Minkowski and de Sitter space-times is still lacking.
A second important limitation is that most of the results on
classical and semi-classical limits we summarized have been carried
out using only one simplex. There is substantial ongoing work that
considers refinements, allowing a large number of simplexes in the
interior, keeping the boundary state peaked on the classical
geometry of interest. These results will either firmly establish the
infrared viability of the specific Spinfoam model that is currently
used, or, suggest better alternatives.

\textbf{$\bullet$ Matter couplings:} In the cosmological truncation
of LQG, matter fields have been incorporated and their effect has
been analyzed in detail \cite{asrev,mbrev,barraurev}. In full
connection dynamics, \index{connection dynamics}
matter couplings have been discussed
exhaustively at the classical level \cite{art,aabook,ttbook} and the
framework has also been extended to incorporate supersymmetry
\cite{ttetal2}. However, at the quantum level, so far only formal
schemes have been laid out in the full theory \cite{ttbook,fermions}. Interestingly, one can arrive at a unification which is `dual' to the Kaluza-Klein
scheme in the following sense: One can continue with 4 spacetime
dimensions but enlarge the \emph{internal} group to a product of the
Lorentz group (associated with gravity) with groups associated with
Yang Mills fields governing other interactions \cite{peldan}.
Whether these ideas are fully compatible with particle physics
phenomenology is, however, still unclear. More generally, constructing 
a detailed quantum theory with matter coupling represents a challenging 
and fertile area for LQG research in coming years.

In the Asymptotic Safety program, by contrast, there is already very
substantial work on incorporating matter. It has provided
interesting constraints on the number of fermions and gauge fields
that can be accommodated within this scenario, constraints that are
satisfied by the standard model of particle physics \cite{percacci}.
Furthermore, insights on the quantum nature of geometry provided by
results to date are likely to have implications to the ultraviolet
issues in field theories of other interactions as well. Indeed,
there are already indications that the coupling to asymptotically
safe gravity might cure certain notorious problems in the matter
sector \cite{QEG+QED,higgs-triv}, 
and it is conceivable that the coupled system is more predictive than the standard model of particle physics without gravity. 
\index{QED} There are
for example scenarios in which the Higgs mass \cite{shaposhnikov} or
the fine structure constant \cite{QEG+QED} are computable
quantities. These promising ideas are likely to be more fully
developed in the coming decade. \index{QEG quantum Einstein gravity}\index{quantum Einstein gravity|see{QEG}}

\textbf{$\bullet$ New Physics:} In both approaches, there is a large
number of avenues that will be pursued to explore new physics. We
will present just a few illustrative examples.

In the Asymptotic Safety program one is naturally led to
an EAA-based `quantum geometry' of spacetime which goes beyond
Riemannian geometry in a specific sense: in general, the metric is
scale dependent. So a single (smooth) manifold is furnished not with
just one metric, but rather a family, $\{\langle g_{\mu\nu}
\rangle_k\,, \,\, 0\leq k< \infty\}$, where $\langle
g_{\mu\nu}\rangle_k$ is a solution of the effective field equation
following from $\Gamma_k$. This  general framework \cite{jan1} was
used, for instance, to demonstrate that under certain conditions the
EAA, while defined in the continuum, can give rise to a dynamically
generated minimum length scale. It was also used to analyze the
fractal-like properties of the `quantum spacetimes' which follow
from the EAA \cite{oliver1,oliver2}. In particular a running
spectral dimension has been defined and computed
\cite{oliverfrac,frankfrac}. One finds that there is a dimensional
reduction from 4 macroscopic to 2 microscopic dimensions.%
\footnote{It is interesting that this general phenomenon also occurs
in LQG, where the 4-dimensional spacetime continuum arises from
coarse graining of a 2 complex representing the evolution of the
fundamental quanta of geometry.} \index{quanta of geometry}
As a consequence, the graviton propagator is modified near the UV
fixed point \cite{oliver1,cosmo1}. These novel features have
interesting implications for the early universe and black holes
\cite{cosmo1,entropy,bh1,evap} which provide interesting avenues for
future research.


Applications of LQG discussed in section \ref{s3} also provide a
number of interesting directions to explore new
physics. 
First, as we discussed in section \ref{s3.1}, there is a small
window in the parameter space where LQC leads to new predictions
\cite{aan,observations,french,barraurev}. It needs to be analyzed in
much greater detail, keeping in mind the planned astronomical
surveys. While the a priori probability that this window is realized
in Nature is small, if the initial observations were to favor it, it
will be possible to use novel avenues to confront the theory with 
observations in detail, precisely because the window is small. On
the conceptual front, there are a number of issue concerning the
specification of initial conditions at the bounce. 
\index{quantum bounce} So far the focus
has been on establishing the \emph{existence} of initial conditions
that lead to a self-consistent extension of standard inflation to
the Planck regime. But the issue of uniqueness is quite open except
for some preliminary ideas involving a quantum extension of
Penrose's Weyl curvature hypothesis \cite{aan}. These will be
explored in detail in the coming years. If one uses inflation,
observations inform us that the entire observable universe should
originate from a ball of radius of less than $10\lp$ at the bounce.
But standard inflation does not explain why there was an
extraordinary homogeneity at this scale. The repulsive force of LQC
that dominates near the bounce provides a novel avenue to explore
this issue. LQC models that have been analyzed in detail indicate
that in the Planck regime the net effect of this repulsion is to
dilute the wrinkles in the curvature and forcing homogeneity and
isotropy at this scale. It is important to translate these physical
ideas into detailed calculations also because they imply that the
repulsive force would wash away the memory of the pre-bounce phase
as far as observations are concerned, making it natural to specify
initial conditions at the bounce. Finally, in the self-consistent
solutions, it has been possible to argue that while the LQG effects
are critical for the background FLRW quantum geometry, they can be
ignored for perturbations since the energy density in perturbations
is so small. It is important to carry out detailed calculations to
investigate new physics that may be emerge in more general
situations from a full LQG treatment of perturbations. \index{quantum 
bounce}

There are similar challenges and opportunities in the investigation
of quantum properties of black hole and cosmological horizons. While 
there is a detailed understanding of the microscopic quantum geometry 
of horizons in equilibrium, \cite{abck}, the relation between the 
number of these microstates and the more familiar semi-classical 
calculations of entropy \cite{semiclassical} via path integrals has 
begun to receive attention only recently. This is a key open
issue. More generally, the intriguing relation between the microscopic
geometry of quantum horizons and the semi-classical ideas
\cite{perez-bhs} discussed in section \ref{s3.2} remains to be
explored in detail. Finally, as we saw in section \ref{s3.2},
recently, a new window has been opened to investigate the Hawking
effect within LQG \cite{gp}. This important development offers many
opportunities for detailed calculations that will lead us to a
deeper understanding of the evaporation process. \index{black hole!evaporation}

\textbf{$\bullet$ Beyond Truncations:} Recall that in both
approaches discussed in this Chapter, concrete progress could be
made by studying the appropriate truncations of the full theory. As
emphasized towards the end of sections \ref{s2.3} and \ref{s3.1},
this is the common situation in fundamental physics: all the
concrete calculations in QED, QCD \index{QCD} \index{QED}
and scenarios of the early
universe, for example, involve truncations. Nonetheless, 
from the conceptual viewpoint, a central question remains: Is there an underlying coherent theory without reference to truncations that is being approximated in these calculations?

In the Asymptotic Safety program, a conceptual framework to address
this question is provided by the infinite dimensional theory space
$\mathcal{T}$. At a fundamental level, one should find the
renormalization group flows in $\mathcal{T}$ and then investigate
whether, in concrete physical problems, finite dimensional
truncations carried out to date provide a trustable approximation.
However, this lofty goal is far too ambitious for now. Progress is
likely to occur by further enlarging the reach of truncations. In
particular, a simplified version of the (particle physics) standard
model in which the gauge fields are assumed to be Abelian has
already been incorporated in the EAA program \cite{percacci}. An
important goal which may be within reach in the foreseeable future
would be to extend these calculations to include the full standard
model.

What is the situation with Spinfoams? Results to date have focused
on finite simplicial decomposition of the spacetime manifold. 
\index{simplicial decomposition}
The key open question is whether one should take a suitable limit by
successively refining the decomposition in a \emph{well-controlled
fashion}, or whether one should sum these contributions,
\emph{appropriately avoiding the obvious redundancy}. Does the final
transition amplitude remain finite in either case? In 3 spacetime
dimensions, the refinement does converge and yields the correct
result. Similarly, one can recast LQC in the Spinfoam framework and
show that the sum converges and yields a result that agrees with the
Hamiltonian theory \cite{ach2}. These calculations are helpful but
do not provide deep insight because these theories do not have local
degrees of freedom. Therefore currently there is a great deal of
activity in the full 4-dimensional theory. In particular,
generalized renormalization group flows are being studied by
Dittrich and others to constrain the refinement procedures and
investigate the phase diagrams that result, and group field theory
\index{group field theory}
is being used by Oriti and others to carry out the sum
systematically. Interestingly, the two procedures have quite
different conceptual underpinnings. In the first, the viewpoint is
more akin to that in the study of condensed matter systems using
statistical mechanics, where the atomic structure is fundamental and
phonon fields are convenient tools to encode collective behavior of
atoms. In LQG, the quanta of geometry play the role of atoms while
continuum quantum fields are the rough analogs of phonons. 
\index{quanta of geometry} In the
second approach, quantum fields are more fundamental as in particle
physics, and one uses well established methods with the goal of
summing a perturbative expansion. But quantum gravity introduces a
key difference: the quantum fields are now defined on a group
manifold rather than spacetime. It is fortunate that the central
issue of whether there is a coherent theory underlying Spinfoam
truncations is being analyzed from very different, if not opposing,
perspectives. Since this central issue is deep and difficult, it is
essential to have variety.\\

\textbf{Acknowledgments:} We thank Alejandro Corichi for permission to use Fig.4. This work was supported in part by the NSF grant
PHY-1205388 and the Eberly research funds of Penn state.

\end{document}